\documentclass[twocolumn]{aastex61}

\usepackage{amsmath,amsopn,amsxtra,txfonts}
\usepackage{comment} 
\usepackage{natbib} 
\defcitealias{2014ApJ...787..147F}{FWB14} 
\defcitealias{2003ApJ...597..948P}{PBV03}
\defcitealias{2012ApJ...749L...2Y}{YKY12}
\defcitealias{2013ApJ...778...58B}{BMS13}
\defcitealias{1996AJ....111.1156W}{WW96}
\defcitealias{2002ASPC..254..256W}{WVW02}
\defcitealias{2003ASSL..281..163W}{W03}

\newcommand{\ha}{H$\alpha$}
\newcommand{\hb}{H$\beta$}
\newcommand{\kms}{ \ifmmode{\rm km\thinspace s^{-1}}\else km\thinspace s$^{-1}$\fi}

\newcommand{\pc}{\ensuremath{ \, \mathrm{pc}}}
\newcommand{\kpc}{\ensuremath{\, \mathrm{kpc}}}
\newcommand{\cm}{\ensuremath{ \, \mathrm{cm}}}

\newcommand{\s}{\ensuremath{ \, \mathrm{s}}}
\newcommand{\yr}{\ensuremath{ \, \mathrm{yr}}}

\newcommand{\sr}{\ensuremath{ \, \mathrm{sr}}}

\newcommand{\dg}{\ifmmode{^{\circ}}\else $^{\circ}$\fi}

\newcommand{\hi}{H\textsc{~i}}
\newcommand{\hii}{H\textsc{~ii}}

\newcommand{\lb}{\ifmmode{(\ell,b)} \else $(\ell,b)$\fi}
\newcommand{\nii}{[N\textsc{~ii}]}
\newcommand{\niiblue}{[N\textsc{~ii}]$~\lambda5755$}
\newcommand{\sii}{[S\textsc{~ii}]}
\newcommand{\oiii}{[O\textsc{~iii}]}
\newcommand{\hei}{He\textsc{~i}}
\newcommand{\oi}{[O\textsc{~i}]}
\newcommand{\oii}{[O\textsc{~ii}]}

\newcommand{\vlsr}{\ifmmode{{\rm v}_{\rm{LSR}}}\else ${\rm v}_{\rm{LSR}}$\fi}
\newcommand{\av}{\ifmmode{A(V)}\else $A(V)$\fi}
\newcommand{\ebv}{\ifmmode{E(B-V)}\else $E(B-V)$\fi}
\newcommand{\iha}{\ifmmode{I_{\rm{H}\alpha}} \else $I_{\rm H \alpha}$\fi}
\newcommand{\ihb}{\ifmmode{I_{\rm{H}\beta}} \else $I_{\rm H \beta}$\fi}
\newcommand{\isii}{\ifmmode{I_{\ion{\rm{S}}{2}}} \else $I_{\rm [S \textsc{ ii}]}$\fi}
\newcommand{\inii}{\ifmmode{I_{\ion{\rm{n}}{2}}} \else $I_{\rm [N \textsc{ ii}]}$\fi}
\newcommand{\ioi}{\ifmmode{I_{\ion{\rm{o}}{1}}} \else $I_{\rm [O \textsc{ i}]}$\fi}

\newcommand{\mha}{\ensuremath{\mathrm{H} \alpha}}

\newcommand{\vgeo}{\ifmmode{v_{\mathrm{GEO}}} \else $v_{\mathrm{GEO}}$\fi}

\shorttitle{Ionized gas in the Magellanic Stream}
\shortauthors{Barger et al.}

\begin{document}

\author{K. A. Barger}

\affiliation{Department of Physics \& Astronomy, Texas Christian University, Fort Worth, TX 76129, USA}
\affiliation{Department of Physics, University of Notre Dame, Notre Dame, IN 46556 , USA}

\author{G. J. Madsen}
\affiliation{Institute of Astronomy, University of Cambridge, Madingley Road, Cambridge CB3 0HA, UK}

\author{A. J. Fox}
\affiliation{Space Telescope Science Institute, Baltimore, MD 21218, USA}

\author{B. P. Wakker}
\affiliation{Supported by NASA/NSF, affiliated with Department of Astronomy, University of Wisconsin-Madison, Madison, WI 53706, USA}

\author{J. Bland-Hawthorn}
\affiliation{Sydney Institute for Astronomy, School of Physics A28, University of Sydney, NSW 2006}

\author{D. Nidever}
\affiliation{National Optical Astronomy Observatory, 950 North Cherry Ave, Tucson, AZ, 85719, USA}

\author{L. M. Haffner}
\affiliation{Department of Astronomy, University of Wisconsin-Madison, Madison, WI 53706, USA}
\affiliation{Space Science Institute, Boulder, CO 80301, USA}

\author{Jacqueline Antwi-Danso}
\affiliation{Department of Physics \& Astronomy, Texas A\&M University, College Station, TX 77843, USA}
\affiliation{Department of Physics \& Astronomy, Texas Christian University, Fort Worth, TX 76129, USA}

\author{Michael Hernandez}
\affiliation{Department of Physics \& Astronomy, Texas Christian University, Fort Worth, TX 76129, USA}

\author{N. Lehner}
\affiliation{Department of Physics, University of Notre Dame, Notre Dame, IN 46556 , USA}

\author{A. S. Hill}
\affiliation{Departments of Physics and Astronomy, Haverford College, Haverford, PA 19041, USA}
\affiliation{CSIRO Astronomy \& Space Science, Marsfield, NSW, Australia, 1710}

\author{A. Curzons}
\affiliation{Sydney Institute for Astronomy, School of Physics A28, University of Sydney, NSW 2006}

\author{T. Tepper-Garc\'{i}a}
\affiliation{Sydney Institute for Astronomy, School of Physics A28, University of Sydney, NSW 2006}
\title{Revealing the Ionization Properties of the Magellanic Stream using Optical Emission}

\begin{abstract}

The Magellanic Stream, a gaseous tail that trails behind the Magellanic Clouds, could replenish the Milky Way with a tremendous amount of gas if it reaches the Galactic disk before it evaporates into the halo. To determine how the Magellanic Stream's properties change along its length, we have conducted an observational study of the \ha\ emission, along with other optical warm ionized gas tracers, toward $39$ sight lines. Using the Wisconsin \ha\ Mapper telescope, we detect \ha\ emission brighter than $30-50~{\rm mR}$ in $26$ of our $39$ sight lines. This \ha\ emission extends more than $2\arcdeg$ away from the \hi\ emission. By comparing  $I_{\rm H\alpha}$ and $I_{\rm [O\textsc{~i}]}$, we find that regions with $\log N_{\rm H\textsc{~i}}/\cm^{-2}\approx19.5-20.0$ are $16-67\%$ ionized. Most of the $I_{\rm H\alpha}$ along the Magellanic Stream are much higher than expected if the primary ionization source is photoionization from Magellanic Clouds, the Milky Way, and the extragalactic background. We find that the additional contribution from self ionization through a ``shock cascade'' that results as the Stream plows through the halo might be sufficient to reproduce the underlying level of \ha\ emission along the Stream. In the sparsely sampled region below the South Galactic Pole, there exists a subset of sight lines with uncharacteristically bright emission, which suggest that gas is being ionized further by an additional source that could be a linked to energetic processes associated with the Galactic center. 

\end{abstract}

\keywords{galaxies: Magellanic Clouds - galaxies: dwarf - Galaxy: evolution - Galaxy: halo - ISM: individual (Magellanic Stream)}

\section{Introduction}\label{section:intro}

Star formation in galaxies is regulated by their ability to accrete and retain gas. Chemical evolution models require the inflow of low-metallicity gas to explain the observed stellar abundance patterns of the Milky Way (MW; e.g., \citealt{2008ASPC..396..113C}). The star-formation rates of $L^*$ galaxies indicates that they will quickly exhaust their gas reservoirs making stars without external sources of gas that they accrete onto their disks \citep{2008ApJ...674..151E, 2008ApJ...682L..13H, 2012ARA&A..50..491P, 2014A&ARv..22...71S}. Our own galaxy will deplete its gas supplies in only $1-2~{\rm Gyrs}$ at its present star-formation rate \citep{1980ApJ...237..692L} of $1-3~{\rm M}_\odot~\yr^{-1}$ (e.g., \citealt{2010ApJ...710L..11R, 2011AJ....142..197C}). With the MW's long history of consistently forming stars (e.g., \citealt{2000A&A...358..869R}), this rapid gas depletion time suggests that our galaxy has sustained itself by acquiring gas from external sources. These sources include primordial material that is left over from the formation of the universe and material that is ripped from nearby dwarf galaxies, such as the tidal remnants of the Magellanic Clouds (MCs). However, the recent results of \citet{2011ApJ...736...84M} show that the MW's star-formation rate could be in decline and that the Galaxy is likely in the process of transitioning from the ``green valley'' to the ``red sequence''. Although numerous clouds surround the MW as detected in neutral (e.g., \citealt{2005A&A...440..775K}, \citealt{2007ApJ...670L.113W, 2008ApJ...672..298W} and \citealt{2012ApJ...758...44S}) and ionized gas (e.g., \citealt{2003ApJS..146..165S}, \citealt{2006ApJS..165..229F}, \citealt{2009ApJ...703.1832H}, \citealt{2011Sci...334..955L}, \citealt{2012ApJ...761..145B}, and \citealt{2012MNRAS.424.2896L}), it is unclear how much of that gas will reach the Galactic disk to provide replenishment. Interactions with high-temperature coronal gas in the halo combined with an ionizing radiation field from a number of sources may heat and ionize inflowing gas, hindering accretion (see \citealt{2012ARA&A..50..491P} for review). 

However, the MW's gas crisis might soon be over as it has recently captured the gas-rich MCs (e.g., \citealt{2007ApJ...668..949B}), which may provide enough gas to sustain or even boost its star formation (\citetalias{2014ApJ...787..147F}: \citealt{2014ApJ...787..147F}). Galaxy interactions have stripped more than $2\times10^9~{\rm M}_\odot$ of neutral and ionized gas from the Large and Small Magellanic Clouds (SMC \& LMC; e.g., \citealt{2005A&A...432...45B, 2007ApJ...670L.109B, 2013ApJ...771..132B, 2005ApJ...630..332F, 2014ApJ...787..147F}). This debris spans over $200\arcdeg$ on the sky \citep{2010ApJ...723.1618N} and is especially concentrated at $b<0\arcdeg$ \citep{1998Natur.394..752P}. The tidal structures known as the Magellanic Stream and Leading Arm could supply the MW with $\sim3-7~{M_{\odot}}~{\rm yr}^{-1}$ (\citetalias{2014ApJ...787..147F}). However, the exact amount of gas contained within these streams is uncertain due to weak constraints on their distance. Additionally, much of this material will likely evaporate into the Galactic halo before reaching the star-forming disk. Although the evaporated gas will build the halo, it could eventually condense, fall to the disk, and replenish the Milky Way's star-formation gas reservoir on longer timescales (e.g., \citealt{2012ApJ...745..148J}).

\begin{figure*}
\begin{center}
\includegraphics[clip,scale=0.90,angle=0]{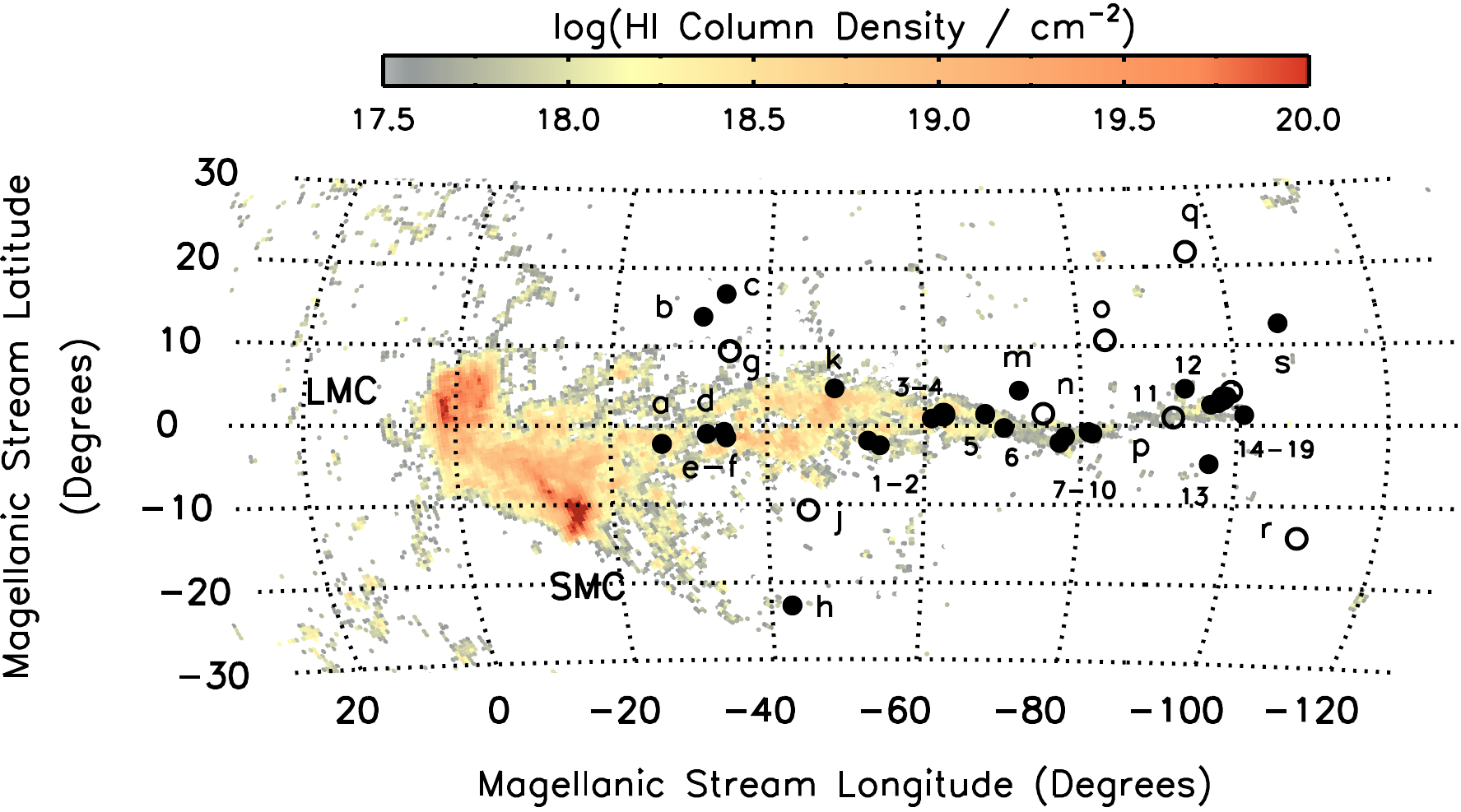}
\end{center}
\figcaption{Locations of pointed WHAM observations along the Magellanic Stream, where the letters and numbers coincide with the sight line identifiers listed in Table~\ref{table:obs}. Filled and open circles denote \ha\ detections and non-detections, respectively. The background H\textsc{~i} column density gas map of the Magellanic System is comprised of the Gaussian decompositions of the Leiden-Argentine-Bonn (LAB) all-sky survey adapted from \citet{2008ApJ...679..432N}.
\label{figure:hi_map}}
\end{figure*}

The trailing tidal debris, known as the Magellanic Stream, contains $\gtrsim10^8~{M_{\odot}}$ in \hi\ gas alone \citep{2005A&A...432...45B} and over three times that in ionized gas (\citetalias{2014ApJ...787..147F}). Its elongated structure substantially increases its surface area, which increases its exposure to the surrounding coronal gas and ionizing radiation field. Further, numerous cloudlets may have splintered off of or evaporated away from the two main filaments of the Stream (\citealt{2003ApJ...586..170P, 2014ApJ...792...43F}; see Figure~\ref{figure:hi_map}), making them even more susceptible to their environment. \citet{2007ApJ...670L.109B} and \citet{2009ApJ...698.1485H} found that clouds with \hi\ masses of less than $10^{4.5}~{\rm M}_\odot$ may become fully ionized through Kelvin-Helmholtz instabilities within $\sim100~{\rm Myr}$ due to interactions with surrounding coronal gas. Although the Stream is quite massive in its entirety, the morphology of the neutral gas is fragmented and the numerous, small offset clouds (\citealt{2008MNRAS.388L..29W} and \citealt{2008ApJ...680..276S}) suggest that it is evaporating.  \citet{2012ApJ...760...48N} examined the \hi\ morphology of one of the small clouds that fragmented off of the Magellanic Stream and found that it has a cold and dense core that is enclosed within a warm extended envelope. They illustrate that its diffuse skin was likely a result of turbulence that was generated by the cloud rubbing against the surrounding coronal gas and not by conductive heating. Additionally, clouds are eroded away by the ionizing radiation field of the MW, nearby galaxies, and extragalactic background (EGB). Determining the properties of both the neutral and ionized gas phases is vital for ascertaining how the Magellanic Stream is affected by its environment. 

The strength of the incident ionizing radiation field and of the ram-pressure stripping effects from the surrounding coronal gas on the Magellanic Stream strongly depend on its position relative to nearby galaxies and its location within the Galactic halo. Its distance is only known where the Magellanic Stream originates at the MCs. Observationally, \citet{2012MNRAS.424.2896L} found that the trailing end of the Stream must lie further than $20~\kpc$ away due to the lack of absorption seen toward stellar targets and from the abundance of absorption detected toward active galactic nuclei (AGN) in the same vicinity \citep{2011Sci...334..955L}. Because the MCs are just past perigalacticon in their orbits (see \citealt{2016ARA&A..54..363D} for a review) and because the Magellanic Stream is a \textit{trailing} Stream, the closest possible distance to the trailing gas is the present distances of the MCs (i.e., $d\gtrsim50~\kpc$). Galaxy interaction models of this system predict that the trailing gas in the Stream could lie as far as $100-200~\kpc$ from the Galactic center (e.g., \citealt{2010ApJ...721L..97B, 2012MNRAS.421.2109B}, \citealt{2011MNRAS.413.2015D, 2012ApJ...750...36D}, and \citealt{2014MNRAS.444.1759G}). The most recent models predict that these galaxies are on a highly elliptical orbit when accounting for (i) an increase in the measured proper motions from Hubble Space Telescope observations of the LMC and SMC by a third \citep{2006ApJ...638..772K} and (ii) decreases in the total mass estimates of the Galactic halo (e.g., \citealt{2012ApJ...761...98K, 2017MNRAS.465...76M}). More work is still needed to produce a physically consistent Stream as only the \citet{2014MNRAS.444.1759G} simulation includes both the neutral and ionized gas contained within the Stream, the existing models exclude either the effect of ram-pressure stripping or photoionization, and only \citet{2010ApJ...721L..97B, 2012MNRAS.421.2109B} includes cooling (radiative only). However, it is also important to note that the further the Magellanic Stream lies from the MW, the less it will be influenced by ram-pressure stripping from the MW's coronal halo gas and its ionizing radiation field. 

Numerous studies have detected ionized gas in the Magellanic Stream along individual lines of sight, including through \ha\ emission (WW96: \citealt{1996AJ....111.1156W}; \citetalias{2002ASPC..254..256W}; WVW02: \citealt{2002ASPC..254..256W}, W03: \citetalias{2003ASSL..281..163W}: \citealt{2003ASSL..281..163W}, \citetalias{2003ApJ...597..948P}: \citealt{2003ApJ...597..948P}, \citetalias{2012ApJ...749L...2Y}: \citealt{2012ApJ...749L...2Y}, and \citetalias{2013ApJ...778...58B}: \citealt{2013ApJ...778...58B}) and UV absorption (e.g., \citealt{2003ApJS..146..165S} , \citealt{2012MNRAS.424.2896L}, \citealt{2013ApJ...772..111R}, and \citealt{2005ApJ...630..332F, 2010ApJ...718.1046F, 2014ApJ...787..147F}). Because some of these \ha\ detections are brighter than expected if the primary source of their ionization is the MW's ionizing radiation field and the EGB, other ionization sources are also thought to contribute. However, previous studies have neglected the ionizing contribution from the MCs themselves, which we account for in this study. 

We describe our multiline observations of the warm ionized gas emission of the Magellanic Stream in Section~\ref{section:obs} and their reduction in Section~\ref{section:reduction}. We compare the neutral gas as traced by \hi\ emission with the ionized gas in Section~\ref{section:compare} by examining their kinematics, emission strengths, distribution, and ionization fraction. We then discuss how well the observed \ha\ emission of the Stream can be reproduced by photoionization alone from the surrounding galaxies (MW, LMC, and SMC) and the EGB, photoionization plus collisional ionization from interactions with the Galactic halo and with the Stream itself, and energetic processes that are associated with the Galactic center in Section~\ref{section:ionization}. We finally summarize our major conclusions in Section~\ref{section:summary}.

\section{Observations}\label{section:obs}

Our $39$ pointed observations of the warm ionized gas ($\sim10^4~{\rm K}$) in the Magellanic Stream span over $100\arcdeg$ along its length. To assess the extent of the ionized gas, many of these observations lie off the \hi\ emission ($\log{N_{\rm H\textsc{~i}}/\cm^{-2}}\lesssim 18$), as illustrated in Figure \ref{figure:hi_map}. We preferentially aligned the observations at positions (a--s; see Figure~\ref{figure:hi_map} and Table~\ref{table:obs}) toward bright background quasars with existing absorption-line observations. In the complementary \citetalias{2014ApJ...787..147F} study, we combined the aligned emission- and absorption-line observations to further explore how ionization conditions of the Stream vary along its length. We aligned the observations at positions (1--19; see Figure~\ref{figure:hi_map} and Table~\ref{table:obs}) with the main \hi\ filamentary structures of the Magellanic Stream.

\startlongtable
\begin{deluxetable*}{cccrrcrrrccl}
\tabletypesize{\scriptsize}
\tablecaption{WHAM Spectroscopic Observations \label{table:obs}}
\tablewidth{0pt}
\tablehead{
\colhead{ } &\colhead{ } &\colhead{ } & \multicolumn{3}{c}{On~Target} & & \multicolumn{3}{c}{Off~Target} & \colhead{} & \colhead{Notes\tablenotemark{a}} \\
\cline{4-6} \cline{8-11}
\colhead{ID} & \colhead{Filter} & \colhead{\vlsr\tablenotemark{b}} & \colhead{$l_{\rm MS},~b_{\rm MS}$\tablenotemark{c}} & \colhead{$l,~b$} &
\colhead{Exposure Time} & \colhead{ } &
\colhead{$l_{\rm MS},~b_{\rm MS}$\tablenotemark{c}} & \colhead{$l,~b$} &
\colhead{Exposure Time} \\
\colhead{} & \colhead{} & \colhead{($\kms$)} & \colhead{(degrees)} & \colhead{(degrees)} &
\colhead{(s)} & \colhead{ } &
\colhead{(degrees)} & \colhead{(degress)} &
\colhead{(s)}
}
\startdata
a & \ha & 	$+80, +280$ & $-26.6$, $-02.3$             	& $295.1$, $-57.8$  & $10\times60$ && $-24.2$, $\phantom{-}00.8$ & $288.5$, $-56.4$ & $12\times60$ && FAIRALL~9	\\	
b & \ha & 	$+45, +245$ & $ -31.5$, $ \phantom{-}13.8$ 	& $262.3$, $-63.9$  & $18\times60$ & & $-33.8$, $\phantom{-}16.7$ & $256.8$, $-02.5$ & $18\times60$ && Near HE0226-4110 \\
c & \ha & 	$+45, +245$ & $ -34.3$, $ \phantom{-}16.8$ 	& $253.9$, $-65.8$  & $15\times60$ & &$-32.7$, $-08.6$ & $311.3$, $-60.7$ & $18\times60$ && HE0226-4110	\\
d & \ha & 	$+5, +205$ & $-32.3$, $-01.0$             	& $296.4$, $-63.7$  & $15\times60$ & & $-32.7$, $-08.6$ & $311.3$, $-60.7$ & $18\times60$ && Near RBS~144 \\
e & \ha & 	$+5, +205$	& $-34.8$, $-01.5$             	& $299.5$, $-65.8$  & $15\times60$ & & $-32.7$, $-08.6$ & $311.3$, $-60.7$ & $18\times60$ && RBS~144	\\
f & \ha & 	$+5, +205$ & $-34.5$, $-00.8$             	& $297.5$, $-65.8$  & $16\times60$ & & $-35.1$, $-02.3$ & $301.5$, $-65.8$ & $16\times60$ && Near RBS~144  	\\
g & \ha & 	$+150, +350$ & $ -35.1$, $ \phantom{-}09.5$ 	& $271.8$, $-68.9$  & $12\times60$ & & $-32.8$, $\phantom{-}13.9$ & $261.4$, $-65.1$ & $14\times60$ && HE0153-4520       	\\
h & \ha & 	$+55, +255$	& $-42.8$, $-23.0$             	& $343.9$, $-56.4$  & $16\times60$ & & $-46.6$, $-20.2$ & $347.9$, $-60.4$ & $15\times60$ && Near RBS~1892	\\
j & \ha & 	$+100, +300$ & $-45.2$, $-10.7$             	& $334.9$, $-68.2$  & $15\times30$ & & $-45.7$, $-08.7$ & $332.9$, $-70.2$ & $16\times30$ && ES0292-G24 \\
k & \ha & 	$+45, +245$ & $-48.6$, $\phantom{-}04.7$   	& $295.7$, $-80.9$  & $16\times30$ & & $-48.5$, $\phantom{-}05.0$ & $293.7$, $-80.9$ &  $15\times30$&& Near HE0056-3622 \\
m & \ha & 	$-245, -45$ & $-72.0$, $\phantom{-}04.4$     & $87.5$, $-75.0$ & $12\times60$ & & $119.9$, $-05.3$ & $93.6$, $-72.4$ & $12\times60$	&& Near PHL2525 	\\
n & \ha & 	$-245, -45$ & $-75.1$, $\phantom{-}01.5$      & $80.7$, $-71.2$ & $10\times60$ & & $119.9$, $-05.3$ & $93.6$, $-72.4$ & $12\times60$ && PHL2525 	\\
o & \ha & 	$-245, -45$ & $ -83.4$, $ \phantom{-}10.8$ 	& $107.7$, $-64.0$ & $13\times60$ & & $-80.9$, $\phantom{-}09.1$ & $104.1$, $-66.6$ & $14\times60$ && UM239 \\
p & \ha & 	$-315, -115$  & $ -90.8$, $ -00.7$           		& $ 86.2$, $-55.6$  & $17\times60$ & & $-91.3$, $\phantom{-}01.5$ & $90.2$, $-55.6$ & $17\times60$ && Near NGC7714	\\
q & \ha & 	$-295, -95$ & $ -95.9$, $ \phantom{-}21.8$ 	& $123.8$, $-50.2$  & $15\times60$ & & $-99.8$, $\phantom{-}19.3$ & $118.8$, $-47.1$ & $15\times60$ && MRK~1502 \\
r & \ha & 	$-295, -95$ & $-109.2$, $-13.9$            		& $ 76.0$, $-34.2$ & $15\times60$ & & $-109.7$, $-12.3$ & $77.8$, $-34.2$ & $16\times60$ && MRK~304 \\
s & \ha & 	$-430, -230$ & $-106.4$, $ \phantom{-}12.6$ 	& $108.8$, $-41.4$ & $15\times60$ && $-104.9$, $\phantom{-}09.2$ & $104.1$, $-42.9$ & $16\times60$ && MRK~335 \\
1 & \ha &	$-184, +14$ & $-52.8$, $-01.9$ & $342.6, -79.6$ & $19\times120$ &  & $-54.6$, $-06.4$ & $357.4$, $-75.8$ & $20\times120$ && \citetalias{2013ApJ...778...58B} \\
2 & \ha &	$-184, +14$ & $-54.3$, $-02.5$ & $351.7, -79.5$ &  $7\times120$ &  & $-55.8$, $-07.2$ & $2.7$, $-75.2$ &  $7\times120$ && \citetalias{2013ApJ...778...58B} \\
3 & \ha &	$-203, -3$ & $-60.9$,  $\phantom{-}00.9$ &  $37.5, -82.5$ &  $4\times300$ &  & $-63.5$, $-03.9$ & $37.0$, $-77.0$ &  $5\times300$ && \citetalias{2013ApJ...778...58B} \\ 
4a & \ha &	$-221, -21$ & $-62.4$,  $\phantom{-}01.4$ &  $48.6, -82.0$ &  $2\times300$ &  & $-63.1$, $\phantom{-}12.9$ & $142.4$, $-82.2$ &  $3\times300$ && \citetalias{2013ApJ...778...58B} \\
4b & \ha &	$-201, -1$ & $-62.5$,  $\phantom{-}01.4$ &  $49.0, -82.0$ &  $3\times300$ &  & $-67.9$,  $\phantom{-}04.3$ & $82.0$, $-79.0$ &  $1\times300$ && \citetalias{2013ApJ...778...58B} \\
5a & \ha & 	$-200, 0$ & $-67.7$,  $\phantom{-}01.5$ &  $68.9, -78.0$ &  $3\times300$ &  & $-67.9$, $\phantom{-}04.3$ &  $82.0$, $-79.0$ &  $6\times300$ && \citetalias{2013ApJ...778...58B} \\
5b & \ha &	$-220, -20$ & $-67.7$,  $\phantom{-}01.4$ &  $68.7, -78.0$ &  $2\times300$ &  & $-63.1$, $\phantom{-}12.9$ & $142.4$, $-82.2$ &  $3\times300$ && \citetalias{2013ApJ...778...58B} \\
6 & \ha &	$-202, -2$ & $-70.1$, $-00.3$ &  $67.5, -75.0$ &  $6\times300$ &  & $-70.6$, $-04.0$ &  $58.0$, $-72.5$ &  $6\times300$ && \citetalias{2013ApJ...778...58B} \\
7a & \ha & 	$-373, -173$& $-77.2$, $-02.1$ & $73.5, -68.0$ &  $4\times300$ &  & $-77.0$, $-05.3$ &  $66.0$, $-66.5$ &  $4\times300$ && \citetalias{2013ApJ...778...58B} \\
7b & \ha & 	$-305, -105$	& $-77.2$, $-02.2$ & $73.3, -67.9$ &  $7\times120$ &  & $-81.0$, $-03.4$ &  $74.5$, $-64.0$ & $13\times120$ && \citetalias{2013ApJ...778...58B} \\ 
8 & \ha & 	$-306, -106$	& $-77.9$, $-01.4$ & $75.8, -67.6$ &  $7\times120$ &  & $-81.0$, $-03.4$ &  $74.5$, $-64.0$ & $13\times120$ && \citetalias{2013ApJ...778...58B} \\
9 & \ha & 	$-372, -172$ & $-80.9$, $-00.8$ & $80.0, -65.0$ &  $3\times300$ &  & $-67.9$, $\phantom{-}04.3$ &  $82.0$, $-79.0$ &  $3\times300$ && \citetalias{2013ApJ...778...58B} \\
10 & \ha & 	$-373, -172$ & $-81.4$, $-01.0$ & $80.0, -64.5$ &  $4\times300$ &  & $-80.2$, $\phantom{-}01.8$ &  $85.5$, $-66.5$ &  $4\times300$ && \citetalias{2013ApJ...778...58B} \\
11 & \ha & 	$-441, -241$ & $-91.8$, $\phantom{-}01.0$ & $89.5, -55.0$ & $10\times300$ &  & $-91.4$, $-02.7$ &  $83.0$, $-54.5$ &  $7\times300$ && \citetalias{2013ApJ...778...58B} \\
12 & \ha & 	$-440, -240$	& $-93.4$, $\phantom{-}04.6$ & $96.0, -54.0$ & $10\times300$ &  & $-91.0$,  $\phantom{-}06.2$ &  $98.5$, $-56.5$ & $10\times300$ && \citetalias{2013ApJ...778...58B} \\
13 & \ha & 	$-441, -241$ & $-96.5$, $-04.8$ & $82.3, -49.0$ &  $9\times300$ &  & $-91.4$, $-02.7$ &  $83.0$, $-54.5$ &  $7\times300$ && \citetalias{2013ApJ...778...58B} \\
14 & \ha & 	$-439, -239$ & $-96.7$, $\phantom{-}02.6$ & $93.5, -50.5$ &  $8\times300$ &  & $-100.9$, $\phantom{-}04.1$ &  $96.5$, $-46.5$ &  $8\times300$ && \citetalias{2013ApJ...778...58B} \\
15 & \ha & 	$-482, -282$ & $-97.7$, $\phantom{-}02.8$ & $94.0, -49.5$ &  $4\times300$ &  & $-100.7$, $\phantom{-}12.0$ & $108.0$, $-47.0$ &  $8\times300$ && \citetalias{2013ApJ...778...58B} \\
16 & \ha & 	$-482, -282$ & $-98.3$, $\phantom{-}03.4$ & $95.0, -49.0$ &  $4\times300$ &  & $-100.7$, $\phantom{-}12.0$ & $108.0$, $-47.0$ & $16\times300$ && \citetalias{2013ApJ...778...58B} \\
17 & \ha & 	$-481, -281$ & $-98.8$, $\phantom{-}03.6$ & $95.5, -48.5$ &  $4\times300$ &  & $-100.7$, $\phantom{-}12.0$ & $108.0$, $-47.0$ & $16\times300$ && \citetalias{2013ApJ...778...58B} \\
18 & \ha & 	$-481, -281$	& $-99.4$, $\phantom{-}04.2$ & $96.5, -48.0$ &  $4\times300$ &  & $-100.7$, $\phantom{-}12.0$ & $108.0$, $-47.0$ & $16\times300$ && \citetalias{2013ApJ...778...58B} \\
19 & \ha & 	$-438, -238$ & $-101.0$, $\phantom{-}01.3$ & $92.5, -46.0$ &  $8\times300$ &  & $-100.9$, $\phantom{-}04.1$ &  $96.5$, $-46.5$ &  $8\times300$ && \citetalias{2013ApJ...778...58B} \\  
\hline
a & \hb & $+80, +280$ & $-26.6$, $-02.3$ & $295.1$, $-57.8$  & $10\times60$ && $-24.3$, $\phantom{-}04.3$ & $282.2$, $-57.1$ & $11\times60$ && FAIRALL~9	\\
e & \hb & $+5, +205$	 & $-34.8$, $-01.5$ & $299.5$, $-65.8$  & $23\times60$ & & $-32.7$, $-08.6$ & $311.3$, $-60.7$ & $23\times60$	&& RBS~144	\\
\hline
a & \sii$~\lambda6716$ &	$+80, +280$ & $-26.6$, $-02.3$ & $295.1$, $-57.8$  & $15\times60$ && $-26.2$, $-01.3$ & $293.1$, $-57.8$ & $16\times60$ && FAIRALL~9	\\
e & \sii$~\lambda6716$ &	 $+5, +205$ & $-34.8$, $-01.5$ & $299.5$, $-65.8$  & $5\times60$ & & $-32.7$, $-08.6$ & $311.3$, $-60.7$ & $6\times60$	&& RBS~144	\\
1 & \sii$~\lambda6716$ &	 $-184, +14$  & $-52.8$, $-01.9$ & $342.6, -79.6$ & $14\times120$ &  & $-54.6$, $-06.4$ & $357.4$, $-75.8$ & $14\times120$ && \citetalias{2013ApJ...778...58B} \\
2 & \sii$~\lambda6716$ & $-184, +14$ & $-54.3$, $-02.5$ & $351.7, -79.5$ & $11\times120$ &  & $-55.8$, $-07.2$ & $2.7$, $-75.2$ &  $4\times120$ && \citetalias{2013ApJ...778...58B} \\
3 & \sii$~\lambda6716$ &	$-175, +25$ & $-60.9$, $\phantom{-}00.9$ &  $37.5, -82.5$ &  $6\times300$ &  & $-63.5$, $-03.9$ &  $37.0$, $-77.0$ &  $6\times300$ && \citetalias{2013ApJ...778...58B} \\
\hline
a & \nii$~\lambda6583$ &	$+80, +280$ & $-26.6$, $-02.3$ & $295.1$, $-57.8$  & $15\times60$ && $-26.2$, $-01.3$ & $293.1$, $-57.8$ & $16\times60$ && FAIRALL~9	\\
e & \nii$~\lambda6583$ &	 $+5, +205$ & $-34.8$, $-01.5$ & $299.5$, $-65.8$  & $15\times60$ & & $-32.7$, $-08.6$ & $311.3$, $-60.7$ & $18\times60$	&& RBS~144	\\
1 & \nii$~\lambda6583$ &	$-184$ to  $+14$  & $-52.8$, $-01.9$ & $342.6, -79.6$ & $14\times120$ &  & $-54.6$, $-06.4$ & $357.3$, $-75.8$ & $14\times120$ && \citetalias{2013ApJ...778...58B} \\
2 & \nii$~\lambda6583$ & $-184, +14$ & $-54.3$, $-02.5$ & $351.7, -79.5$ &  $5\times120$ &  & $-55.8$, $-07.2$ & $2.7$, $-75.2$ &  $5\times120$ && \citetalias{2013ApJ...778...58B} \\ 
3 & \nii$~\lambda6583$ & $-175, +25$ & $-60.9$,  $\phantom{-}00.9$ &  $37.5, -82.5$ &  $8\times300$ &  & $-63.5$, $-03.9$ &  $37.0$, $-77.0$ &  $8\times300$ && \citetalias{2013ApJ...778...58B} \\
\hline
a & \niiblue & $+80, +280$ & $-26.6$, $-02.3$ & $295.1$, $-57.8$  & $11\times60$ && $-25.0$, $\phantom{-}00.6$ & $289.2$, $-57.1$ & $12\times60$ && FAIRALL~9	\\
e & \niiblue &  $+5, +205$ & $-34.8$, $-01.5$ & $299.5$, $-65.8$  & $15\times60$ & & $-32.7$, $-08.6$ & $311.3$, $-60.7$ & $16\times60$	&& RBS~144 \\
\hline
a & \oi$~\lambda6300$ &	$+80, +280$ & $-26.6$, $-02.3$ & $295.1$, $-57.8$  & $11\times60$ && $-25.0$, $\phantom{-}00.6$ & $289.2$, $-57.1$ & $12\times60$ && FAIRALL~9	\\
e & \oi$~\lambda6300$ &	 $+5, +205$ & $-34.8$, $-01.5$ & $299.5$, $-65.8$  & $14\times60$ & & $-32.7$, $-08.6$ & $311.3$, $-60.7$ & $16\times60$	&& RBS~144	\\
1 & \oi$~\lambda6300$ &	$-184$ to  $+14$  & $-52.8$, $-01.9$ & $342.6, -79.6$ & $25\times120$ &  & $-54.6$, $-06.4$ & $357.4$, $-75.8$ & $25\times120$ && \citetalias{2013ApJ...778...58B} \\
\hline
a & \oii$~\lambda7320$ & $+80, +280$ & $-26.6$, $-02.3$ & $295.1$, $-57.8$  & $11\times60$ && $-25.0$, $\phantom{-}00.6$ & $289.2$, $-57.1$ & $12\times60$ && FAIRALL~9	\\
e & \oii$~\lambda7320$ &	 $+5, +205$ & $-34.8$, $-01.5$ & $299.5$, $-65.8$  & $15\times60$ & & $-32.7$, $-08.6$ & $311.3$, $-60.7$ & $16\times60$	&& RBS~144	\\
1 & \oii$~\lambda7320$ & $-184$ to  $+14$ & $-52.8$, $-01.9$ & $342.6, -79.6$ & $12\times120$ &  & $-54.6$, $-06.4$ & $357.4$, $-75.8$ & $12\times120$ && \citetalias{2013ApJ...778...58B} \\
\hline
a & \oiii$~\lambda5007$ &  $+5, +205$ & $-26.6$, $-02.3$ & $295.1$, $-57.8$  & $11\times60$ && $-25.0$, $\phantom{-}00.6$ & $289.2$, $-57.1$ & $12\times60$ && FAIRALL~9	\\
e & \oiii$~\lambda5007$ &  $+5, +205$ & $-34.8$, $-01.5$ & $299.5$, $-65.8$  & $15\times60$ & & $-32.7$, $-08.6$ & $311.3$, $-60.7$ & $16\times60$	&& RBS~144	\\
\hline
a & \hei$~\lambda5876$ & $+80, +280$ & $-26.6$, $-02.3$ & $295.1$, $-57.8$  & $10\times60$ && $-25.0$, $\phantom{-}00.6$ & $289.2$, $-57.1$ & $11\times60$ && FAIRALL~9	\\
e & \hei$~\lambda5876$ &  $+5, +205$ & $-34.8$, $-01.5$ & $299.5$, $-65.8$  & $15\times60$ & & $-32.7$, $-08.6$ & $311.3$, $-60.7$ & $16\times60$	&& RBS~144	\\
\enddata 
\tablecomments{Only the sight lines with $[\rm S\textsc{~ii}]$ and $[\rm N\textsc{~ii}]$ in the (a--s) set are listed for brevity. Although all of these sight lines were observed at these wavelengths for $10-15~{\rm minutes}$ each, no emission was detected above the $\sim30~{\rm mR}$ sensitivity of the observations.}
\tablenotetext{a}{Lists either the background object that aligns with the sight line or the reference of the WHAM observation.} 
\tablenotetext{b}{Observed velocity range.} 
\tablenotetext{c}{Magellanic Stream Coordinate system defined in \citet{2008ApJ...679..432N}, which positions $l_{\rm MS}=0\arcdeg$ at the center of the LMC at $(l,b)=(280\fdg47,-32\fdg75)$ and $b_{\rm MS}=0\arcdeg$ along the center of the Magellanic Stream.} 
\end{deluxetable*}       

\begin{figure*}
\begin{center}
\includegraphics[scale=0.45,angle=0]{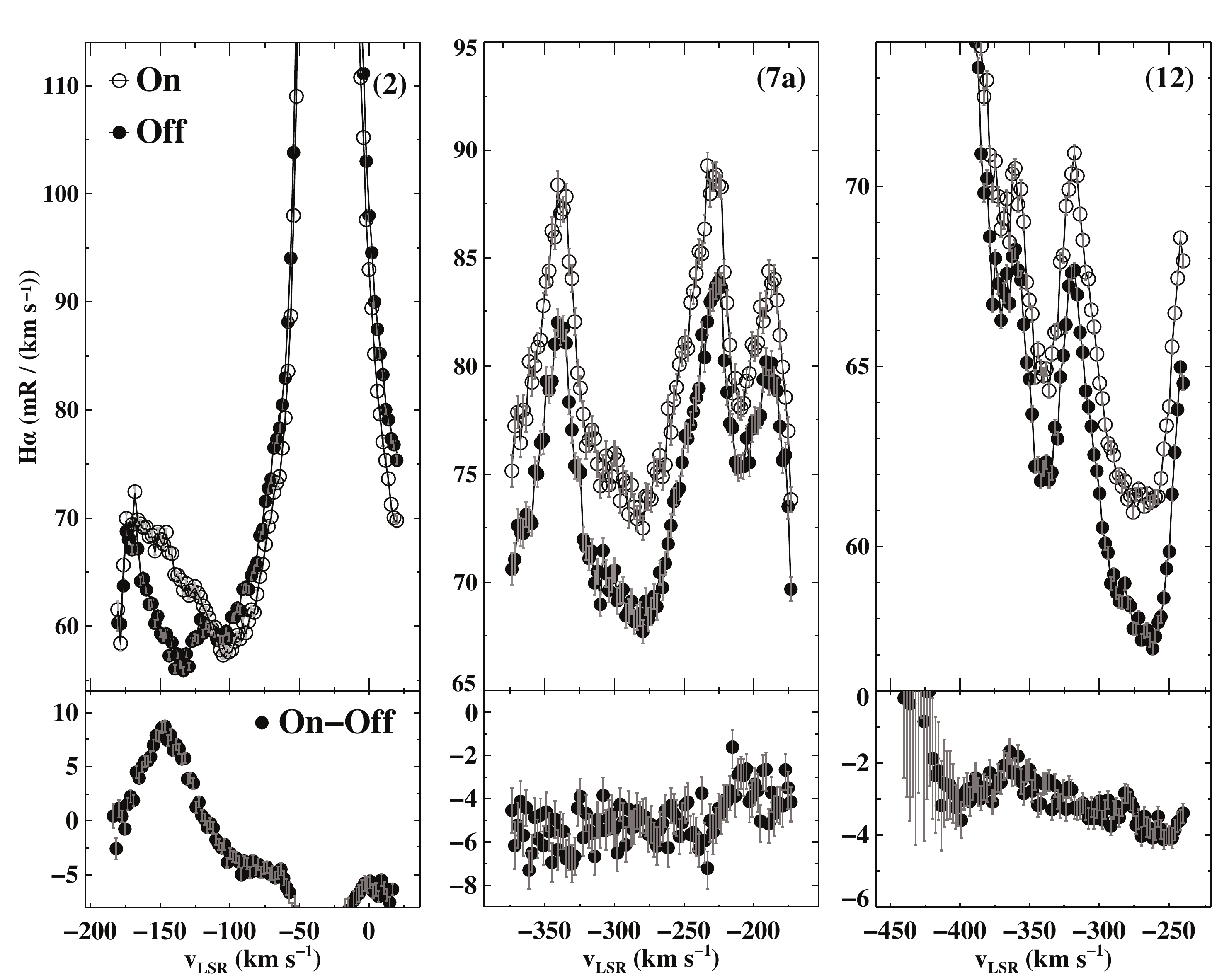}
\end{center}
\figcaption{Sample on- and off-target WHAM spectra. The labels in the top-right corners coincide with the IDs used in Figure~\ref{figure:hi_map} and Table~\ref{table:obs}. The top half of the panels show an average of the on-target spectra as open circles and the average of the off-target spectra as filled circles, where the on and off observations were spaced $4\fdg9$, $3\fdg2$, and $2\fdg9$ apart for the~(2), (7a), and~(12) sight lines, respectively. The bottom half of each panel shows the subtraction of the off-target spectra from the on-target spectra. Some of the atmospheric lines are many times brighter than the weaker emission from the Stream, such as the \ha\ geocoronal line located at $\vlsr\approx-25~\kms$ for sight line~(2).
\label{figure:on_off}}
\end{figure*}

The detection of diffuse optical-emission lines from the Stream, or from other high-velocity clouds (HVCs), requires very high surface brightness sensitivity and sufficient spectral resolution to differentiate it  from local Galactic emission. The WHAM telescope is optimized to detect faint emission from diffuse, ionized sources with its high-throughput, $15~\cm$ diameter, dual-etalon Fabry-P{\'e}rot spectrometer that is coupled to a $0.6~{\rm m}$ objective lens \citep{1998PASA...15...14R}. We achieve a sensitivity of $\sim30~{\rm mR}$,\footnote{$1~\rm{Rayleigh} = 10^6 / 4 \pi\,\textrm{photons}\, \cm^{-2}\, \sr^{-1}\, \s^{-1}$, which is $\sim2.41\times10^{-7}\,{\rm erg} \cm^{-2}\s^{-1}\sr^{-1}$ at \ha.} assuming a $30~\kms$ line width. Each individual exposure produces one spatially averaged spectrum that is $200~\kms$ wide with a spectral resolution of $12~\kms$ (${\rm R}\approx 25,000$) within the telescope's $1\arcdeg$ beam. The ${\rm S F_6}$ pressure-controlled etalons and interference filters enable the spectra to be centered at any wavelength between $4800~\AA$ and $7300~\AA$. The WHAM telescope and its capabilities are described further in \citet{2003ApJS..149..405H}. Despite the large declination range, all of our observations were collected with the same facility. Positions (1)--(19) were targeted while WHAM was located at Kitt Peak National Observatory (KPNO; 1997--2007). It was then relocated to Cerro Tololo Inter-American Observatory (CTIO; 2009--present). From this current site, we observed the southern-hemisphere targets, (a)--(s).

\begin{figure}
\begin{center}
\includegraphics[scale=0.5,angle=0]{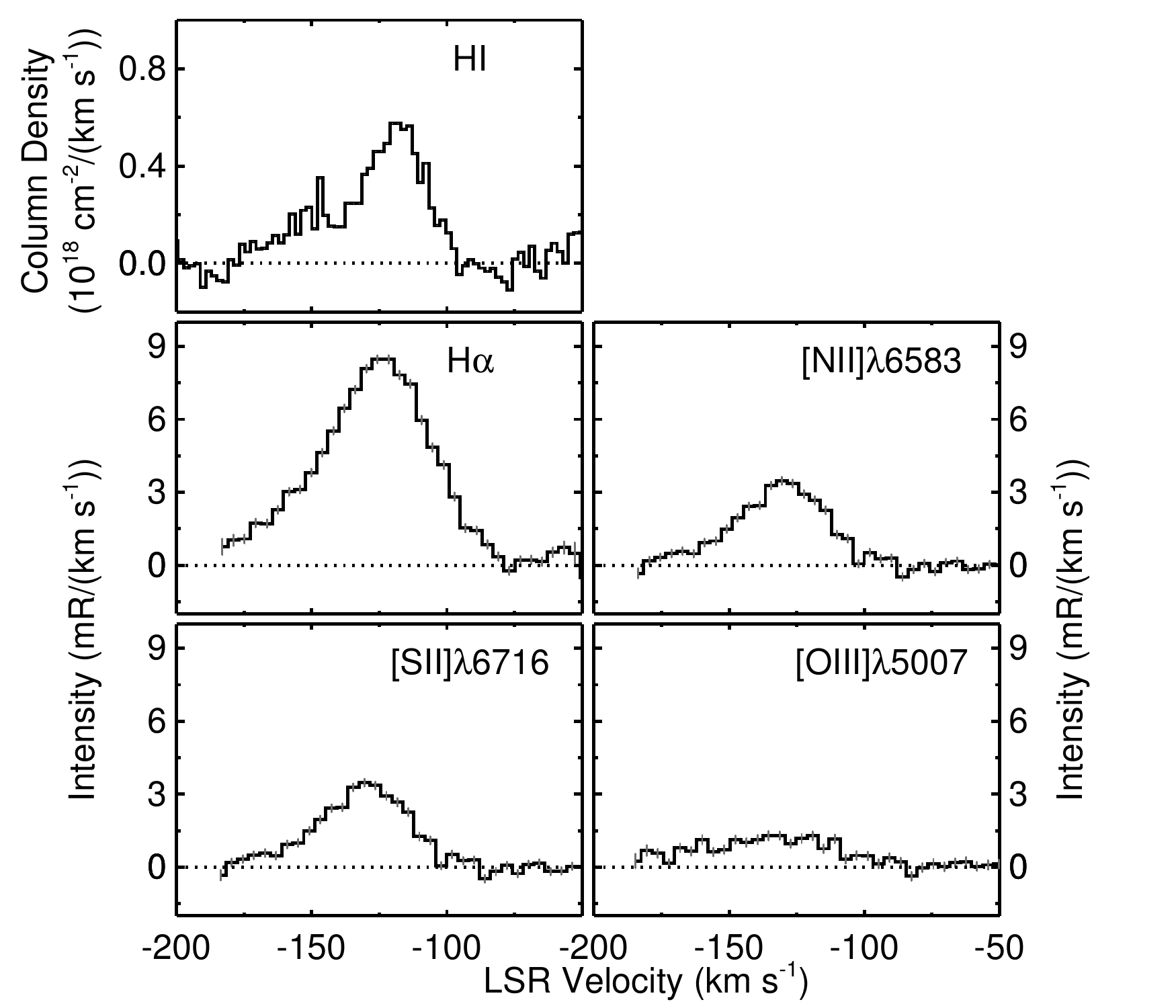}
\end{center}
\figcaption{Multiline WHAM spectra along the \citetalias{2013ApJ...778...58B} sight line labeled as (1) in Tables~\ref{table:obs},~\ref{table:intensities}, and~\ref{table:multiline_intensities}. The \hi\ spectrum is from the LAB Survey and was produced by averaging all the \hi\ spectra within the WHAM beam. 
\label{figure:greg1}}
\end{figure}

\begin{figure}
\begin{center}
\includegraphics[scale=0.5,angle=0]{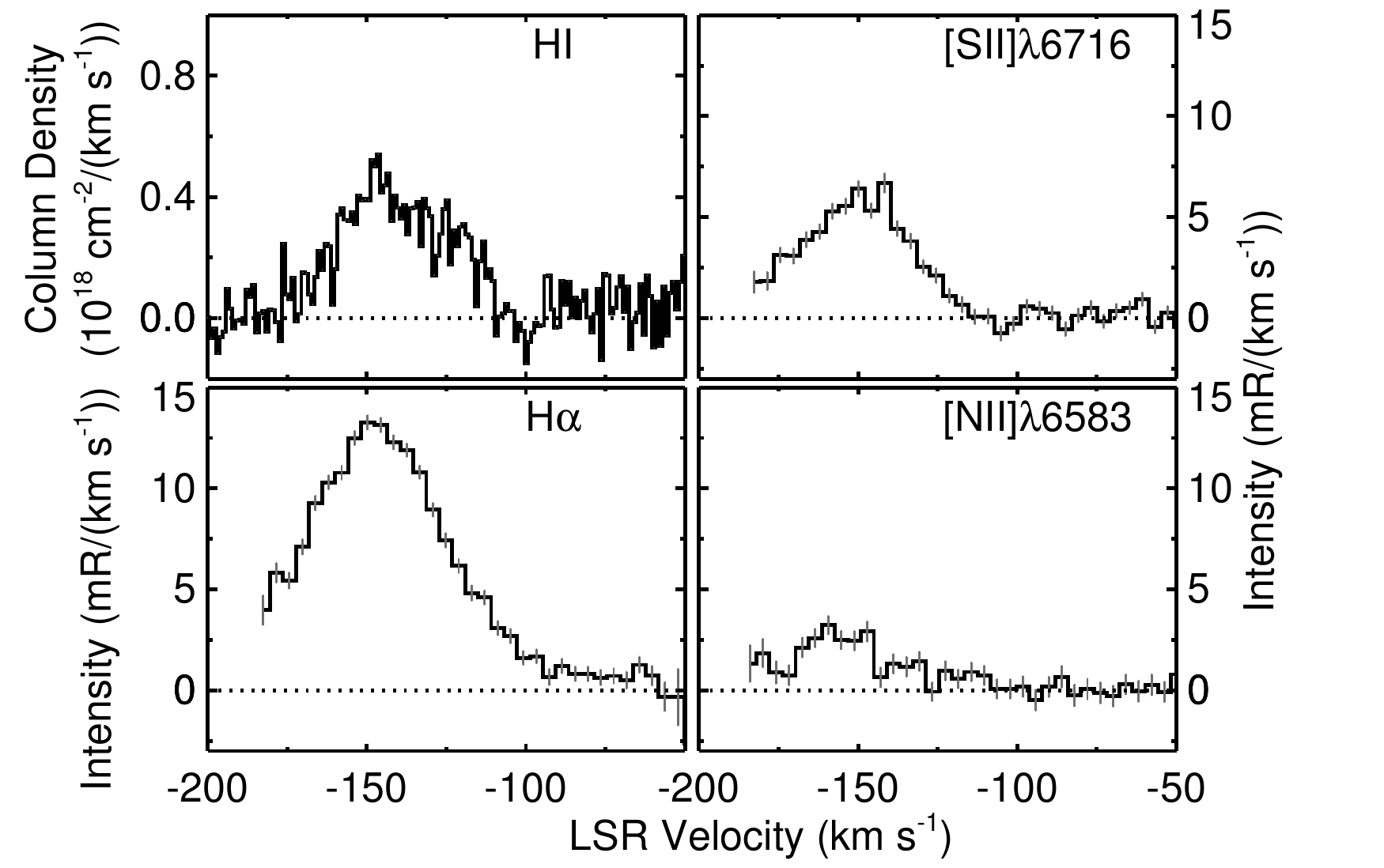}
\end{center}
\figcaption{Same as Figure~\ref{figure:greg1}, but for the \citetalias{2013ApJ...778...58B} sight line labeled as~(2). 
\label{figure:greg2}}
\end{figure}

\begin{figure}
\begin{center}
\includegraphics[scale=0.5,angle=0]{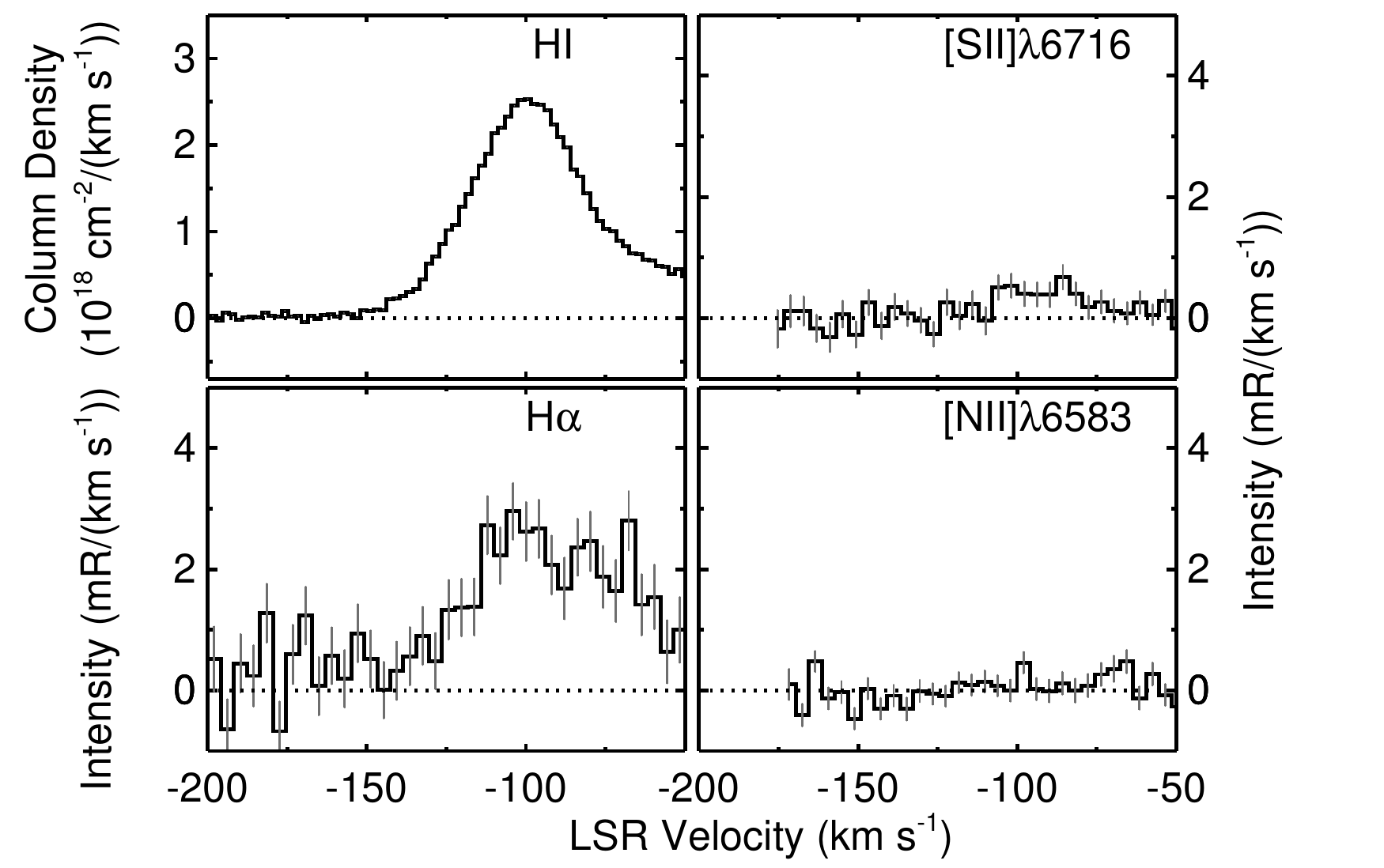}
\end{center}
\figcaption{Same as Figure~\ref{figure:greg1}, but for the \citetalias{2013ApJ...778...58B} sight line labeled as~(3). 
\label{figure:greg3}}
\end{figure}

From 2001--2008 at KPNO, we acquired $22$~targeted \ha\ observations and complimentary off-target observations of the Magellanic Stream  $50\arcdeg$ to $100\arcdeg$ from the MCs. These sight lines are labeled ($1-19$) under the ID column in Tables~\ref{table:obs} and~\ref{table:intensities}, which includes $3$ sight line pairs labeled ``a'' and ``b'' that substantially overlap. Figure~\ref{figure:on_off} illustrates the on- and off-target observations along three sight lines and the results of their subtraction. Toward three of these $22$ sight lines, we also observed \sii$~\lambda6716$, and \nii~$\lambda6583$ as well as \oii$~\lambda7320$ and \oiii$~\lambda5007$ toward one sight line (see Table~\ref{table:multiline_intensities}). These multiline spectra are shown in Figures~\ref{figure:greg1}--\ref{figure:greg3}. The \ha\ intensities for these $22$ sight lines were first presented in \citetalias{2013ApJ...778...58B} in their Figure~2. We have since reanalyzed their dataset and list updated values for the emission-line fits in Table~\ref{table:intensities}.

\begin{figure}
\begin{center}
\includegraphics[scale=0.5,angle=0]{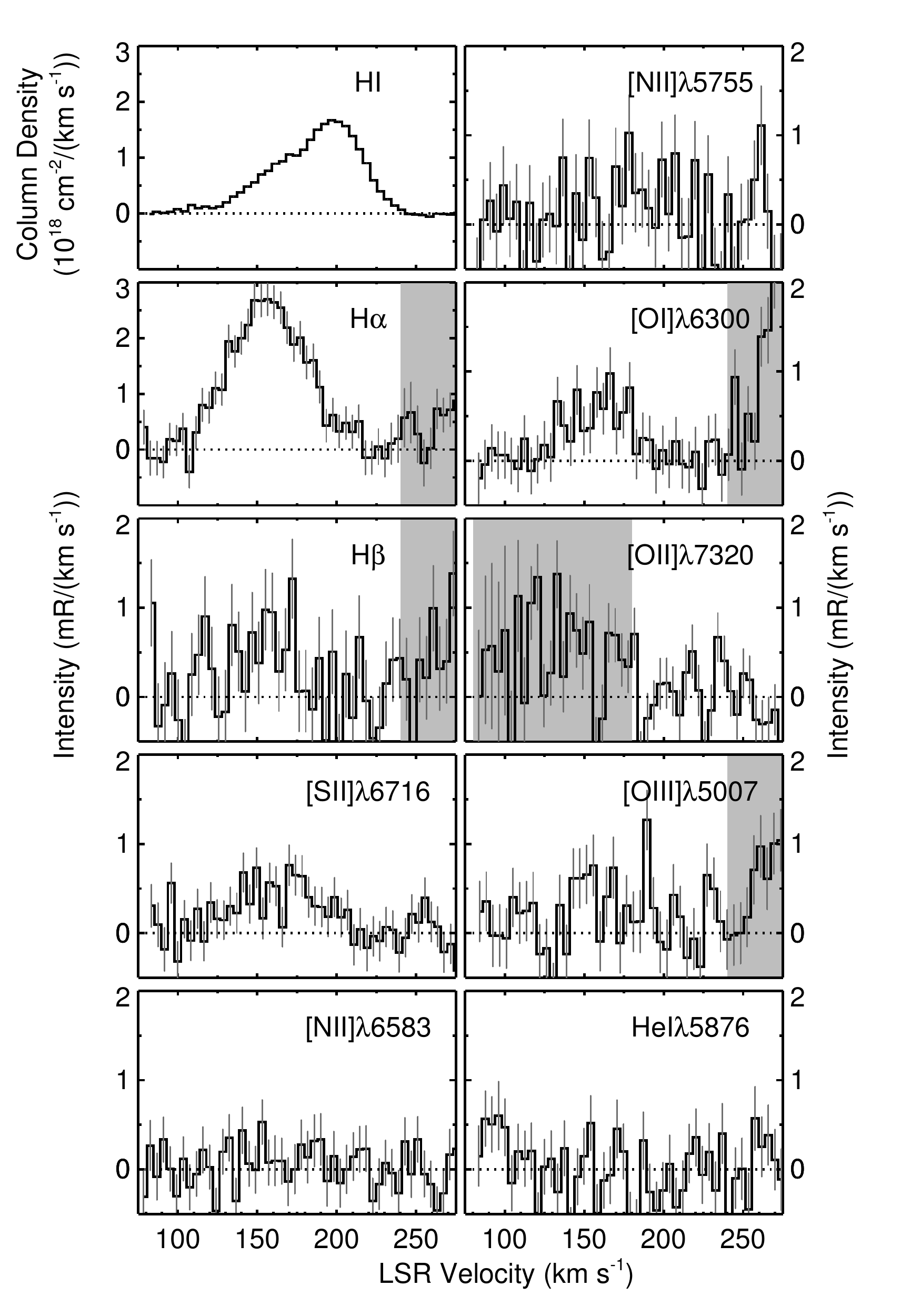}
\end{center}
\figcaption{Same as Figure~\ref{figure:greg1}, but for the FAIRALL~9 sight line labeled as~(a). Spectral regions highlighted in grey have diminished sensitivity due to increased residuals associated with bright atmospheric emission lines. 
\label{figure:Fairall9}}
\end{figure}

\begin{figure}
\begin{center}
\includegraphics[scale=0.5,angle=0]{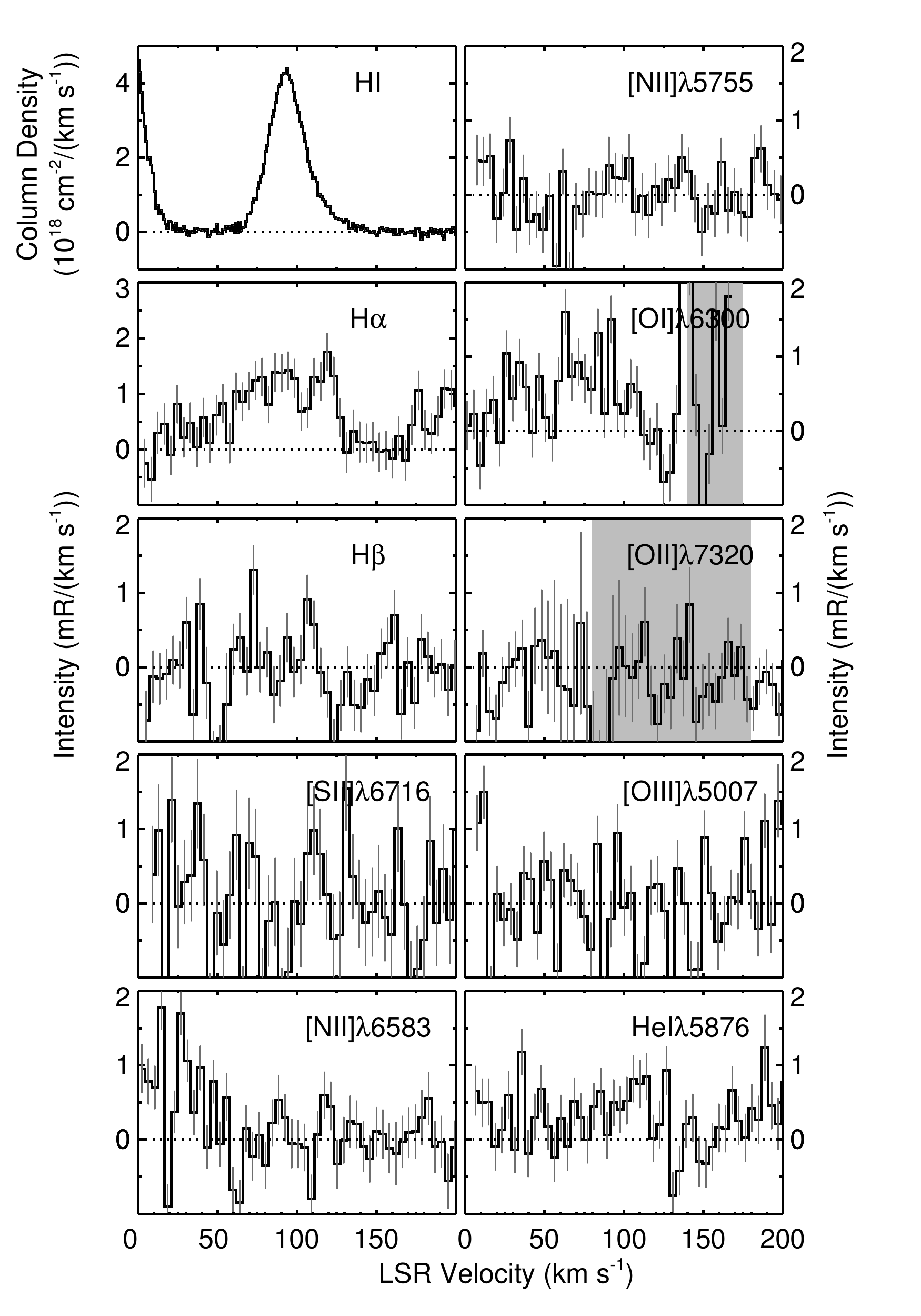}
\end{center}
\figcaption{Same as Figure~\ref{figure:greg1}, but for the RBS~144 sight line labeled as~(e). Spectral regions highlighted in grey have diminished sensitivity due increased residuals associated with bright atmospheric emission lines. 
\label{figure:RBS144}}
\end{figure}

We collected $17$~more targeted \ha, \sii$~\lambda6716$, and \nii~$\lambda6583$ observations of the Stream from 2011--2013 at CTIO, which are labeled with IDs (a--s). Toward positions (a) and (e)---located near the MCs and only 8.2-degrees apart---we also observed the \hb, \nii$~\lambda5755$, \oi$~\lambda6300$, \oii$~\lambda7320$, \oiii$~\lambda5007$, and \hei$~\lambda5876$ lines; these two sight lines probe two separate filaments of the Magellanic Stream with distinct velocities \citep{2008ApJ...679..432N} and different metallicities \citep{2000AJ....120.1830G, 2010ApJ...718.1046F, 2013ApJ...772..110F, 2013ApJ...772..111R}. These multiline spectra are shown in Figures~\ref{figure:Fairall9} and~\ref{figure:RBS144}.

We observed each sight line for a total-integrated-exposure time of $5-50~{\rm minutes}$ with each filter. Each of the individual exposures lasted $60-300$~seconds (see Table~\ref{table:obs}). We alternated between on-target exposures with off-target exposures of the same length. These brief, individual exposures minimize the subtle changes in atmospheric emission that occur over short time scales. Although off-target sight lines that are positioned close to the on-target direction minimize the atmospheric differences between observations, \citetalias{2014ApJ...787..147F} found that the ionized gas of the Stream can extend as much as $30\arcdeg$ off the \hi\ emission. We therefore chose $2-3$ off-target directions per on-target, each positioned within $2-10~{\rm degrees}$ of the target and the \hi\ emission; because our emission-line observations are less sensitive to low column density gas than the \citetalias{2014ApJ...787..147F} absorption-line observations (our sensitivity scales as the square of the density and theirs scales linearly with density), this offset tended to be sufficient. To ensure that all of the chosen off-target directions were off of the \ha\ emitting regions of the Stream, we paired and subtracted all nearby off-target observations from each other. In a few of the sight lines that were originally selected to be off positions, we detected \ha\ emission with kinematics consistent with the \hi\ gas in that region of the Stream. These serendipitous on-target observations are labeled with the ``Near'' prefix in the Notes column of Table~\ref{table:obs}. Additionally, all of these observations were positioned at least $0\fdg55$ away from bright foreground stars ($m_V < 6~{\rm mag}$) to avoid the distortion they cause to the continuum.

\section{Reduction}\label{section:reduction}

\subsection{Velocity and Intensity Calibration}
We reduced the data with the WHAM pipeline (see~\citealt{2003ApJS..149..405H}). This is the same reduction procedure employed by \citet{2012ApJ...761..145B} in their HVC Complex~A study that also utilized the WHAM telescope. Corrections to intensities were performed using synchronous observations of calibration targets, such as the North America Nebula (NGC~7000) that has an $I_{H\alpha} = 850\pm50~{\rm R}$ within a $50\arcmin$ beam (\citealt{1981ApJ...243..644S} and \citealt{2003ApJS..149..405H}). For the \ha\ observations taken at KPNO, we also used these calibration target observations to correct for atmospheric transmission and to register the velocity frame. The procedures described below outline the velocity calibration and atmospheric line subtraction for all other data. To check for changes in instrumental throughput and to calibrate the surface brightness of the observations taken at CTIO, we compared the intensity of the \hii\ region surrounding $\lambda$~Orionis with data taken at both observing sites.  

Roughly $20\%$ of our observations are contaminated by either the bright geocoronal \ha\ line or strong OH molecular lines. These bright atmospheric lines lie at constant and well established velocities in the geocentric (GEO) frame (e.g., \citealt{2002ApJ...565.1060H}) with the geocoronal line at $\vgeo=-2.3~\kms$ and a bright OH line lies at $\vgeo=+272.44~{\kms}$ relative to the \ha\ recombination line at $6562.8~\AA$. The presence of these lines enables us to calibrate the velocity extremely accurately. These bright lines also introduce more Poisson noise and typically lower our sensitivity by $10~{\rm mR}$, assuming a line width of $30~\kms$.

For spectra devoid of bright atmospheric contamination, we adopt an alternative method for their velocity calibration. To tune the wavelength center of WHAM's spectroscopic window, we adjust the index of refraction of the gas located between WHAM's Fabry-P{\'e}rot etalons by adjusting the gas pressure. The  $\Delta \lambda$ varies linearly with the gas pressure \citep{1997PhDT........15T}. By monitoring the gas pressure, we can therefore determine the radial velocity of the emission by measuring its Doppler shift as this linear relationship is well-measured for the WHAM spectrograph. This is essentially the reverse of the tuning process described by \citet{2003ApJS..149..405H} and is accurate to $\lesssim5~\kms$ \citep{2004PhDT........24M}, though the relative velocities calibration of observations taken with different pressures but a the same interference order will agree within $0.1~\kms$ of each other. 

\subsection{Atmospheric Subtraction}

When present, the bright geocoronal \ha\ and OH molecular atmospheric lines dominate over the emission of the Magellanic Stream and faint atmospheric lines, with strengths that are often more than an order of magnitude greater (see Figure~3 of \citealt{2013ApJ...771..132B}). These bright lines are produced in Earth's upper atmosphere from interactions with solar radiation. Therefore, their strength varies with both direction on the sky and the time of the observations. The difference in emission strength between these bright lines in the on- and off-target observations is much larger than the strength of the faint diffuse astronomical targets in this study. For these reasons, we first removed these bright lines from each exposure before subtracting the off-target observations from the on-target observations by modeling their shape as a single Gaussian convolved with the instrument profile of WHAM. 

Once the bright atmospheric lines have been removed, we combine each set of individual ${\rm on}-{\rm off}$ target pairs by weighting their intensities and uncertainties with  the number of observations in each velocity bin. Figure~\ref{figure:on_off} illustrates average on- and off-target spectra as well as the resultant spectra after the on-off subtraction. Most of the observations were only affected by faint atmospheric emission lines. The surface brightness of these faint atmospheric lines exhibit spatial and temporal variations of up to $\sim50\%$ throughout a single night, but are typically on the order of $\sim0.1~{\rm R}$. \citet{2013ApJ...771..132B} list the atmospheric lines that contaminate the \ha\ spectra at CTIO over the $-40~\kms\lesssim\vgeo\lesssim+310~\kms$ velocity range in their Table~1 and they illustrate them in their Figure~3.  \citet{2002ApJ...565.1060H} list these lines for KPNO over a shifted velocity window in their Table~3 and illustrate them in their Figures~1 and~2. Similar faint atmospheric lines contaminate all of the observations taken over the other wavelengths observed in this study. We isolate the astronomical emission  by subtracting off-target observations from on-target observations, which substantially removes the fainter atmospheric contamination. We then remove the background continuum level by assuming a flat or linear background over all velocities. Slight differences in the continuum level in the reanalysis of the \citetalias{2013ApJ...778...58B} dataset has resulted in a $\sim10-20\%$ reduction in $I_{\rm H\alpha}$ from the values found in Figure~2 of their study. 

\subsection{Line Fitting}

To measure the line strength and characteristics, we employed the Levenberg-Marquardt iteration technique (see \citealt{More78}) to calculate the fit parameters of a Gaussian convolved with the WHAM instrument profile using $\widetilde{\chi}^2$ minimization with the MPFIT IDL routine \citep{2009ASPC..411..251M}. A second Gaussian was used only when the data could not be reasonably fit with one Gaussian such that the reduced $\widetilde{\chi}_{\rm min}^2$ exceeded $2.5$. For example, although a second Gaussian fit could be imposed to align with the sharp peak at $\sim120~\kms$ along sight line (e), at an offset of only $\sim15~\kms$ from the \hi\ center; however, this peak is quite narrow and uncharacteristic of typical warm ionized gas emission line profiles. Although we could improve our fits further through the inclusion of more Gaussians, each spectrum represents an average of all component structure contained within the observed $1\arcdeg$ beam, which corresponds to a physical diameter of $\sim1~\kpc$ near the Magellanic Clouds and up to $\sim3.5~\kpc$ at the trailing tip of the Stream where its distance its predicted to lie between $55-200~\kpc$ from the Galactic Center (see Section~\ref{section:intro}). Additional components would not necessarily result in a disentanglement of that structure. This means that kinematically the resultant spectral lines become substantially broadened by unresolved non-thermal motions (e.g., bulk motion, larger scale turbulence, shear, rotation, etc.) of the gas contained within the large WHAM beam. 

The results of the \ha\ and \hi\ data fits and their corresponding $\widetilde{\chi}_{\rm min}^2$ values and their corresponding $1\sigma$ statistical uncertainties are listed in Table~\ref{table:intensities} and in Table~\ref{table:multiline_intensities} for the multiline observations. In Table~\ref{table:intensities}, we also include a spread of intensities that have a $\widetilde{\chi}^2$ that lie within $10\%$ of the $\widetilde{\chi}_{\rm min}^2$. This was included because for some of our spectra there was insufficient continuum surrounding the emission line to adequately anchor its level and slope, which translated to a wider range of parameters that could describe the line profile well. We represented this as ${\rm best~fit~I_{\rm H\alpha}}\pm{\rm 1\sigma}^{+\rm spread}_{-\rm spread}$ in Table~\ref{table:intensities}. For example, the best fit \ha\ intensity for sight line (a) is $165\pm8~{\rm mR}$ with ${\widetilde{\chi}}_{\rm min}^2=1.0$ and the range of intensities with ${\widetilde{\chi}}^2\le1.1 \widetilde{\chi}_{\rm min}^2$ is $104\le I_{\rm H\alpha}\le190~{\rm mR}$.

We report upper limits for the multiline WHAM observations and \hi\ column densities in Tables~\ref{table:intensities} and~\ref{table:multiline_intensities} if the line strength does not exceed a $3\sigma$ detection. For the \hi\ upper limits, we equate this significance as three times the standard deviation of the scatter in the background over a fixed line width of $30~\kms$. For the WHAM dataset, we calculate this upper limit using the width of the \hi\ emission or as $30~\kms$ in the absence of a \hi\ detection. Note that because the gas phase traced by the WHAM observations is generally warmer than the \hi\ gas, their actual line widths are expected to be larger than the \hi. We find that FWHM of the \ha\ emission in the Magellanic Stream is typically $\sim55~\kms$ compared to $\sim35~\kms$ for the \hi\ (Table~\ref{table:intensities}). Although the \ha\ width is typically $1.6$ times that of the \hi\ emission, there is likely unresolved multi-component substructure in the \ha\ emission within the 1-degree WHAM beam ($\sim1~\kpc$ at 55~\kpc) that would broadens the emission as suggested in spectra of sight lines (e) and (k). 

\subsection{Extinction Correction}

Lastly, we apply the extinction correction procedure used by \citet{2013ApJ...771..132B} to correct the intensity attenuation caused by the foreground dust in the MW. \citet{2013ApJ...772..110F} found that the depletion of the low- and high-metallicity filaments of the Magellanic Stream primarily varies with the strength of the \hi\ column density ($N_{\rm H\textsc{~i}}$) and not with the metallicity. Therefore, the self extinction is negligible in low \hi\ column density regions of the Stream. Among our sample, only  $5/39$ sight lines have $\log N_{\rm H\textsc{~i}}/\cm^{-2}\geq20$ and for those sight lines, the self extinction correction is only $\lesssim1-2\%$ for these sight lines. 

This extinction correction is smallest when the Stream lies significantly above or below the Galactic plane ($\lvert b\rvert\gtrsim25\arcdeg$). The correction substantially increases near the Galactic disk ($\lvert b\rvert\lesssim25\arcdeg$). To correct the attenuated intensities, we use the excess color presented by \citet{1994ApJ...427..274D} for a warm diffuse medium:
\begin{equation}
E(B-V)=\frac{\langle N_{\rm H{~\textsc{i}}}\rangle}{4.93\times10^{21}~\mathrm{atoms\cdot  cm^{-2}\cdot mag^{-1}}}
\end{equation}
where $N_{\rm H{~\textsc{i}}}$ only includes the foreground Galactic \hi\ emission \citep{1978ApJ...224..132B} and is calculated using the average of Leiden/Argentine/Bonn Galactic H{\sc~i}\ survey (LAB: \citealt{2005A&A...440..775K, 1997agnh.book.....H}) spectra within the WHAM 1-degree beam over the $-150\le\vlsr\le+150~\kms$ velocity range. We assume that the extinction follows the $\langle A(\mha)/A(V)\rangle=0.909-0.282/R_v$ 
optical curve presented in \citet{1989ApJ...345..245C} for a diffuse interstellar medium, where R$_v\equiv A(V)/E(B-V)=3.1$, so the expression for the total extinction becomes
\begin{equation}\label{eq:extinction}
A(\mha)=6.3\times10^{-22}~\langle N_{\rm H{~\textsc{i}}}\rangle~\mathrm{cm^2\cdot atoms^{-1} \cdot mag}.
\end{equation} The extinction corrected intensity for Galactic attenuation is then $I_{\mha,~corr} = I_{\mha,~obs}~e^{\ A(\mha)/2.5}$, where $1.03\le e^{\ A(\mha)/2.5}\le1.14$ for all sight lines in this study. 

Only for the sight line toward Fairall~9 at position~(a), we include a correction for Magellanic Stream attenuation but only for the $I_{\rm H\alpha}/I_{\rm H\beta}$ ratio as these lines have a large difference in wavelength. Along this sight line, we detect both \ha\ and \hb\ at $165\pm8~{\rm mR}$ and $33\pm9~{\rm mR}$ (non-extinction corrected; see Figure~\ref{figure:Fairall9}) yielding an extinction corrected ratio of $I_{\rm H\alpha}/I_{\rm H\beta}=5.0\pm1.4$. As the FAIRALL~9 sight line lies along the high metallicity filament \citep{2013ApJ...772..111R}, we corrected for self extinction using the average LMC extinction curves of \citet{2003ApJ...594..279G}. As these lines trace the same ionization conditions (i.e., density and temperature) and do not depend on metallicity, their ratio is often used to measure the amount of reddening due to dust extinction. Although this value is larger than the nominal value of $\sim3$, implying the presence of dust, slow shocks could also elevate the \ha\ emission and increase this ratio beyond $4$ \citep{1978ApJ...225L..27C, 1979ApJ...227..131S, 2007ApJ...670L.109B}. This large ratio suggests that this gas is also being collisionally ionized. Unfortunately, due to the low signal-to-noise ratio of the \hb\ detection and the lack of detections toward other sight lines, we are unable to confidently determine the source of this enhancement. In the analysis below, we only correct our observations for extinction due to the foreground dust in the MW.

\startlongtable
\begin{deluxetable*}{clllrrlllrll}
\tabletypesize{\scriptsize}
\tablecaption{H$\alpha$ and H\textsc{~i} Line Fitting Results \label{table:intensities}}
\tablewidth{0pt}
\tablehead{
\colhead{ } & \multicolumn{4}{c}{\ha} & & \multicolumn{4}{c}{H\textsc{~i}\tablenotemark{d}} & \colhead{position\tablenotemark{e}}  \\
\cline{2-5} \cline{7-10} 
\colhead{ID\tablenotemark{a}} & \colhead{\iha\tablenotemark{b,~c}} & \colhead{$\vlsr$} & \colhead{FWHM}  & \colhead{$\widetilde{\chi}^2$} & & 
                         \colhead{$\log {\rm N}_{\rm H\textsc{~i}}/\cm^{-2}$\tablenotemark{c}}	&\colhead{$\vlsr$} & \colhead{FWHM} & \colhead{$\widetilde{\chi}^2$}\\
\colhead{} & \colhead{(mR)} & \colhead{($\kms$)} & \colhead{($\kms$)} &  & & \colhead{} & \colhead{($\kms$)} & \colhead{($\kms$)} & 
}
\startdata
a&$165\pm8^{+25}_{-61}$&$158\pm1$&$67\pm2$&$1.1$ &&$19.48\pm0.01^{+0.06}_{-0.11}$&$158\pm1$&$40\pm1$&$1.0$ & on, FAIRALL~9\\
a&&&&&&$19.85\pm0.01^{+0.07}_{-0.02}$&$200\pm1$&$38\pm1$&$1.0$& on, FAIRALL~9\\
b&$69\pm5^{+26}_{-65}$&$135\pm5$&$79\pm5$&$1.0$& & $<18.67$ & $+135$\tablenotemark{g} & \nodata & \nodata & edge, Near~HE0226-4110\\  
c&$53\pm9^{+24}_{-19}$&$150\pm6$&$74\pm12$&$1.4$&& $17.05\pm0.04$\tablenotemark{f}       & $+165$\tablenotemark{f} & \nodata & \nodata & off, HE0226-4110 \\  
d&$53\pm10^{+15}_{-24}$&$102\pm3$&$54\pm6$&$1.6$&&$19.38\pm0.01^{+0.01}_{-0.01}$&$135\pm1$&$28\pm1$&$1.0$& on-small, Near~RBS~144\\  
e&$101\pm9^{+19}_{-22}$&$87\pm3$&$79\pm5$&$1.0$&&$19.93\pm0.01^{+0.01}_{-0.01}$&$93\pm1$&$26\pm1$&$1.0$& off, RBS~144\\ 
e&&&&&&$18.95\pm0.02^{+0.12}_{-0.07}$&$117\pm1$&$25\pm1$&$1.0$& off, RBS~144\\ 
f&$57\pm11^{+16}_{-12}$&$99\pm3$&$50\pm3$&$1.2$&&$19.06\pm0.02^{+0.01}_{-0.01}$&$94\pm1$&$29\pm1$&$1.1$& edge,  Near~RBS~144 \\  
g & $<45$\tablenotemark{h}        	& \nodata  & \nodata  & \nodata && $<18.59$ & $+210$\tablenotemark{f} & \nodata & \nodata & off, HE0153-4520 \\  
h&$105\pm21^{+42}_{-40}$&$109\pm2$&$61\pm6$&$1.9$&&$18.63\pm0.04^{+0.01}_{-0.01}$&$92\pm2$&$32\pm2$&$1.1$& off, Near~RBS~1892\\  
j  & $<49$\tablenotemark{h}        	& \nodata & \nodata  & \nodata & & $<18.78$ & $+200$\tablenotemark{f} & \nodata & \nodata & off,~ESO292-G24 \\  
k&$101\pm22^{+35}_{-48}$&$117\pm6$&$80\pm7$&$1.0$  &  & $18.76\pm0.01$\tablenotemark{i} & $+125\pm1$\tablenotemark{h} & $34\pm7$\tablenotemark{i} & $1.3$ & edge, Near~HE0056-3622\\  
m&$69\pm3^{+16}_{-21}$&$-133\pm2$&$52\pm1$&$1.0$&&$18.74\pm0.04^{+0.01}_{-0.01}$&$-132\pm2$&$47\pm3$&$1.0$ & off, Near~PHL2525\\  
n & $<37$\tablenotemark{h}   	&  \nodata & \nodata  & \nodata & & $<18.83$ & $-135$\tablenotemark{f} & \nodata  & \nodata & edge-small, PHL2525\\  
o  & $<25$\tablenotemark{h}  		& \nodata  & \nodata & \nodata && $<18.83$       & $-150$\tablenotemark{f} & \nodata  & \nodata & off, UM239 \\  
p  & $<30$ 	&  \nodata  &  \nodata  &  \nodata &  & $<18.82$ & $-250$\tablenotemark{e} & \nodata & \nodata & edge, NEAR~NGC7714 \\ 
q  & $<31$\tablenotemark{h}   		& \nodata  & \nodata  & \nodata && $<18.85$ & $-200$\tablenotemark{f} & \nodata & \nodata & off, MRK~1502\\  
r  & $<34$\tablenotemark{h}   		& \nodata  & \nodata  & \nodata && $<18.76$ & $-200$\tablenotemark{f} & \nodata & \nodata & off, MRK~304\\  
s&$162\pm31^{+62}_{-84}$&$-328\pm3$&$92\pm8$&$1.9$ &  & $16.67\pm0.02$\tablenotemark{f} & $-300$\tablenotemark{f} & \nodata & \nodata & off, MRK~335 \\  
1&&&&&&$18.78\pm0.01^{+0.11}_{-0.09}$&$-146\pm1$&$26\pm1$&$1.0$ & on\\ 
1&$417\pm1^{+17}_{-14}$&$-121\pm1$&$50\pm1$&$2.1$&&$19.14\pm0.01^{+0.10}_{-0.05}$&$-118\pm1$&$30\pm1$&$1.0$ & on\\ 
2&$574\pm17^{+1}_{-21}$&$-141\pm1$&$52\pm1$&$1.1$ &&$19.17\pm0.02^{+0.09}_{-0.05}$&$-148\pm1$&$30\pm1$&$1.3$& edge\\ 
2&&&&&&$18.83\pm0.02^{+0.12}_{-0.09}$&$-121\pm1$&$18\pm1$&$1.1$& edge\\ 
3&$108\pm27^{+53}_{-102}$&$-100\pm4$&$64\pm4$&$1.0$ &&$20.03\pm0.01^{+0.01}_{-0.01}$&$-101\pm1$&$40\pm1$&$1.0$& on\\ 
4a & $<50$\tablenotemark{h}  & \nodata & \nodata & \nodata&&$20.02\pm0.01^{+0.01}_{-0.04}$&$-109\pm1$&$34\pm1$&$1.0$ & on \\ 
4b&$66\pm18^{+17}_{-12}$&$-107\pm3$&$44\pm6$&$1.0$ &&$20.01\pm0.01^{+0.03}_{-0.04}$&$-110\pm1$&$34\pm1$&$1.0$& on\\
5a&$138\pm23^{+50}_{-45}$&$-163\pm2$&$82\pm9$&$1.1$&&$19.65\pm0.01^{+0.09}_{-0.04}$&$-155\pm1$&$45\pm1$&$1.0$& on\\
5b&$122\pm6^{+19}_{-27}$&$-173\pm2$&$45\pm1$&$1.0$&&$19.56\pm0.01^{+0.04}_{-0.12}$&$-159\pm1$&$34\pm1$&$1.0$& on\\
6&$110\pm11^{+26}_{-24}$&$-134\pm2$&$75\pm6$&$1.4$&&$19.73\pm0.01^{+0.08}_{-0.04}$&$-158\pm1$&$56\pm1$&$1.0$& on\\
7a&$77\pm16^{+17}_{-13}$&$-202\pm2$&$45\pm3$&$1.5$&&$19.96\pm0.01^{+0.01}_{-0.01}$&$-211\pm1$&$29\pm1$&$1.5$& edge\\
7a&&&&&&$19.06\pm0.02^{+0.07}_{-0.08}$&$-179\pm1$&$27\pm1$&$1.5$& edge\\
7b&$112\pm19^{+19}_{-21}$&$-189\pm2$&$72\pm3$&$1.2$&&$19.93\pm0.01^{+0.01}_{-0.01}$&$-211\pm1$&$31\pm1$&$1.0$& edge\\
7b&&&&&&$19.49\pm0.01^{+0.05}_{-0.03}$&$-182\pm1$&$35\pm1$&$1.5$& edge\\
8&$50\pm6^{+24}_{-36}$&$-197\pm5$&$78\pm8$&$1.0$&&$19.58\pm0.02^{+0.01}_{-0.02}$&$-217\pm1$&$29\pm1$&$1.9$& on\\ 
8&&&&&&$18.97\pm0.02^{+0.06}_{-0.03}$&$-219\pm3$&$27\pm2$&$1.0$& on\\
9&$104\pm22^{+23}_{-47}$&$-238\pm2$&$49\pm1$&$1.0$&&$19.60\pm0.01^{+0.13}_{-0.01}$&$-227\pm1$&$63\pm1$&$1.0$& on\\ 
10&$152\pm5^{+32}_{-69}$&$-231\pm1$&$61\pm1$&$1.0$&&$19.68\pm0.01^{+0.08}_{-0.06}$&$-231\pm1$&$45\pm1$&$1.0$& on\\ 
11 & $<30$\tablenotemark{h}       & \nodata  & \nodata  & \nodata    && $19.22\pm0.01^{+0.07}_{-0.09}$&$-300\pm1$&$32\pm1$&$1.4$& on\\ 
12&$64\pm10^{+9}_{-11}$&$-354\pm3$&$66\pm5$&$1.3$&&$19.49\pm0.01^{+0.05}_{-0.05}$&$-361\pm1$&$35\pm1$&$1.5$& on\\
13&$30\pm5^{+15}_{-30}$&$-317\pm4$&$49\pm6$&$1.0$&&$19.07\pm0.02^{+0.09}_{-0.16}$&$-317\pm2$&$64\pm3$&$1.0$& on\\
14&$59\pm5^{+25}_{-27}$&$-332\pm4$&$74\pm5$&$1.0$&&$19.17\pm0.02^{+0.06}_{-0.10}$&$-322\pm1$&$24\pm1$&$1.0$ & on \\
14&&&&&&$19.14\pm0.01^{+0.08}_{-0.09}$&$-354\pm1$&$22\pm1$&$1.0$ & on\\
15&$34\pm6^{+13}_{-39}$&$-319\pm4$&$46\pm2$&$1.0$&&$19.23\pm0.01^{+0.06}_{-0.03}$&$-323\pm1$&$27\pm1$&$1.2$ & on \\
15&&&&&&$18.77\pm0.03^{+0.06}_{-0.11}$&$-351\pm1$&$27\pm1$&$2.4$ & on\\
16 & $<50$\tablenotemark{h}       & \nodata  & \nodata  &  \nodata     & &  $18.92\pm0.01^{+0.08}_{-0.04}$&$-321\pm1$&$31\pm1$&$1.0$& edge\\
17 & $<34$\tablenotemark{h}   & \nodata & \nodata &  \nodata & & $<18.85$ & $-320$\tablenotemark{g} & \nodata & \nodata & off\\
18 & $<50$\tablenotemark{h}      &   \nodata        &   \nodata        &  \nodata    & &       $<18.82$     & $-320$\tablenotemark{g} & \nodata & \nodata  & off \\ 
19&$60\pm9^{+12}_{-18}$&$-331\pm2$&$59\pm2$&$1.0$&&$19.13\pm0.01^{+0.10}_{-0.06}$&$-335\pm1$&$26\pm1$&$1.0$& on\\
\enddata
\tablenotetext{a}{IDs (a--s) correspond to new WHAM observations acquired at CTIO. IDs (1-19) correspond to WHAM observations acquired at KPNO. The \ha\ intensities for the (1-19) sight lines were first presented in \citetalias{2013ApJ...778...58B} and have been reanalyzed here. Above we list new values for the line fit parameters for the \citetalias{2013ApJ...778...58B} dataset.}
\tablenotetext{b}{Non-extinction corrected intensity.} 
\tablenotetext{c}{Represented as ${\rm best~fit~I_{\rm H\alpha}}\pm{\rm 1\sigma}^{+\rm spread}_{-\rm spread}$, where $1\sigma$ is the statistical uncertainty and a spread in intensity values within ${\widetilde{\chi}}^2\le1.1 \widetilde{\chi}_{\rm min}^2$ is indicated.} 
\tablenotetext{d}{Calculated using averaged LAB \hi\ Survey spectra of the sight lines within the WHAM 1-degree beam.} 
\tablenotetext{e}{Position of the sight line on, off, or on the edge of the main H\textsc{~i} filaments or when the sight line pierces a small, offset cloud and lists background objects that align with the sight line.}
\tablenotetext{f}{Measured from H\textsc{~i} Lyman series absorption lines by \citet{2005ApJ...630..332F} for sight line (c) and by \citet{2010ApJ...718.1046F} for sight line (s).} 
\tablenotetext{g}{Measured from nearest sight line with detectable \hi\ emission using LAB \hi\ Survey.} 
\tablenotetext{h}{Assumes a line width equal to that of the \hi\ emission or of $30~\kms$ if no \hi\ exists.} 
\tablenotetext{i}{Measured from Galactic All Sky Survey (GASS) \hi\ observations that have been averaged to match the $1\arcdeg$ angular resolution of WHAM; not detected in the averaged LAB survey data.} 

\end{deluxetable*}

\begin{deluxetable}{cllllr}
\tabletypesize{\scriptsize}
\tablecaption{WHAM Multiline Fitting Results \label{table:multiline_intensities}}
\tablewidth{0pt}
\tablehead{
\colhead{ID\tablenotemark{a}} & \colhead{Line} & \colhead{I\tablenotemark{b}} & \colhead{$\vlsr$} & \colhead{FWHM}  & \colhead{$\widetilde{\chi}^2$}  \\
\colhead{}  & \colhead{} & \colhead{(mR)} & \colhead{($\kms$)} & \colhead{($\kms$)}  & 
}
\startdata
a &	$\rm H\beta$		& $33\pm9$ 		& $+158\pm27$ & $55\pm25$ & $0.9$ \\  
a &	\sii$~\lambda6716$	& $36\pm6$ 		& $+164\pm12$ & $70\pm14$ & $85$ \\  
a &	\nii$~\lambda6583$	& $<22$ 			& \nodata & \nodata & \nodata \\
a &	\nii$~\lambda5755$	& $<43$ 			& \nodata & \nodata & \nodata \\
a &	\oi$~\lambda6300$	& $33\pm8$ 		& $+163\pm14$ & $56\pm17$ & $0.71$ \\  
a &	\oii$~\lambda7320$	& $<39$ 			& \nodata & \nodata & \nodata \\
a &	\oiii$~\lambda5007$	& $<36$ 			& \nodata & \nodata & \nodata \\
a &	He\textsc{~i}	& $<31$ 			& \nodata & \nodata & \nodata \\
b--d & \sii$~\lambda6716$	& $\lesssim40$ 			& \nodata & \nodata & \nodata \\
b--d & \nii$~\lambda6583$	& $\lesssim40$ 			& \nodata & \nodata & \nodata \\
e &	$\rm H\beta$	& $<47$ 			& \nodata & \nodata & \nodata \\
e &	\sii$~\lambda6716$	& $<53$ 			& \nodata & \nodata & \nodata \\
e &	\nii$~\lambda6583$	& $<28$ 			& \nodata & \nodata & \nodata \\
e &	\nii$~\lambda5755$	& $<24$ 			& \nodata & \nodata & \nodata \\
e &	\oi$~\lambda6300$	& $45\pm8$ 		& $+81\pm15$ & $54\pm30$ & $1.7$ \\  
e &	\oii$~\lambda7320$	& $<38$ 			& \nodata & \nodata & \nodata \\
e &	\oiii$~\lambda5007$	& $<52$ 			& \nodata & \nodata & \nodata \\
e &	\hei$~\lambda5876$	& $<33$ 			& \nodata & \nodata & \nodata \\
f--s & \sii$~\lambda6716$	& $\lesssim40$ 			& \nodata & \nodata & \nodata \\
f--s & \nii$~\lambda6583$	& $\lesssim40$ 			& \nodata & \nodata & \nodata \\
1 &	\sii$~\lambda6716$	& $140\pm7$ 		& $-130\pm2$ & $51\pm3\phantom{0}$ & $1.8$ \\  
1 &	\nii$~\lambda6583$	& $74\pm10$ 		& $-136\pm6$ & $67\pm12$ & $1.5$ \\  
1 &	\oi$~\lambda6300$	& $<40$ 			& \nodata & \nodata & \nodata \\  
1 &	\oiii$~\lambda5007$	& $<25$ 			& \nodata & \nodata & \nodata \\  
2 &	\sii$~\lambda6716$	& $230\pm21$ 		& $-151\pm2$ & $51\pm3\phantom{0}$ & $1.8$ \\  
2 &	\nii$~\lambda6583$	& $77\pm16$ 		& $-156\pm4$ & $45\pm6\phantom{0}$ & $1.0$ \\
3 &	\sii$~\lambda6716$	& $<28$ 			& \nodata & \nodata & \nodata \\
3 &	\nii$~\lambda6583$	& $<22$ 			& \nodata & \nodata & \nodata \\
\enddata 
\tablenotetext{a}{IDs a--s correspond to new WHAM observations acquired at CTIO; IDs 1--19 correspond to WHAM observations acquired at KPNO (intensities first presented in \citetalias{2013ApJ...778...58B}).}
\tablenotetext{b}{Non-extinction corrected.}
\end{deluxetable}

\section{Comparison of Neutral and Ionized Gas}\label{section:compare}      

In this section, we compare the strength and velocity distribution of the \hi\ and \ha\ emission. The large and small scale similarities and differences between the neutral and ionized gas phases provide clues as to which astrophysical processes are affecting the Magellanic Stream. Some of these processes include photoionization from the surrounding radiation field and ram-pressure stripping from the Galactic halo, among others. Understanding their influence is critical to predict whether the neutral gas contained within the Stream can reach the disk and replenish the MW before it evaporates into the halo.

The Magellanic Stream extends over $165\arcdeg$ evident by its widespread \hi\ emission that decreases steadily along its length (see Figures~\ref{figure:hi_map} and Figure~10 of \citealt{2010ApJ...723.1618N}).  This Stream is composed of two spatially \citep{2003ApJ...586..170P} and kinematically (see Figure~13b of \citealt{2008ApJ...679..432N}) distinct filaments. Each of these filaments has a different metallicity with one measured at $50\%~{\rm solar}$ toward only the FAIRALL~9 sight line (S\textsc{~ii}/H=0.27~solar: \citealt{2000AJ....120.1830G}; S/H=0.5 and N/H=0.07~solar: \citealt{2013ApJ...772..111R}) and the other at $10\%~{\rm solar}$ toward eight sight lines \citep{2000AJ....120.1830G, 2010ApJ...718.1046F, 2013ApJ...772..110F}. A gap between these two filaments is visible right above the (e-f) label in Figure~\ref{figure:hi_map}. 

Our targets lie on and off these two \hi\ filaments. The right panels of Figures~\ref{figure:RBS144_map}--\ref{figure:greg_1_2} include examples of the \hi\ distribution surrounding $9$ of our $39$ sight lines. The locations of the WHAM observations within these representative gas distribution maps are marked with black hollow circles that are the size of our $1\arcdeg$~beam. The corresponding \ha\ and \hi\ spectra at these locations are displayed in the left panels of these figures. The Appendix includes the \ha\ and \hi\ spectra along all of our sight lines and two ${N}_{\rm H\textsc{~i}}$ maps showing distribution of the neutral gas around most of the sight lines that are not included in Figures~\ref{figure:RBS144_map}--\ref{figure:greg_1_2}.

\subsection{$\rm H\alpha$ and $\rm H\textsc{~i}$ Emission Strengths and Kinematics}\label{section:hi_ha_emission_velocity}

The strength of the \ha\ emission along the Magellanic Stream does not vary with the \hi\ column density (Figure~\ref{figure:column_intensity}). Differences in path lengths or filling factors between the neutral and ionized gas, as well as changes in the ionization fraction along these sight lines, could result in uncorrelated \hi\ and \ha\ emission strengths. On the other hand, the line centers of the \hi\ and the \ha\ emission are tightly correlated with an average separation of only $\langle |\Delta {\rm v}|\rangle=9.3~\kms$, though the \ha\ emission tends to be $1.6\times$ broader than the \hi\ emission (Figure~\ref{figure:compare_vel}). This correlation in their line centers suggests that the warm ionized and warm neutral gas responsible for the \ha\ and \hi\ emission are physically associated. If a single ionization process along the Stream were producing mixed, co-spatial regions of ionized and neutral gas, we would expect to see a correlation between the \ha\ and \hi\ emission. In this scenario, differences between the path lengths and/or densities along each sight lines would produce a range of intensities but the emission lines would be generally correlated. However, in regions studied so far, the strength of the \ha\ and \hi\ emission do not appear correlated despite having similar velocity centroids. In this case, the neutral and ionized phases are likely not well-mixed but are occurring within a distinct kinematic (and likely physical) region of the Stream. This behavior is also seen in IVCs and HVCs in the Milky Way that tend to have uncorrelated \ha\ and \hi\ emission strengths and correlated line centers (e.g., \citealt{2001ApJ...556L..33H}, \citealt{2003ApJ...597..948P}, \citealt{2005ASPC..331...25H}, \citealt{2009ApJ...703.1832H}, \citealt{2012ApJ...761..145B, 2013ApJ...771..132B}). One straightforward configuration that may give rise to a kinematic association and lack of intensity correlation is if the ionization is occurring on the outside of a neutral cloud or on a side of neutral sheets. We discuss the source of the ionization in more detail in Section~\ref{section:ionization} below. 

\begin{figure*}
\begin{center}
\includegraphics[scale=0.325,angle=0]{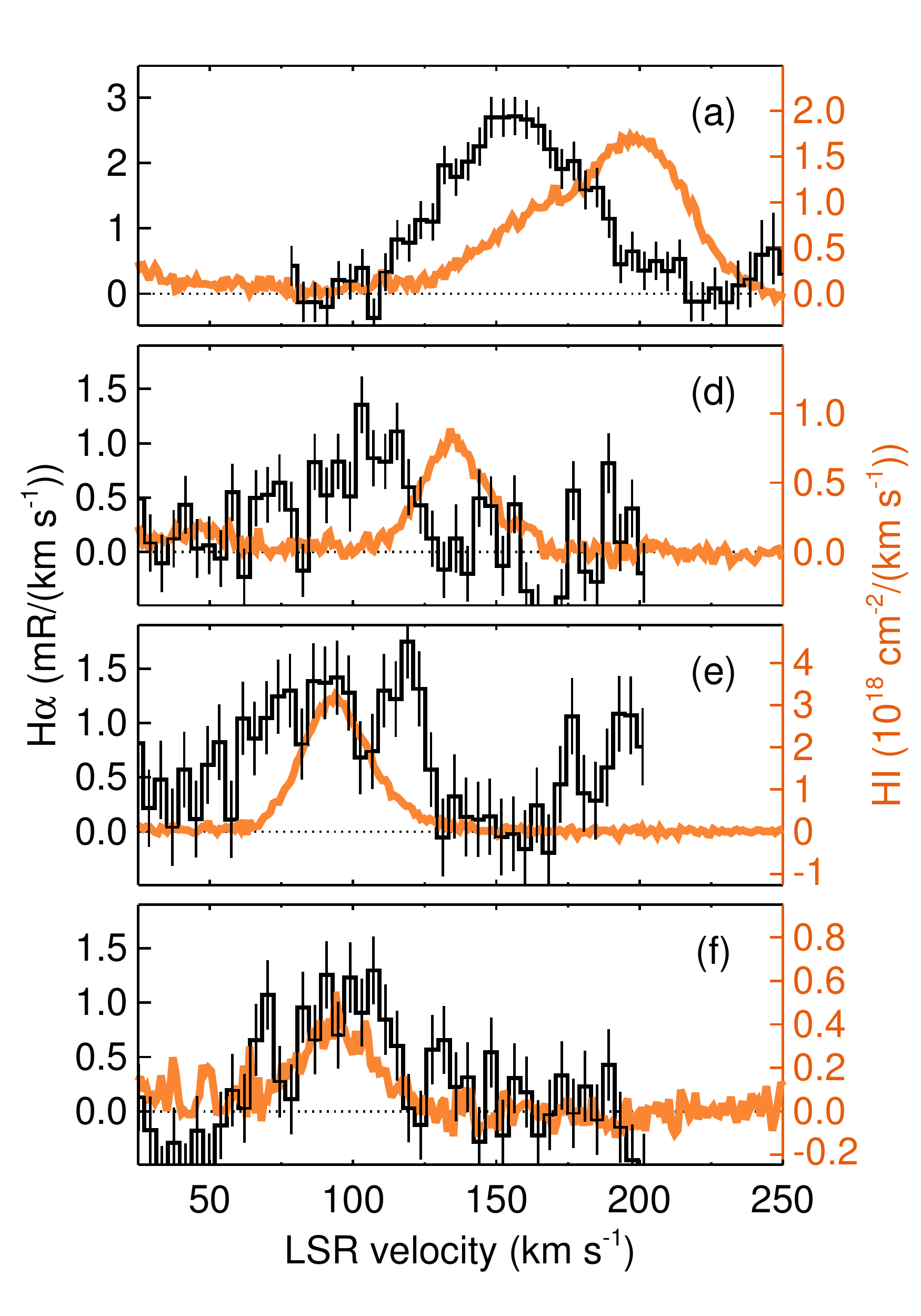}
\includegraphics[scale=0.455,angle=90]{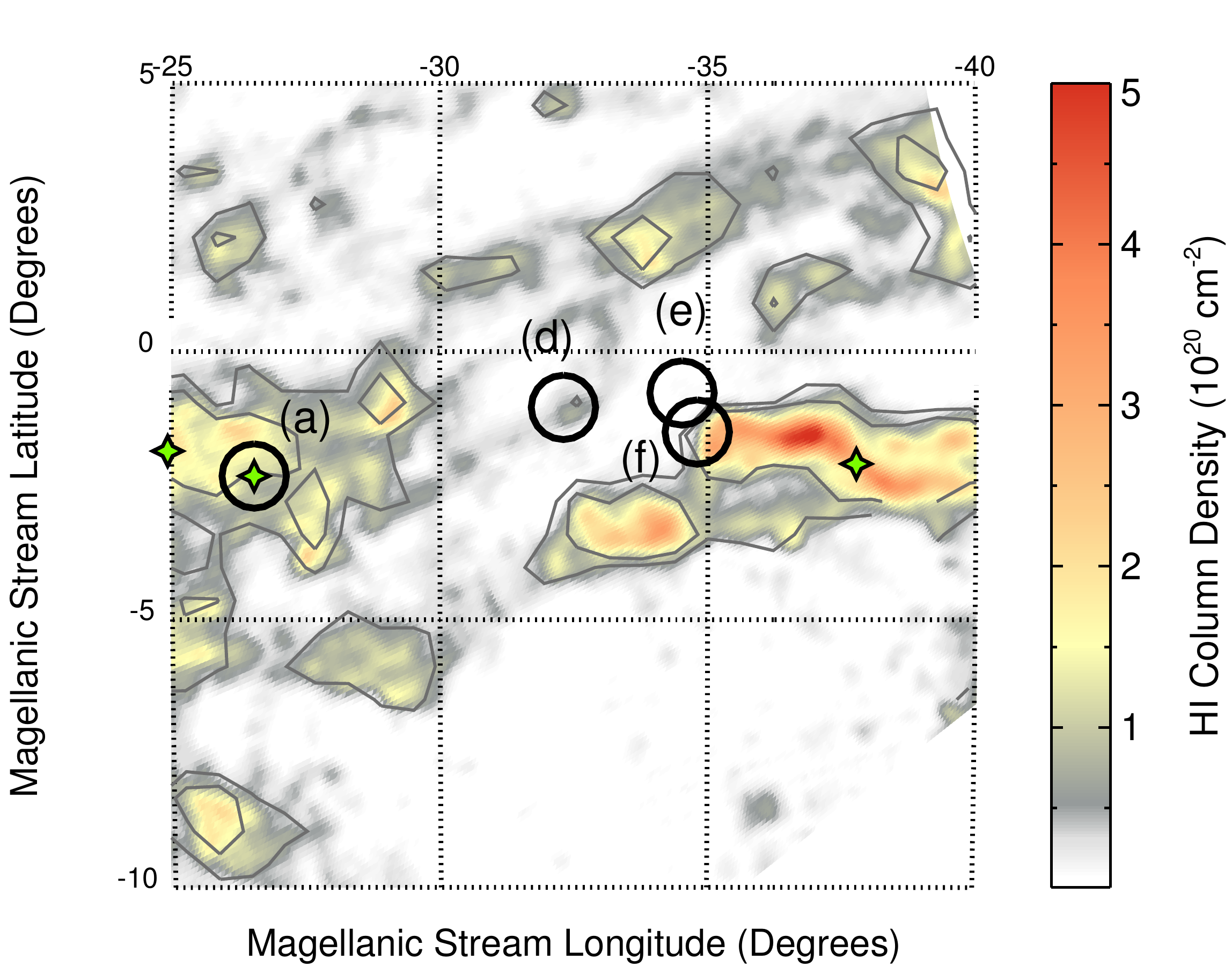}
\end{center}
\figcaption{The H\textsc{~i} column density map on the right shows the neutral gas distribution for the regions surrounding the FAIRALL~9 (a) and RBS~144 (e) sight lines. The \hi\ (orange) and \ha\ (black) spectra along the three targeted sight lines (a) and (d--f) are included on the left. The map and spectra labels coincide with the ones used in Figure~\ref{figure:hi_map} and Table~\ref{table:obs}. The binning of the \ha\ spectra has been reduced from $2~\kms$  to $4~\kms$ to reduce residual systematics from very faint atmospheric lines in the spectra. The black circles in the H\textsc{~i} map (right panel) represent the positions and coverage area of pointed WHAM observations used to produce the spectra in the left panel. The H\textsc{~i} spectra within $1\arcdeg$ have been averaged together to match the WHAM observations. The H\textsc{~i} spectra displayed in the left figure is from the LAB survey and was produced by averaging all the spectra within the WHAM beam. The map in the right figure displays the \hi\ column density over the $+50\le\vlsr\le+250~\kms$ range from the GASS H\textsc{~i} Survey with contour levels at $5\times10^{19}$ and $10^{20}~{\cm^{-2}}$. The three green stars mark the locations of \ha\ observations from \citetalias{2003ApJ...597..948P}. 
\label{figure:RBS144_map}}
\end{figure*}

The sight lines with the largest offsets in \hi\ and \ha\ line centers tend to be toward regions faint in \hi\ as is the case with sight line (d), which is illustrated in Figure~\ref{figure:RBS144_map}; however, regions with low \hi\ column densities do not always equate to large offsets as is the case with the sight line (b) as shown in Figure~\ref{figure:he0226_4110_map}. Below we discuss a few representative examples of the differences and similarities between the \hi\ and \ha\ emission strengths and line centers along high and low \hi\ column density regions. 

\begin{figure*}
\begin{center}
\includegraphics[scale=0.325,angle=90]{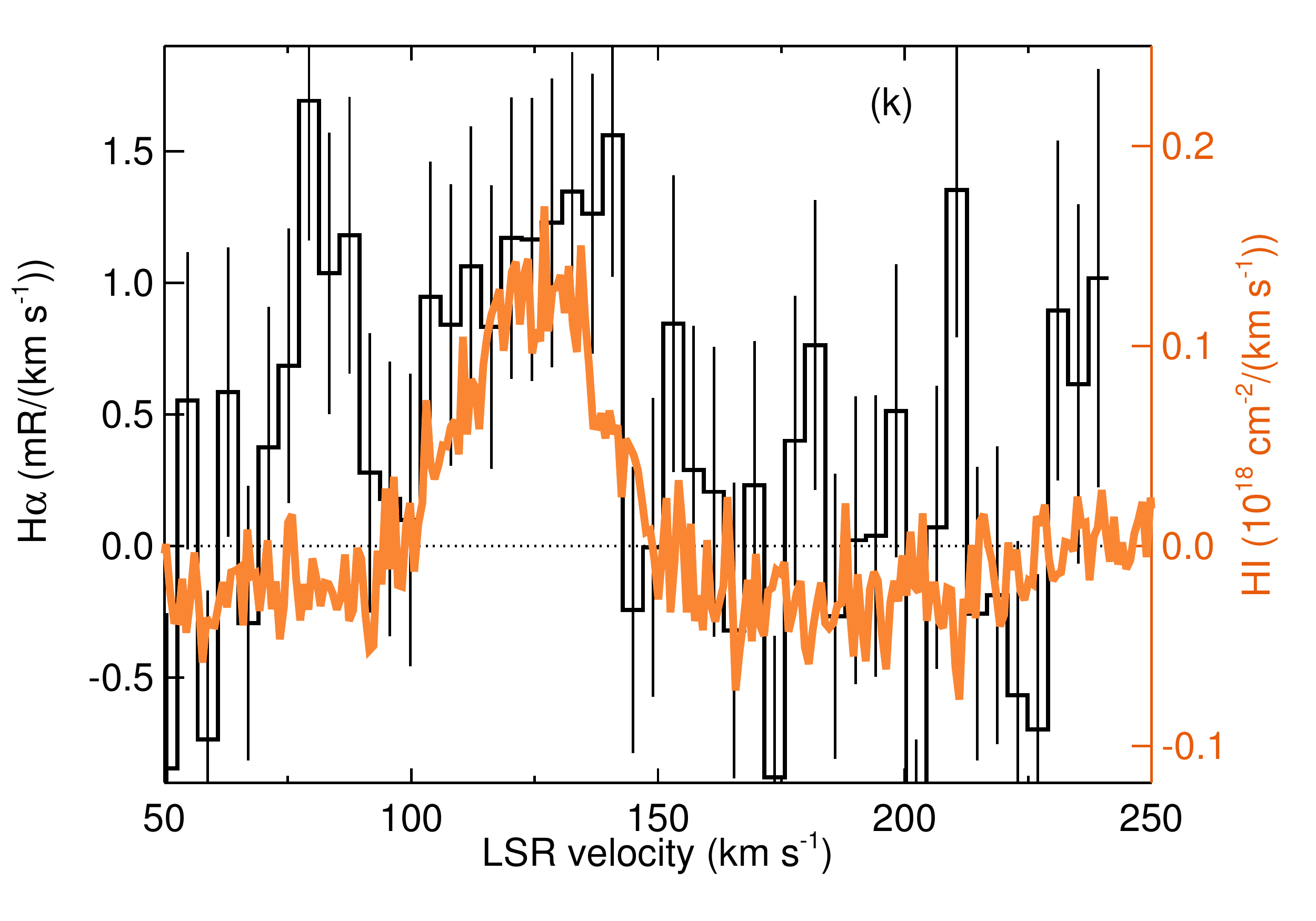}
\includegraphics[scale=0.375,angle=90]{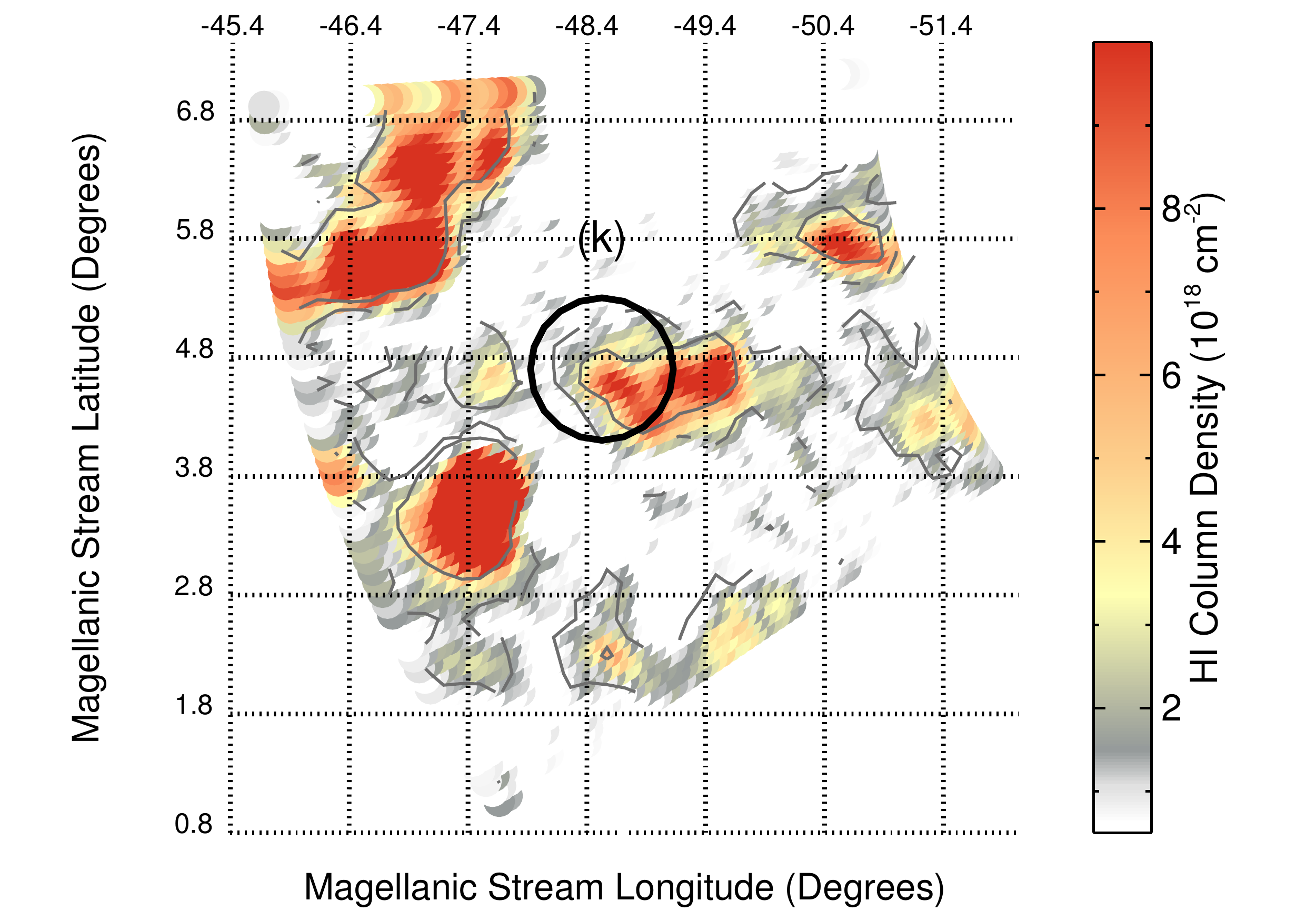}
\end{center}
\figcaption{Same as Figure~\ref{figure:RBS144_map}, but for the region surrounding the Near HE0056-3622 sight line labeled as (k). The map displays the \hi\ column density from the GASS H\textsc{~i} Survey over the $+100\le\vlsr\le+150~\kms$ range with contour levels at $10^{18}$ and $5\times10^{18}~{\cm^{-2}}$.
\label{figure:he0056_3622_map}}
\end{figure*}

\begin{figure*}
\begin{center}
\includegraphics[scale=0.325,angle=0]{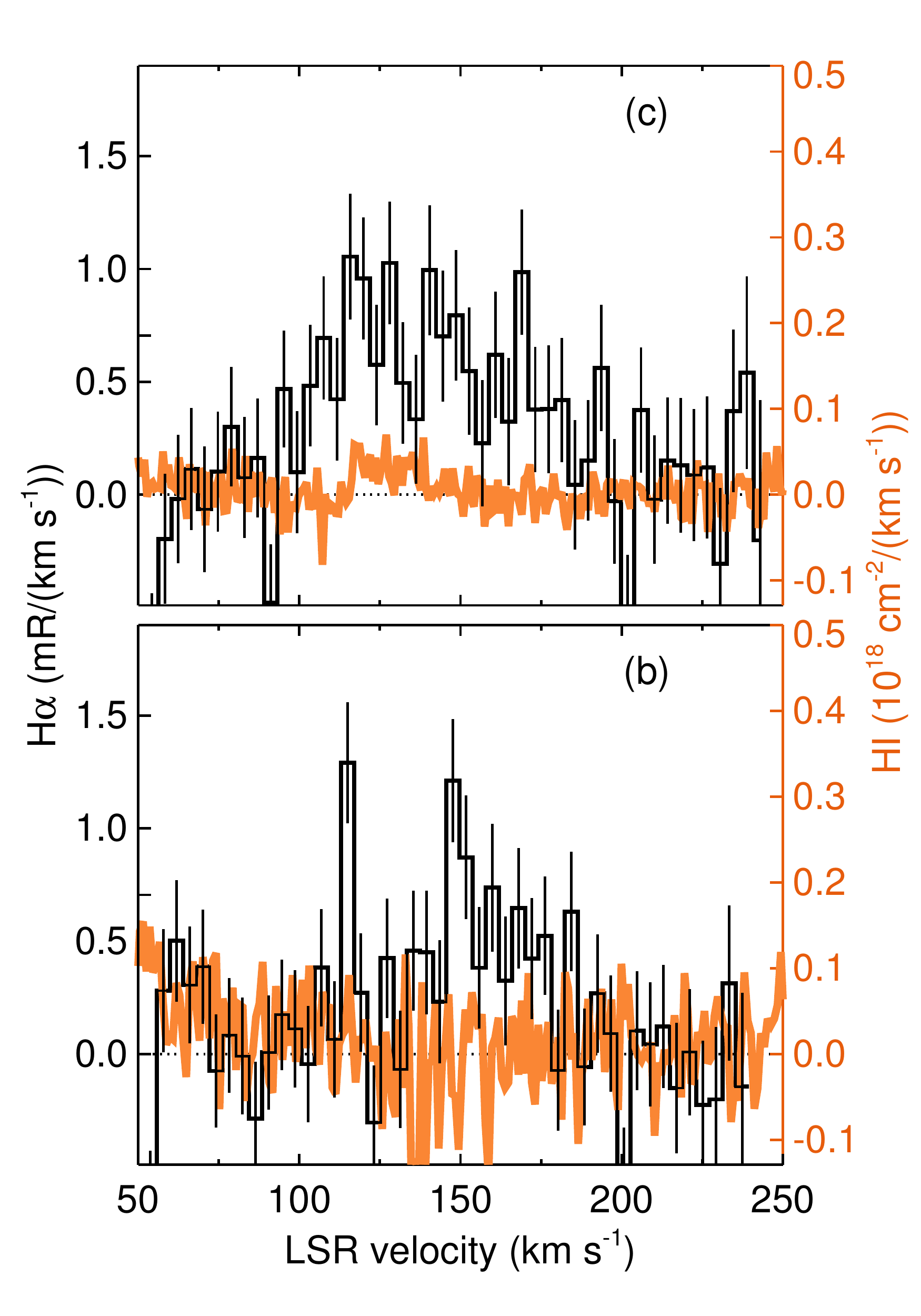}
\includegraphics[scale=0.455,angle=90]{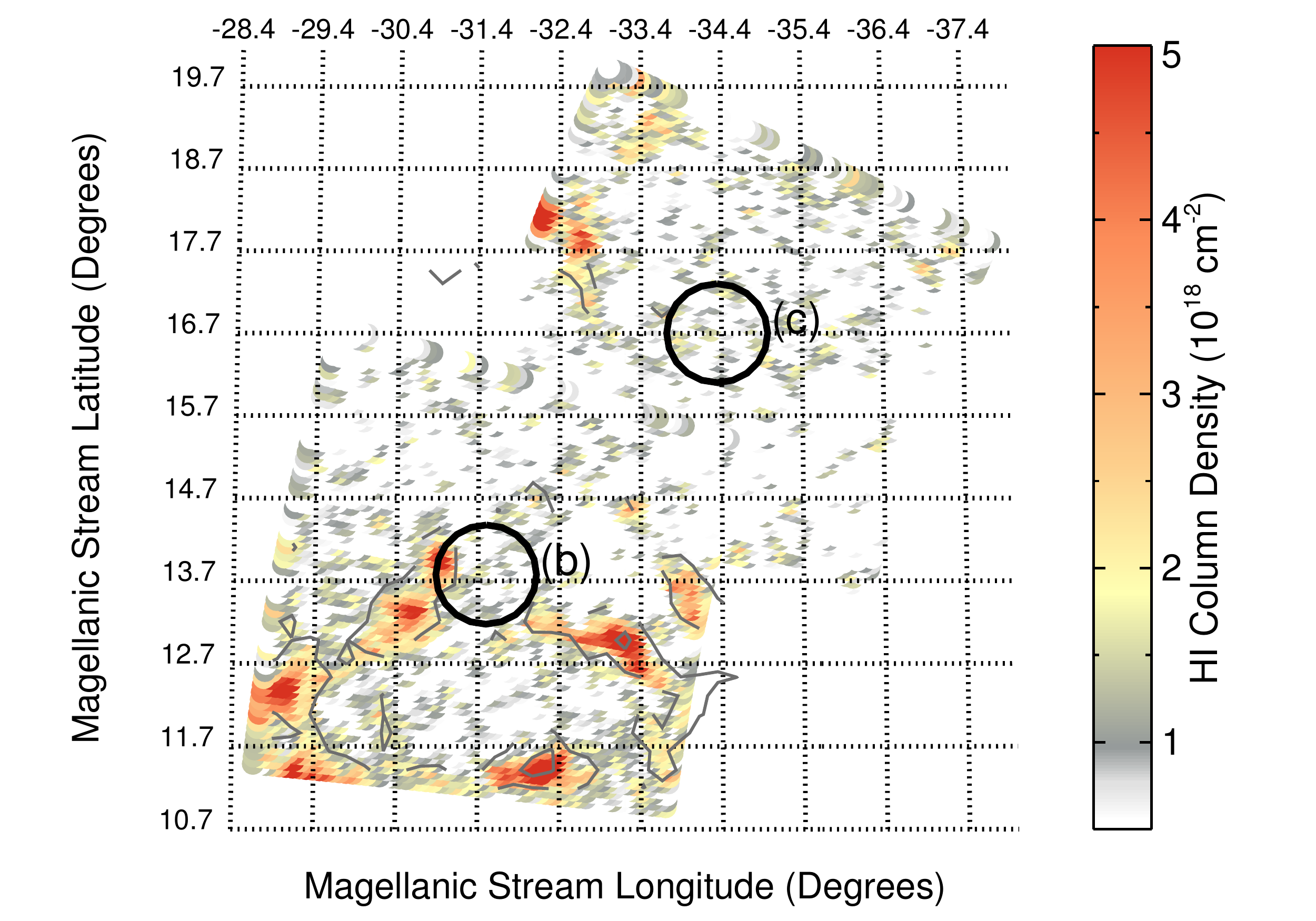}
\end{center}
\figcaption{Same as Figure~\ref{figure:RBS144_map}, but for the region surrounding the HE0226-4110 sight line labeled as~(c). The map displays the \hi\ column density from the GASS H\textsc{~i} Survey over the range $+100\le\vlsr\le+200~\kms$ with contour levels at $1.5\times10^{18}$ and $7\times10^{18}~{\cm^{-2}}$. Although there is no detectable H\textsc{~i} emission along these sight lines when averaged together to match the angular resolution of WHAM, only $2\arcdeg$ from the HE0226-4110 sight line~(c) at $(-32\fdg6, -17\fdg9)$ there is emission with strength $4.5\times10^{18}~\cm^{-2}$ centered at $\vlsr\approx+165~\kms$ and emission with strength $6\times10^{18}~\cm^{-2}$ only $0.6\arcdeg$ away from sight line~(b) at $(-30\fdg9, -13\fdg9)$ centered at $\vlsr\approx+135~\kms$ at the angular resolution of the GASS observations. 
\label{figure:he0226_4110_map}}
\end{figure*}

\begin{figure*}
\begin{center}
\includegraphics[scale=0.325,angle=0]{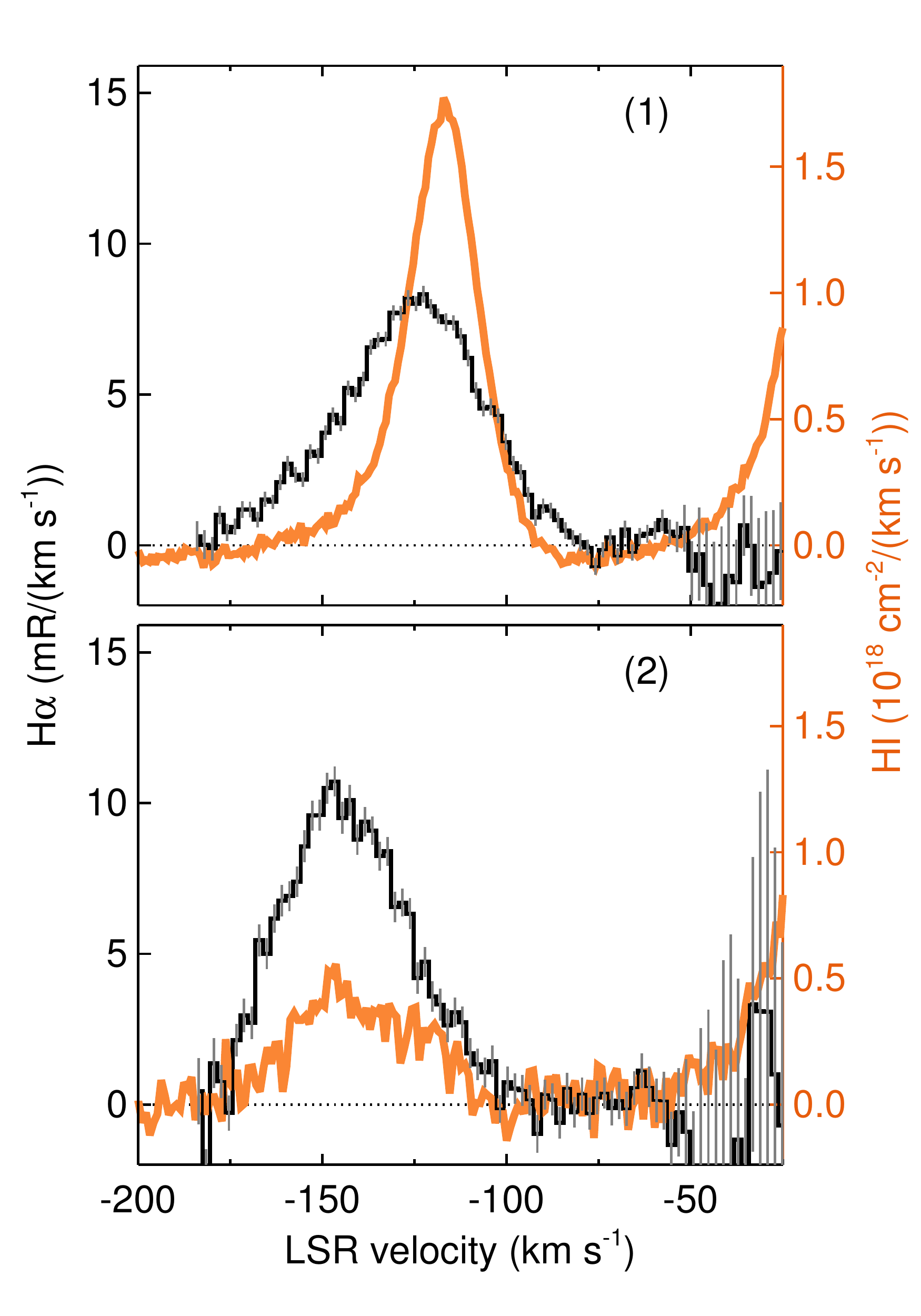}
\includegraphics[scale=0.475,angle=90]{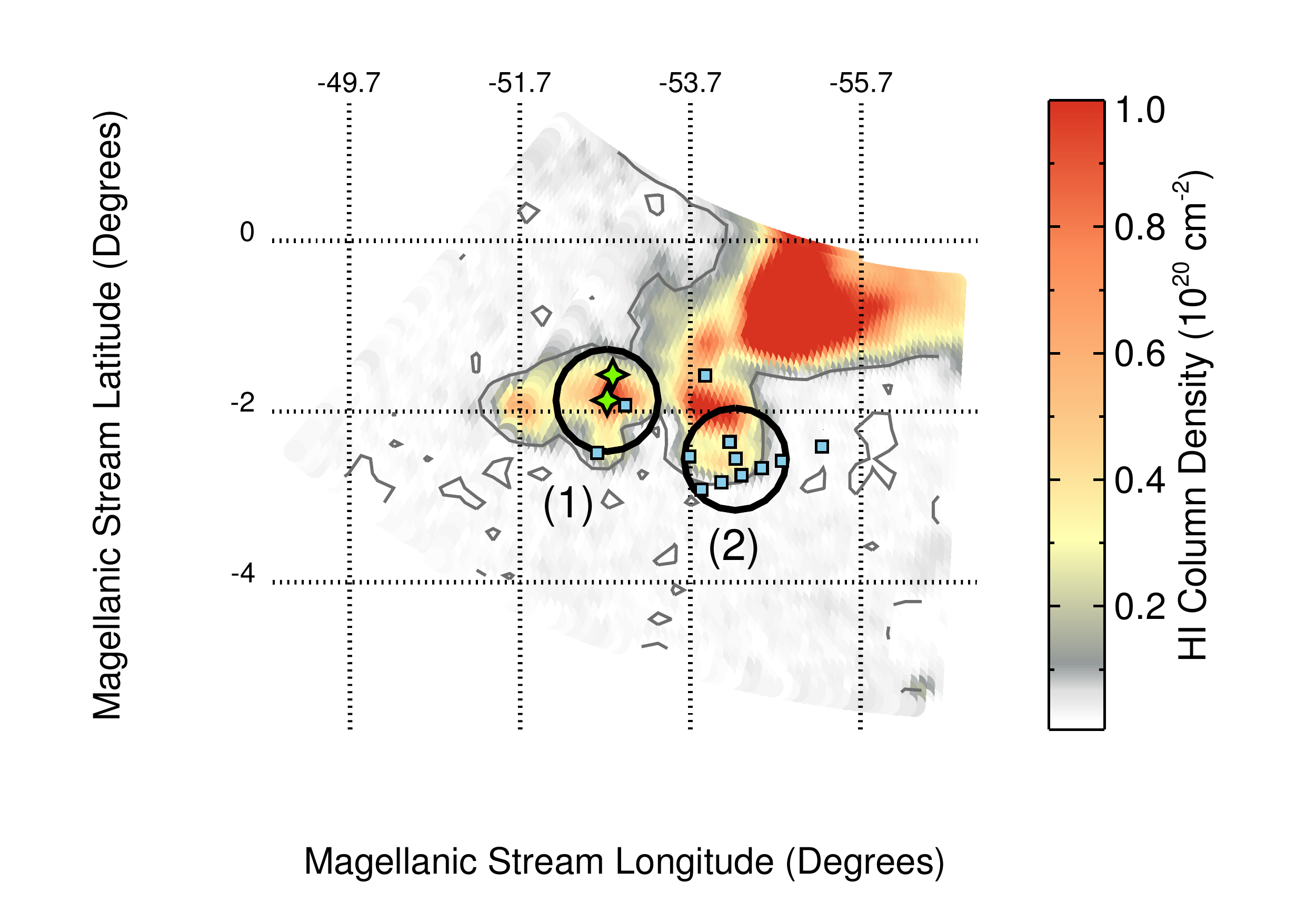}
\end{center}
\figcaption{Same as Figure~\ref{figure:RBS144_map}, but for the region surrounding the~(1) and (2)~\citetalias{2013ApJ...778...58B} sight lines. The two green stars mark the locations of \ha\ observations from \citetalias{2003ApJ...597..948P} and the small light blue squares indicate the positions of the \citetalias{2002ASPC..254..256W} and \citetalias{2003ASSL..281..163W} observations. The map displays the \hi\ column density from the GASS H\textsc{~i} Survey over the $-250\le\vlsr\le-50~\kms$ range with contour levels at $10^{18}$ and $10^{19}~{\cm^{-2}}$.
\label{figure:greg_1_2}}
\end{figure*}

\begin{figure}
\begin{center}
\includegraphics[scale=0.35,angle=90]{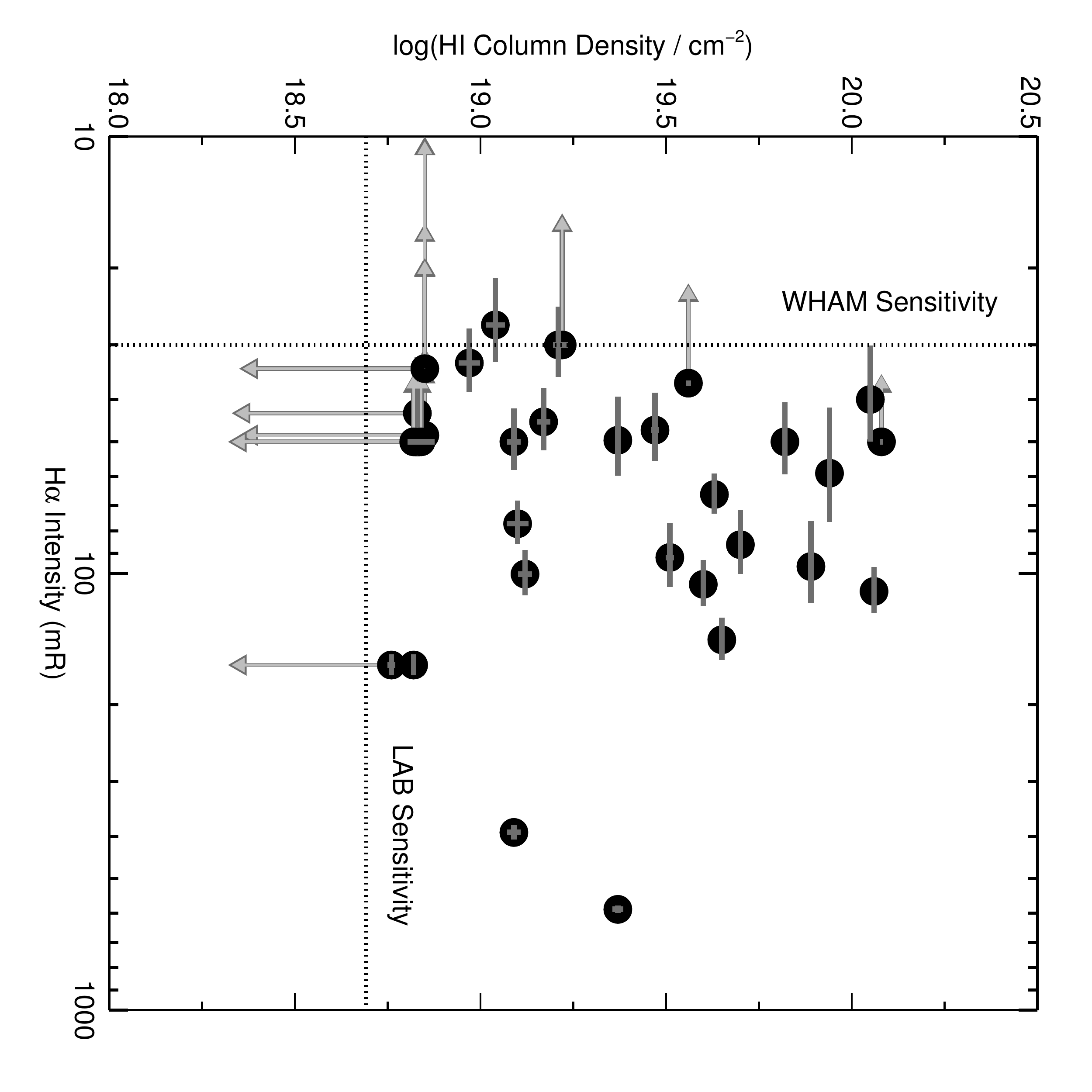}
\end{center}
\figcaption{Comparison of the \ha\ and \hi\ (LAB \hi\ Survey) emission, where \hi\ spectra located within the 1-degree beam have been averaged together to yield same angular resolution. Error bars are shown in gray, with an arrow denoting the upper limit of a non-detection. The dashed lines denote the sensitivity of each survey to emission having $width = 30~\kms$: WHAM, $I_{\rm H\alpha}\approx30~{\rm mR}$ and $N_{\rm H{\textsc{~i}}}\approx4.9\times10^{18}~\cm^{-2}$.
\label{figure:column_intensity}}
\end{figure}

Near the LMC at positions~(e) and (f), we detect bright \ha\ emission along the edge of the low metallicity Stream filament (see Figure~\ref{figure:RBS144_map}). The \hi\ and \ha\ emission along these sight lines peak at roughly the same velocity. Sight line~(e) lies off of the $\log{\rm N_{\rm H\textsc{~i}}/\cm^{-2}}=19$ contour and has an ${\rm N_{\rm H\textsc{~i}}}$ that is factor of seven less than sight line~(f), which straddles the edge of this filament; however, the \ha\ intensity decrease much more gradually, reducing only by a factor of $2$ (see Table~\ref{table:intensities}). This small change in \ha\ intensity with such a large change in \hi\ column density suggests that the ionized gas extends much further than the neutral gas. This is consistent with the results of the \citet{2014ApJ...787..147F}, which found that the ionized gas extends up to $30\arcdeg$ of the $\log{\rm N_{\rm H\textsc{~i}}/\cm^{-2}}=18$ gas when traced by UV absorption. 

Numerous \hi\ cloudlets are offset and moving with the main Magellanic Stream filaments, including one that intersects with sight line~(d). The \ha\ emission along this sight line lags behind the \hi\ emission by roughly $30~\kms$ (see Figure~\ref{figure:RBS144_map} and Table~\ref{table:intensities}), whereas the neutral and ionized gas at the nearby sight line~(f), only 2-degrees away and on the edge of the \hi\ filament, have a very similar velocity. However, as the kinematic span of the \ha\ emission along the~(d) and~(f) sight lines is very similar, the large offset in the \hi\ and \ha\ line centroids along sight line~(d) may indicate that the ionized and neutral gas are not well mixed. 

\begin{figure}
\begin{center}
\includegraphics[trim=0 25 0 25,clip,scale=0.275,angle=0]{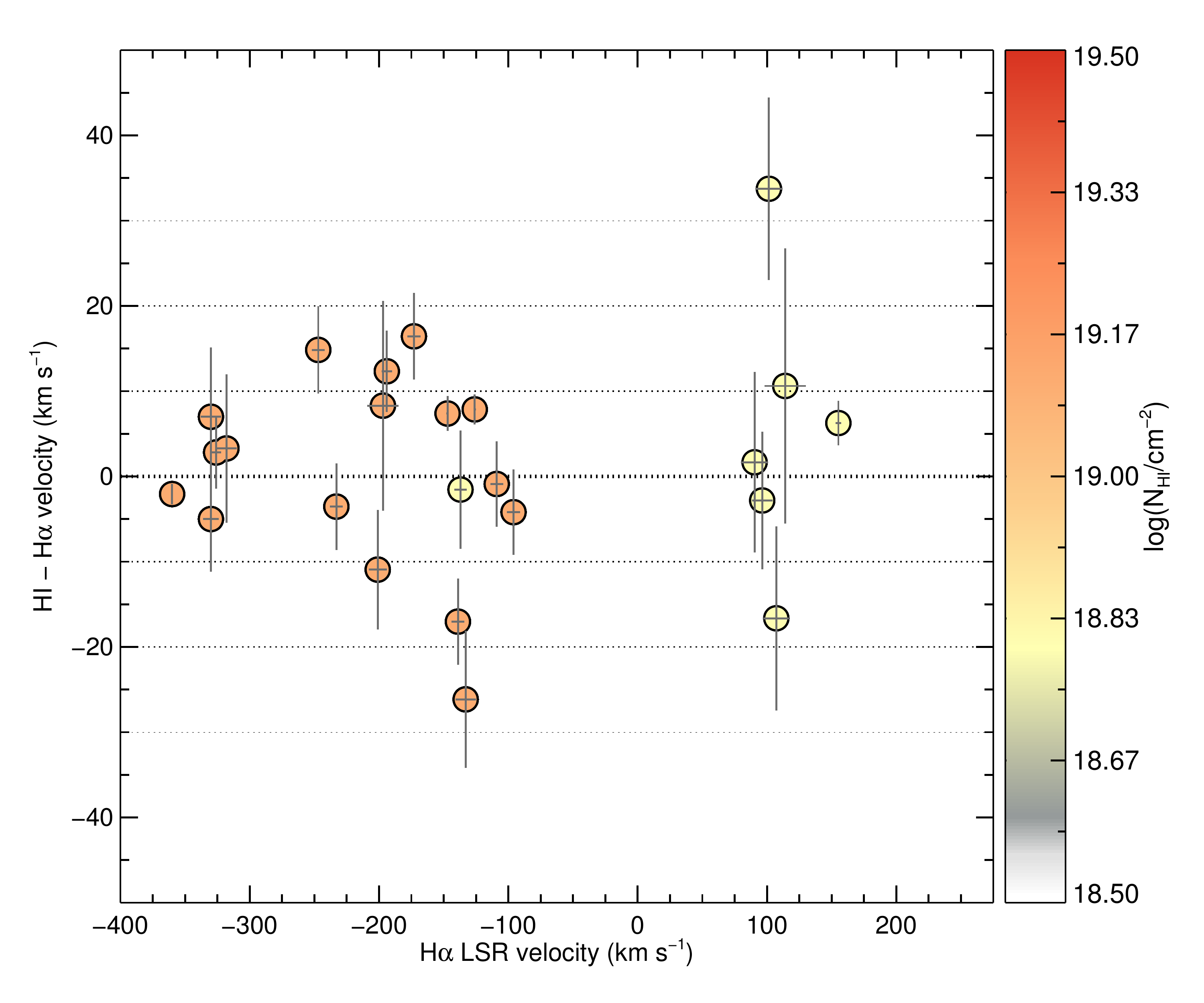} \\
\includegraphics[trim=0 25 0 25,clip,scale=0.275,angle=0]{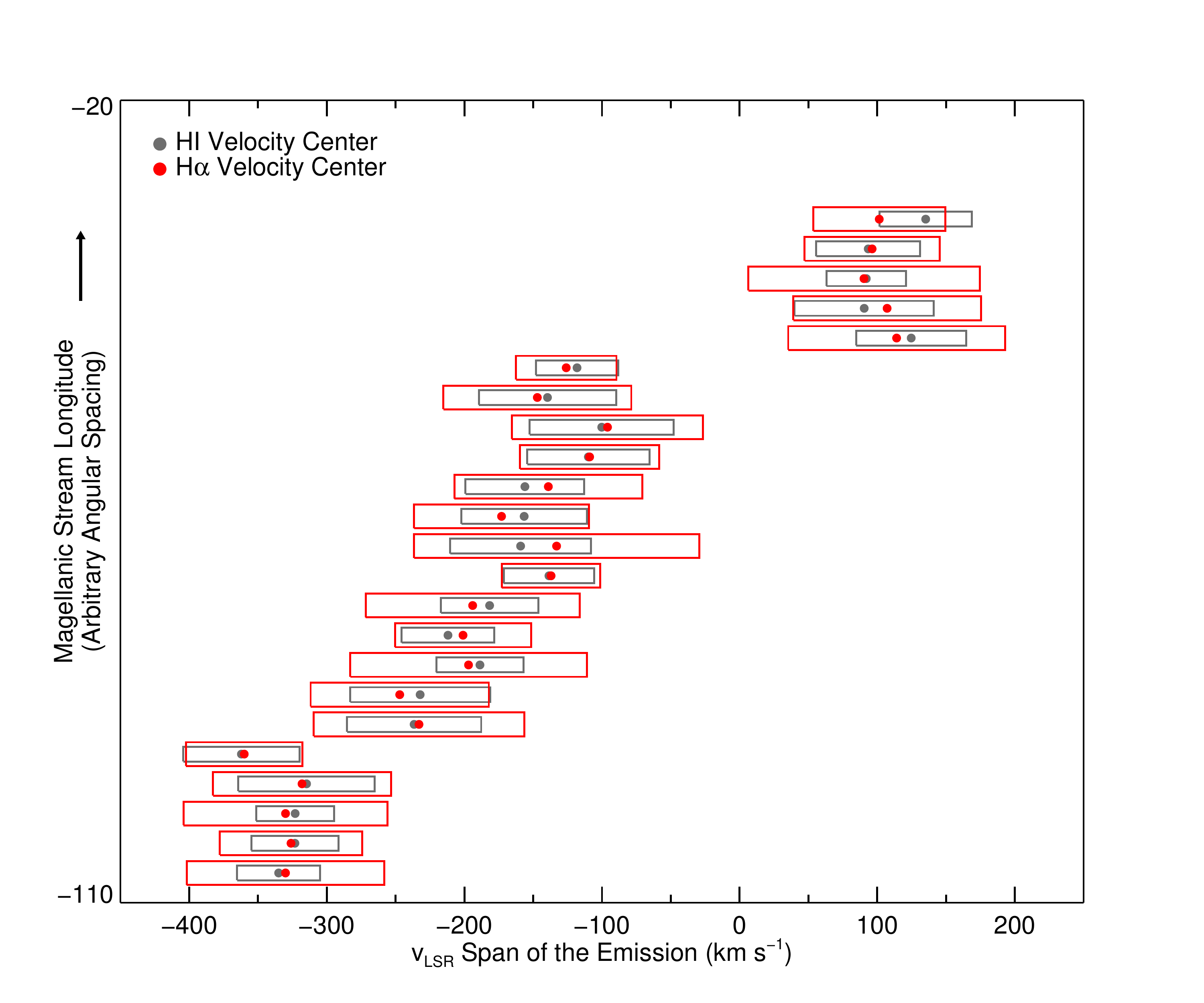} \\
\end{center}
\figcaption{\ha\ and \hi\ kinematics. 
\textit{(Top)} The offset of the \ha\ line center from the \hi\ with the symbol color representing the strength of the $N_{\rm H\textsc{~i}}$ emission. The horizontal lines at $0,~\pm10,~\pm20, and~\pm30~\kms$ are provided for reference. \textit{(Bottom)} The \ha\ and \hi\ LSR centroid position and the velocity spread of the lines. The vertical displacement follows the Magellanic Stream longitude at an arbitrary angular separation. The vertical sizes of the \ha\ and \hi\ rectangles are different only so that they are easier to visualize. The fit parameters of the \hi\ lines were measured from LAB \hi\ Survey spectra that were averaged together to match the WHAM angular resolution.
\label{figure:compare_vel}}
\end{figure}

Toward a similarly detached cloud probed by sight line (k) positioned roughly $15\arcdeg$ downstream, we find that the kinematics of the neutral and ionized gas phase track each other well (see Figures~\ref{figure:he0056_3622_map} and~\ref{figure:hi_map}). Though, it is also important to note that the cloud at position (k) might not actually be associated with the Magellanic Stream. The Sculptor Group of galaxies also lies in this general direction and at a similar velocity (e.g., \citealt{2003ApJ...586..170P}). \citet{2013ApJ...772..110F} also detected this cloud  at  $\vlsr=+150~\kms$ through UV absorption and measured its oxygen abundance to be roughly a tenth solar; this abundance is very similar to the values they measure throughout the low metallicity filament of the Magellanic Stream. The $\vlsr=+125~\kms$ \ha\ and \hi\ emission along this sight line could therefore be associated with either the Magellanic Stream or the Sculptor Group.

Like the cloud at position~(k), the sight lines at positions~(1) and~(2) also lie near $l_{\rm MS}=-50\arcdeg$, but on an offset cloud positioned at the low Magellanic Stream Latitude edge of an \hi\ filament (see Figures~\ref{figure:greg_1_2} and~\ref{figure:hi_map}). These two sight lines have an $I_{\rm H\alpha}$ that is $\sim4-6$ times greater than sight line (k) positioned $8\arcdeg$ away and are over $200~{\rm mR}$ brighter than any of the other (a-s) and (3-19) sight lines presented in this study. Kinematically resolved mapped observations of the warm ionized gas in the Magellanic Stream are needed to ascertain why this portion of the Stream is so bright compared to the rest of the Stream and how well the neutral and ionized gas phases are mixed.

Although we generally detect \ha\ emission on or near bright \hi\ structures, we also detect ionized gas many degrees from the \hi. Sight line~(b) lies on the edge of multiple small \hi\ clouds (Figure~\ref{figure:he0226_4110_map}). At this location, the \hi\ column density is below the sensitivity of the GASS \hi\ survey, which is $N_{\rm H{\textsc{~i}}}=1.6\times10^{18}~\cm^{-2}$ at a width of $30~\kms$ \citep{2009ApJS..181..398M}, but we detect warm ionized gas emission with $I_{\rm H\alpha}=69\pm5~{\rm mR}$. Although this sight line lies $0\fdg6$ from an \hi\ cloud at $(l_{\rm MS},~b_{\rm MS})=(-30\fdg9, -13\fdg9)$, the \ha\ emission lines up exceedingly well with the nearest sight line with detectible \hi\ emission at $\vlsr\approx+135~\kms$ (see Table~\ref{table:intensities}). Even further off these \hi\ clouds, we detect \ha\ emission along sight line~(c) at $I_{\rm H\alpha}=53\pm9~{\rm mR}$. Using \hi\ Lyman series absorption,  \citet{2005ApJ...630..332F} measured $\log{N_{\rm H\textsc{~i}}/\cm^{-2}}=17.05\pm0.10$ over the $+80\le\vlsr\le+230~\kms$ velocity range using Far Ultraviolet Space Explorer (\textit{FUSE}) observations with a pencil beam angular resolution; through photoionization modeling, they found that the Stream is more than $97\%$ ionized along this direction. At $2\arcdeg$ from this sight line at $(l_{\rm MS},~b_{\rm MS})=(-32\fdg6, -17\fdg9)$, the \hi\ emission peaks $10~\kms$ from \ha\ emission (${\rm v}_{\rm LSR,~H\alpha}\approx+155$ and ${\rm v}_{\rm LSR,~H\textsc{~i}}\approx+165~\kms$). 

Overall, we find that the \ha\ emission spatially tracks the \hi\ emission of the Magellanic Stream well. In $4$~sight lines, we only detect \ha\ emission, which indicates that the ionized gas has a larger cross section at the sensitivity of the WHAM ($\sim30~{\rm mR}$) observations and the \hi\ surveys (LAB: $4.9\times10^{18}~\cm^{-2}$; GASS: $1.6\times10^{18}~\cm^{-2}$). We also find that the strengths of these lines are not correlated (Figure~\ref{figure:column_intensity}), which suggests that the warm ionized gas phase ($10^4~{\rm K}$) is predominantly photoionized (see the discussion in this section above). The  \hi\ and \ha\ velocity centroids agree well on the main \hi\ Stream filaments, but less so for sight lines positioned off or on the cloudlets away  from the main body of the Stream. This may indicate that clouds on the edge of the Stream are less shielded from their environment. However, the overall kinematic difference between the \hi\ and \ha\ centroids is centered at $\sim0~\kms$ (see Figure~\ref{figure:compare_vel}).

\subsection{Properties of the Ionized Gas}\label{section:ionfrac}

\begin{deluxetable*}{ccccccccc}
\tabletypesize{\scriptsize}
\tablecaption{Ionization Fraction from $N_{\rm H\textsc{~i}}$ and $I_{\rm H\alpha}$ \label{table:ion_frac_hi_ha}}
\tablewidth{0pt}
\tablehead{
\colhead{ } & \multicolumn{2}{c}{Observed} & \colhead{} & \multicolumn{3}{c}{Modeled\tablenotemark{a}} & \colhead{} & \colhead{Derived\tablenotemark{b}} \\
\cline{2-3} \cline{5-7} \cline{9-9} 
\colhead{ID} & \colhead{$\log N_{\rm H\textsc{~i}}/\cm^{-2}$} & \colhead{$I_{\rm H\alpha}$} & \colhead{} & \colhead{$L$} & \colhead{$\chi_{\rm H\textsc{~ii}}$} & \colhead{} & \colhead{} & \colhead{$f_{\rm H\textsc{~ii}}$}   \\
\colhead{} & \colhead{} & \colhead{(mR)}  & \colhead{} & \colhead{(\kpc)}  & \colhead{} & \colhead{}  & \colhead{}
}
\startdata
a	&	$19.56\pm0.01$ ($20.01\pm$0.01)\tablenotemark{c}	&	$165\pm8$  	&& $4$		& $0.36$	&&& $0.08-0.11$ $(0.15-0.22)\tablenotemark{c}$ \\
e	&	$19.89\pm0.01$ ($19.98\pm$0.02)\tablenotemark{c}	&	$101\pm9$ 	&& $23$	& $0.47$ 	&&& $0.08-0.11$ $(0.09-0.13)\tablenotemark{c}$ \\
\enddata
\tablenotetext{a}{CLOUDY model solutions for electron temperatures of $T_4=1$ with the \hi\ column densities and Si\textsc{~iii}/Si\textsc{~ii} ratios constrained from observations \citep{2013ApJ...772..110F}.} 
\tablenotetext{b}{Assumes $T_4=[0.8-1.2]$.} 
\tablenotetext{c}{Using the $N_{\rm H\textsc{~i}}$ that only align with the \ha\ emission; the values enclosed within the parentheses include all H\textsc{~i} components consistent with Magellanic Stream velocities (see Table~\ref{table:intensities}).} 
\end{deluxetable*}             

To constrain how much H\textsc{~ii} gas fills the WHAM beam and the fraction of ionized gas in the Magellanic Stream, we compared the \hi, \ha, and \oi\ emission with the UV absorption from the \citetalias{2014ApJ...787..147F} study. 

\subsubsection{Filling Factor of the H\textsc{~ii} Gas}

The intensity of the \ha\ emission is directly proportional to the rate of the recombination (\citealt{1991ApJ...372L..17R}):
\begin{equation}\label{eq:IHa}
I_{\rm H\alpha}=\frac{1}{2.75}~T_4^{-0.924}\left(\frac{\int{n_e^2}dl}{\rm cm^{-6}\ \pc}\right)~R, 
\end{equation}

\noindent where $T_4=\left(T_e/10^4~{\rm K}\right)$ with $T_e$ being the electron temperature of the gas. The $2.75^{-1} T_4^{-0.924}$ factor accounts for the rate of recombination and the subsequent probability of producing an \ha\ photon in a gas optically thick to Lyman Continuum radiation (see \citealt{1988ApJS...66..125M} for these recombination coefficients). The integral term is known as the emission measure ($EM\equiv\int n_e^2 dl$), where we have assumed that $n_{\rm H\textsc{~ii}}\approx n_e$. Here $n_e$ is the electron density and $dl$ is the line-of-sight path length over which the electrons are recombining. 

Unfortunately, distribution of the electrons along the line of sight is unknown. A common approach is to parameterize the emission measure as $EM = n_c^2 f_{\rm H\textsc{~ii}} L$, where $n_c$ now represents a characteristic electron density and $L$ is the path length of the emitting gas along the line of sight (e.g., \citealt{1991ApJ...372L..17R}). The fraction of the beam that is filled with gas is known as the filling factor ($f$) and is a dimensionless quantity that varies between $0 \le f \le 1$. The product $fL$ is known the occupation length, which is the average portion of the line-of-sight depth that harbors gas. For a sight line in which the \hi\ column density is known, we define the ionization fraction of hydrogen as $\chi_{\rm H\textsc{~ii}} = N_{\rm H\textsc{~ii}} / (N_{\rm H\textsc{~i}}+N_{\rm H\textsc{~ii}})$. Here, the column density of ionized hydrogen is then $N_{\rm H\textsc{~ii}} = \sqrt{EM f_{\rm H\textsc{~ii}}L}$, or $N_{\rm H\textsc{~ii}} = \sqrt{2.75 I_{\rm H\alpha} T_4^{+0.924} f_{\rm H\textsc{~ii}}L}$ with equation~(\ref{eq:IHa}) and assuming that $n_p\approx n_e$.  
\begin{figure}
\begin{center}
\includegraphics[scale=0.35,angle=90]{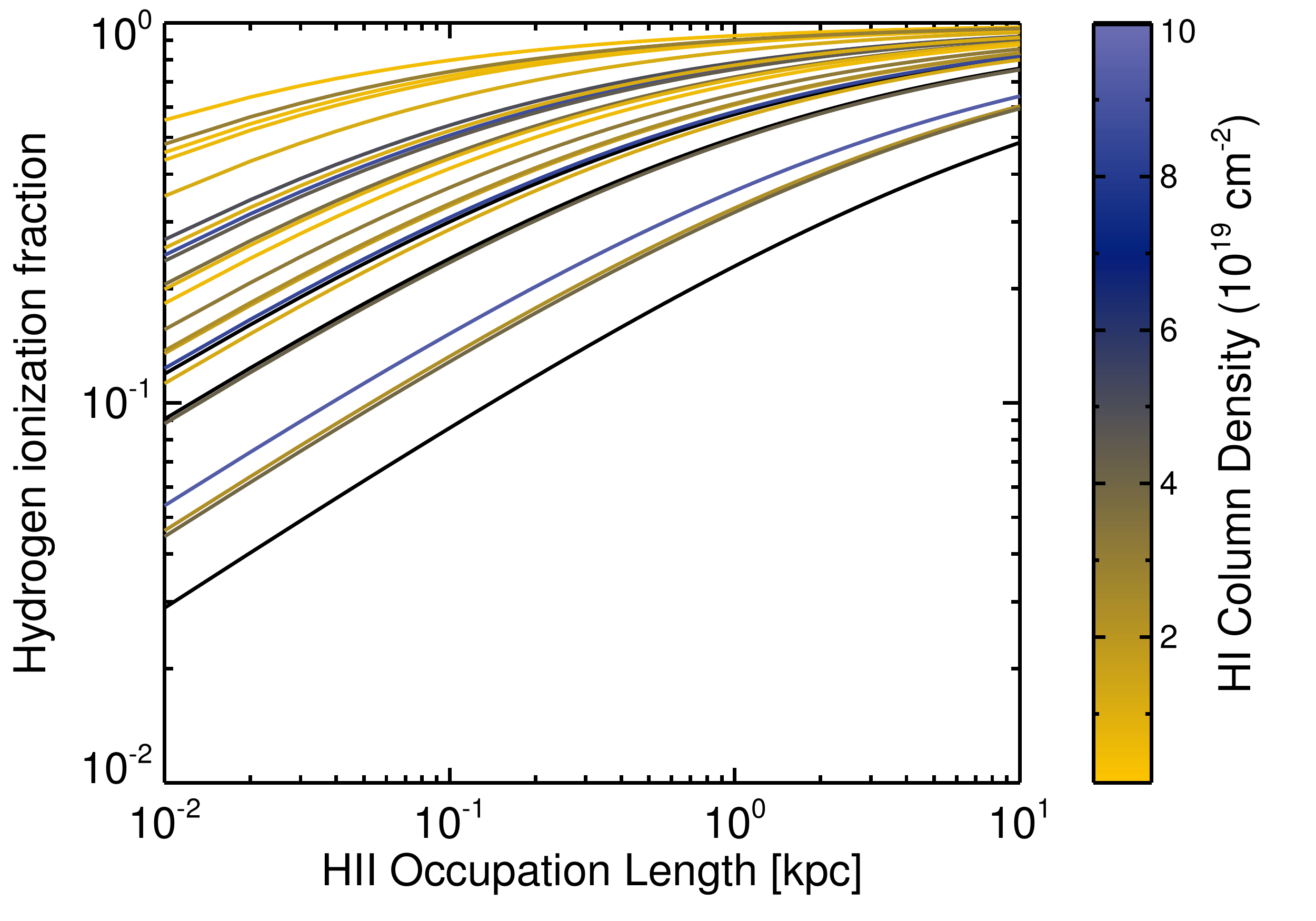}
\end{center}
\figcaption{The $\chi_{\rm H\textsc{~ii}}$ trends (yellow-blue lines) for the $25$ sight lines (out of $39$) with detected \ha\ and \hi\ emission as a function of the H\textsc{~ii} occupation length ($fL$), assuming $T_4=1$. The color of these lines illustrates the strength of the ${N_{\rm H\textsc{~i}}}$. 
\label{figure:ion_frac}}
\end{figure}

With the $N_{\rm H\textsc{~i}}$ and $I_{\rm H\alpha}$ detections alone, the occupation length is unconstrained. To illustrate how the occupation length varies with hydrogen ionization faction for our measured $N_{\rm H\textsc{~i}}$ and $I_{\rm H\alpha}$ along the Magellanic Stream, we assume $-2\le \log\left(f_{\rm H\textsc{~ii}}L\right)\le1$ in Figure~\ref{figure:ion_frac} such that the values span the $f_{\rm H\textsc{~ii}}=0.01$ and $L=1~\kpc$ to $f_{\rm H\textsc{~ii}}=1$ and $L=10~\kpc$ parameter space. For reference, at the distance of the Magellanic Clouds ($d_{\odot}\approx55~\kpc$) the width of the $1\arcdeg$ WHAM beam corresponds to a projected physical width of $\sim1~\kpc$; at the tip, the Stream might lie as far as $d_{\odot}\approx100-200~\kpc$ as discussed in Section~\ref{section:intro}, which would correspond to a projected width of $\sim1.7-3.5~\kpc$ for the WHAM beam. We find that over this $f_{\rm H\textsc{~ii}}L$ range, the $\chi_{\rm H\textsc{~ii}}$ would vary by factor of $10$ or more for small occupation lengths of $f_{\rm H\textsc{~ii}}L\lesssim0.1~\kpc$ (Figure~\ref{figure:ion_frac} and Table~\ref{table:intensities}). The hydrogen ionization fraction begins to converge to $\chi_{\rm H\textsc{~ii}}\approx1$ at $f_{\rm H\textsc{~ii}}L\gtrsim100~\kpc$. Therefore, the combination of $N_{\rm H\textsc{~i}}$ and $I_{\rm H\alpha}$ is especially insensitive to the hydrogen ionization fraction at small occupation lengths. This is not too surprising that the combination of \hi-21~\cm\ and \ha\ emission is inadequate for determining $f_{\rm H\textsc{~ii}}$ as the \hi\ and \ha\ emission are uncorrelated as illustrated in Figure~\ref{figure:column_intensity}. For comparison, \citetalias{2014ApJ...787..147F} found that the hydrogen ionization fraction tends to increase with decreasing $N_{\rm H\textsc{~i}}$ for the low-ionization gas phase.

However, by combining our measurements of the \ha\ and \hi\ emission with UV absorption-line results on the Stream's $L$ and $\chi_{\rm H\textsc{~ii}}$ from the \citetalias{2014ApJ...787..147F} study, we can constrain the filling fraction of the ionized hydrogen: 
\begin{equation}
f_{\rm H\textsc{~ii}}=\frac{1}{2.75 I_{\rm H\alpha} T_4^{+0.924} L}\left(\frac{N_{\rm H\textsc{~i}}~\chi_{\rm H\textsc{~ii}}}{1- \chi_{\rm H\textsc{~ii}}}\right)^2
\end{equation}
The FAIRALL~9 and RBS~144 sight lines at positions (a) and (e) overlap with the \citetalias{2014ApJ...787..147F} sample. Table~\ref{table:ion_frac_hi_ha} lists the $\chi_{\rm H\textsc{~ii}}$ and $L$ found in the \citetalias{2014ApJ...787..147F} study and the measured $I_{\rm H\alpha}$ and $N_{\rm H\textsc{~i}}$ values from this study for sight lines (a) and (e). Assuming a typical electron temperature of $[0.8-1.2]\times10^4~{\rm K}$ in \ha\ emitting regions (e.g., \citealt{2005ApJ...630..925M, 2006ApJ...652..401M}; \citetalias{2014ApJ...787..147F} found that $T_4\approx1$ through CLOUDY radiative transfer modeling), we find that $f_{\rm H\textsc{~ii}}=0.08-0.11$ along both sight lines. Table~\ref{table:ion_frac_hi_ha} summarizes these results.

\begin{deluxetable*}{cccccccc}
\tabletypesize{\scriptsize}
\tablecaption{Ionization Fraction from $I_{\rm O\textsc{~i}}/I_{\rm H\alpha}$ \label{table:ion_frac_ha_oi}}
\tablewidth{0pt}
\tablehead{
\colhead{ } & \multicolumn{3}{c}{Observed} &\colhead{} &\colhead{Derived\tablenotemark{a}} & \colhead{} &\colhead{Modeled\tablenotemark{b}}\\
\cline{2-4} \cline{6-6} \cline {8-8}
\colhead{ID} & \colhead{$\log N_{\rm H\textsc{~i}}/\cm^{-2}$} & \colhead{$I_{\rm H\alpha}$} & \colhead{$I_{\rm O\textsc{~i}}$} & \colhead{} & \colhead{$\chi_{\rm H\textsc{~ii}}$\tablenotemark{a}} & \colhead{} & \colhead{$\chi_{\rm H\textsc{~ii}}$} \\
\colhead{} & \colhead{} & \colhead{(mR)} & \colhead{(mR)} & \colhead{} 
}
\startdata
a	&	$19.48\pm0.01$	&	$165\pm8$ 	& $33\pm8$	&& $0.16$, $0.32$, $0.47$\tablenotemark{d} && $0.30$	\\
e	&	$19.93\pm0.01$	&	$101\pm9$ 	& $45\pm8$	&& $0.31$, $0.51$, $0.67$\tablenotemark{e} && $0.43$	 \\
1	&	$19.14\pm0.02$	&	$417\pm1$ 	& $<40$		&& $<0.67$, $<0.83$, $<0.90$\tablenotemark{e}	&& \nodata \\
\enddata
\tablenotetext{a}{Derived from the observed $I_{\rm O\textsc{~i}}$ and $I_{\rm H\alpha}$.} 
\tablenotetext{b}{CLOUDY model solutions for $T_4=1$ with the \hi\ column densities and Si\textsc{~iii}/Si\textsc{~ii} ratios constrained from observations \citep{2013ApJ...772..110F}.} 
\tablenotetext{c}{At $T_4=[0.8,~1,~1.2]$.} 
\tablenotetext{d}{Assumes $0.5$ solar metallicity as measured by \citet{2000AJ....120.1830G, 2013ApJ...772..111R}.} 
\tablenotetext{e}{Assumes $0.1$ solar metallicity as measured by \citet{2000AJ....120.1830G, 2010ApJ...718.1046F, 2013ApJ...772..110F}.} 
\end{deluxetable*}  

\subsubsection{Ionization Fractions through $\rm H\alpha$ and $[\rm O\textsc{~i}]$}\label{section:oi_ha}

Unlike the $N_{\rm H\textsc{~i}}$ and $I_{\rm H\alpha}$ combination, the $\chi_{\rm H\textsc{~ii}}$ does not depend on the geometry of the emitting cloud when solved for using \ha\ and \oi$~\lambda6300$. This is because the \ha\ recombination line and collisionally excited \oi$~\lambda6300$ emission lines both trace the $n_e^2$ along the line of sight (i.e., $EM$). Assuming that this emission is from the same gas, their $f_{\rm H\textsc{~ii}}L$ and $n_e^2$ dependency cancels when taking their ratio (see \citealt{1998ApJ...494L..99R, 2002ApJ...565.1060H}). As oxygen and hydrogen have very similar first ionization potentials, there is a  strong charge-exchange reaction between the ground and first excited state of these elements. This couples their emission-line ratio to the ionization fraction of hydrogen (\citealt{1971ApJ...166...59F}, \citealt{1987A&A...184..337F}, \citealt{1994ApJ...428..647D}) such that $\chi_{\rm H\textsc{~ii}}$ is roughly inversely proportional to $I_{\rm [O\textsc{~i}]}/I_{\rm H\alpha}$ (\citealt{1998ApJ...494L..99R} and \citealt{2002ApJ...565.1060H}): 

\begin{align}\label{eq:oi_ha}
	I_{\rm [O\textsc{~i}]}/I_{\rm H\alpha}=&2.74\times10^4 \frac{T_4^{1.854}}{1+0.605~T_4^{1.105}} e^{-2.284/T_4} \nonumber \\
	&\times \left(\frac{\rm O}{\rm H}\right) \frac{1+n_{\rm H\textsc{~ii}}/n_{\rm H\textsc{~i}}}{[1+8/9 n_{\rm H\textsc{~ii}}/n_{\rm H\textsc{~i}}]~n_{\rm H\textsc{~ii}}/n_{\rm H\textsc{~i}}}.
\end{align}   

We detected both \oi\ and \ha\ along the low and high metallicity filament (see Figures~\ref{figure:Fairall9} and \ref{figure:RBS144}). To determine the $\chi_{\rm H\textsc{~ii}}$ of the Stream, we assume that the oxygen gas phase abundance is $0.5~{\rm solar}$ for the FAIRALL~9 sight line at position~(a) \citep{2000AJ....120.1830G, 2013ApJ...772..111R} and $0.1~{\rm solar}$ for the RBS~144 sight line at position~(e) and position~(1) \citep{2000AJ....120.1830G, 2010ApJ...718.1046F, 2013ApJ...772..110F} and electron gas temperatures of $T_4=[0.8, 1, 1.2]$. We also use the solar photospheric abundances of \citet{2009ARA&A..47..481A}. Using equation~\ref{eq:oi_ha}, we find that $0.16\le\chi_{\rm H\textsc{~ii}}\le0.47$ and $0.31\le\chi_{\rm H\textsc{~ii}}\le0.67$ for FAIRALL~9 and RBS~144, respectively; these results are summarized in Table~\ref{table:ion_frac_ha_oi}. These $\chi_{\rm H\textsc{~ii}}$ values closely agree with those found by \citetalias{2014ApJ...787..147F} if $T_4\approx1$, resulting in differences of only $6\%$ and $17\%$, suggesting the drastically different beam sizes sample similar ionization fractions even though the $N_{\rm H\textsc{~i}}$ varies within the WHAM beam. 

\section{Ionization Source of the Stream}\label{section:ionization}

The Magellanic Stream contains a substantial amount of ionized gas (e.g., \citealt{1996AJ....111.1156W}, \citealt{2003ApJ...597..948P}, and \citealt{2005ApJ...630..332F,2014ApJ...787..147F}). Both photoionization and collisional ionization processes might contribute to the ionization of this structure. Sources of photoionization include the EGB \citep{1999ApJ...510L..33B, 2001ApJ...561..559W, 2012ApJ...746..125H}, hot stars in the Milky Way and the Magellanic Clouds (i.e., MW:~\citealt{2005ApJ...630..332F}; MCs:~\citealt{2013ApJ...771..132B}), and potentially the Galactic Center. Sources of collisional ionization may include shocks, turbulent mixing, and conductive heating, which all may occur as the Stream interacts with the surrounding halo gas (e.g., \citealt{2007ApJ...670L.109B, 2011MNRAS.418.1575P, 2012ApJ...745..148J}). The region of the Stream that is below the South Galactic Pole (SGP) could also be susceptible to energetic events associated with the Galactic center, such as Fermi Bubbles (e.g., \citealt{2010ApJ...724.1044S, 2014ApJ...793...64A, 2015ApJ...799L...7F}) and short lived Seyfert activity (see \citetalias{2013ApJ...778...58B} and J.~Bland-Hawthorn et al. 2017, in preparation).

\subsection{Photoionization}\label{section:photoionization}

\begin{figure}
\begin{center}
\includegraphics[trim=30 0 30 0,clip,scale=0.5,angle=0]{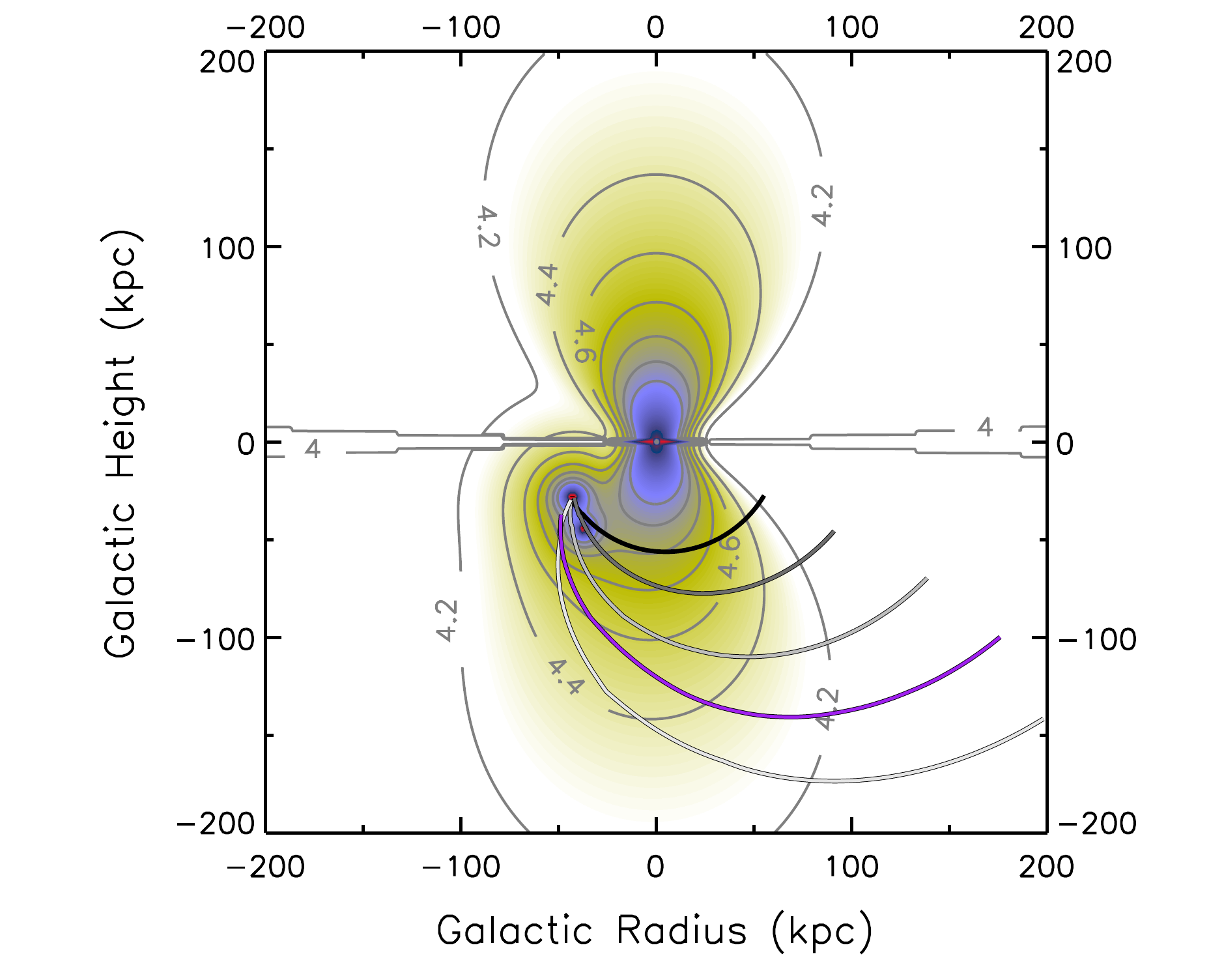}
\end{center}
\figcaption{Ionizing radiation field ($\log{\left[\phi_{\rm{LC}}/{\rm photons}\ {\rm cm}^{-2}\ {\rm s}^{-1}\right]}$) for the MW \citep{2005ApJ...630..332F} and the MCs \citep{2013ApJ...771..132B} for a $400 \times 400~\kpc$ slice through the center of the MW and LMC. The grayscale tracks assume that the Magellanic Stream is anchored at the center of the LMC ($d=50~\kpc$, $l_{\rm MS}=0\arcdeg$, $b_{\rm MS}=0\arcdeg$) and that it extends to $(d=50,~75,~100,~150~\kpc,~l_{\rm MS}=-57\fdg6,~b_{\rm MS}=0\arcdeg)$ below the SGP with a constant $\Delta d/ \Delta l_{\rm MS}$ along its length. The purple line traces the Model~2 track from \citet{2012MNRAS.421.2109B}, which places the Stream at $(l_{\rm MS},~b_{\rm MS},~d)=(-57\fdg6,~0\arcdeg,~63~\kpc$) below the SGP and at $(-120\arcdeg,~0\arcdeg,~210~\kpc$) its tip.
\label{figure:photoionization_model}}
\end{figure}

As \ha\ emission arises from the recombination of electrons and protons, the $I_{\rm H\alpha}$ is directly proportional to the rate of hydrogen ionizations per surface area of the emitting gas in local thermal dynamic equilibrium conditions (see \citealt{2012ApJ...761..145B} for more details): 
\begin{equation}\label{eq:FHalpha}
\phi_{\rm H\textsc{~i}\rightarrow H\textsc{~ii}} = 2.1\times10^5\left(\frac{\iha}{0.1{\rm R}}\right)~T_4^{+0.118}~{\rm cm}^{-2}\ {\rm s}^{-1}.
\end{equation}
If the ionization of an optically thick cloud is dominated by photoionization, then the incident Lyman Continuum flux ($\phi_{\rm LC}$) will be approximately equal to $\phi_{\rm H\textsc{~i}\rightarrow H\textsc{~ii}}$.

\citet{2001ApJ...561..559W} predict that the background of ionizing radiation escaping from distant galaxies has a strength of $\phi_{\rm EGB}\approx10^{4}~{\rm ionizing~photons}~\cm^{-2}~\s^{-1}$ at $z=0$. We approximate this ionizing background radiation as a constant flux along the entire Magellanic Stream, which produces enough ionization to elevate the \ha\ emission of the Stream by $I_{\rm H\alpha}\approx5~{\rm mR}$ (see Equation~\ref{eq:FHalpha}). 

The amount of incident ionizing radiation along the Stream from nearby galaxies varies with its position with respect to those galaxies. The intensity of the ionizing radiation field of disk galaxies is generally expected to be greatest at their center where their stellar production is typically most intense. The amount of ionizing radiation escaping from the centers of these galaxies tends to decrease as the polar angle increases because the photons must travel through more interstellar medium to escape the galaxy. Therefore clouds positioned directly above or below the poles will experience the strongest ionizing flux. To estimate the contribution of ionizing photons that the Galaxy irradiates the Stream with, we use the \citet{2005ApJ...630..332F} model (an updated version of the \citealt{1999ApJ...510L..33B, 2001ApJ...550L.231B} model), which has a vertical escape fraction ($f_{\rm esc}$) of $\sim6\%$ over the poles and $\sim1-2\%$ when averaged over a sphere. Although we also incorporated updated parameters for the Galactic corona from \citet{2015ApJ...800...14M}, the UV emission from the halo remains negligible (i.e., a few percent at most) compared to the Galactic disk ($\phi_{\rm{LC}} \approx 10^5~{\rm photons}\ {\rm cm}^{-2}\ {\rm s}^{-1}$ at $50~\kpc$ along the SGP, corresponding to $I_{\rm H\alpha}\approx50~{\rm mR}$ for $T_4=1$). 

Near the MCs ($l_{\rm MS}> -30\arcdeg$), the ionizing radiation from the SMC and LMC cannot be neglected. We use the \citet{2013ApJ...771..132B} model of the ionizing radiation field emitted by the SMC and LMC to estimate their contribution. Based on an \ha\ survey of the Magellanic Bridge, \citet{2013ApJ...771..132B} showed that the UV radiation from the Magellanic Clouds is sufficient to ionize the Magellanic Bridge, estimating $f_{\rm esc}\le5.5\%$ for the SMC and $f_{\rm esc}\le4.0\%$ for the LMC. Most of the ionizing flux that the Stream receives from these galaxies is from the LMC. 

Due to the poorly constrained distance of the Magellanic Stream, its relative position with respect to the LMC, SMC, and MW is uncertain except for where the Stream originates (see Section~\ref{section:intro}). As the strength of the incident ionizing radiation field onto the Stream from these galaxies greatly depends its position, we compare the $I_{\rm H\alpha}$ with the Lyman continuum radiation field models over a wide range of distances. We therefore explore five different position tracks for the Stream that  position it between $50-150~\kpc$ below the SGP at $l_{\rm MS}=-57\fdg3$. Although it is well known that the Magellanic Stream originates from the MCs, it is uncertain where amongst these galaxies the longer of the two \hi\ filament begins. \citet{2008ApJ...679..432N} kinematically identified these two coherent filaments through component fitting of the \hi\ emission spectra, but they were unable to determine if the longer filament traced back to the SMC, Magellanic Bridge, or the LMC. They were, however, able to determine that the shorter filament traces back to $30$~Doradus. We anchor four of our position tracks at the LMC ($d=50~\kpc$, $l=283\fdg3$, $b=-32\fdg4$), or ($d=50~\kpc$, $l_{\rm MS}=0\arcdeg$, $b_{\rm MS}=0\arcdeg$). For these tracks, we assume that the distance along the Stream changes linearly with angular distance; this assumption agrees well with galaxy interaction models for scenarios where the MCs are on their first-infall or are bound to the MW (e.g., Figure~3: \citealt{2006MNRAS.371..108C}; Figure~10: \citealt{2012MNRAS.421.2109B}). We include a fifth track that follows the Stream positions that resulted from the $2^{\rm nd}$ LMC and SMC interaction model in \citet{2012MNRAS.421.2109B}, which has the beginning of the Stream anchored at the Magellanic Bridge and not the LMC. The combined ionizing radiation field models for the MW, MCs, and EGB is shown in Figure~\ref{figure:photoionization_model} along with these five different position tracks for the Stream. 

\begin{figure*}
\begin{center}
\includegraphics[trim=0 0 0 0,clip,scale=0.55,angle=90]{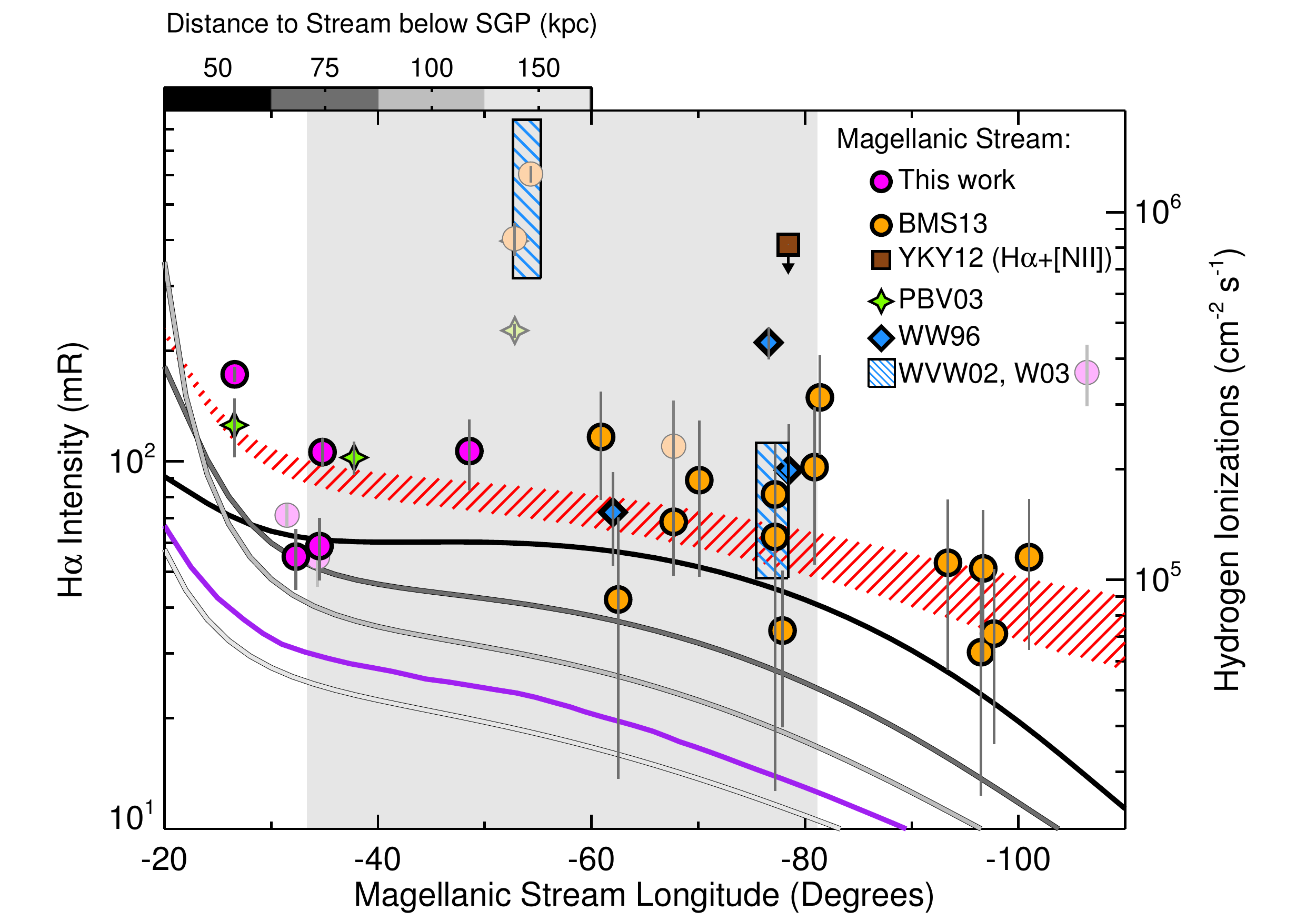} 
\end{center}
\figcaption{The \ha\ intensities along the Magellanic Stream from this and the \citetalias{2013ApJ...778...58B}, \citetalias{2012ApJ...749L...2Y}, \citetalias{2003ApJ...597..948P}, WW96, \citetalias{2002ASPC..254..256W}, and \citetalias{2003ASSL..281..163W} studies \textit{(detections only)}. The blue shaded polygons mark the ${\rm median}\pm {\rm average~deviation~from~the~median}$ for the \citetalias{2002ASPC..254..256W} and \citetalias{2003ASSL..281..163W} detections. Sight lines off of the \hi\ filaments are colored in a lighter shade than indicated by the legend. The grey shaded region marks the portion of the Stream that lies within a $\theta_{1/2}=25\arcdeg$ cone that projects out from the Galactic Center along the SGP. The right-hand vertical axis marks the rate of hydrogen ionizations per surface area of the Stream that are needed to reproduce the \ha\ emission (see Equation~\ref{eq:FHalpha} for photoionization). The four black to light grey lines trace predicted photoionization due to the MW \citep{2005ApJ...630..332F}, LMC \citep{2013ApJ...771..132B}, and EGB \citep{2001ApJ...561..559W} for a Stream positioned at $(d=50,~75,~100,~150~\kpc,~l_{\rm MS}=-57\fdg6,~b_{\rm MS}=0\arcdeg)$ distance tracks as shown in Figure~\ref{figure:photoionization_model}.  The purple trace shows the predicted photoionization for Model~2 of \citet{2012MNRAS.421.2109B}. The combined contribution from MW, MCs, and EGB photoionization and \textit{negligible} halo-gas interactions (maximum contribution of $3~{\rm mR}$; \citetalias{2013ApJ...778...58B}) and shock-cascade self interactions \citep{2015ApJ...813...94T} is shown as a dashed red envelope for $n_{\rm halo}=2\times10^{-4}~\cm^{-3}$ for the $d=75~\kpc$ track.
\label{figure:position_ha}}
\end{figure*}

\begin{figure*}
\begin{center}
\includegraphics[scale=0.65,angle=0]{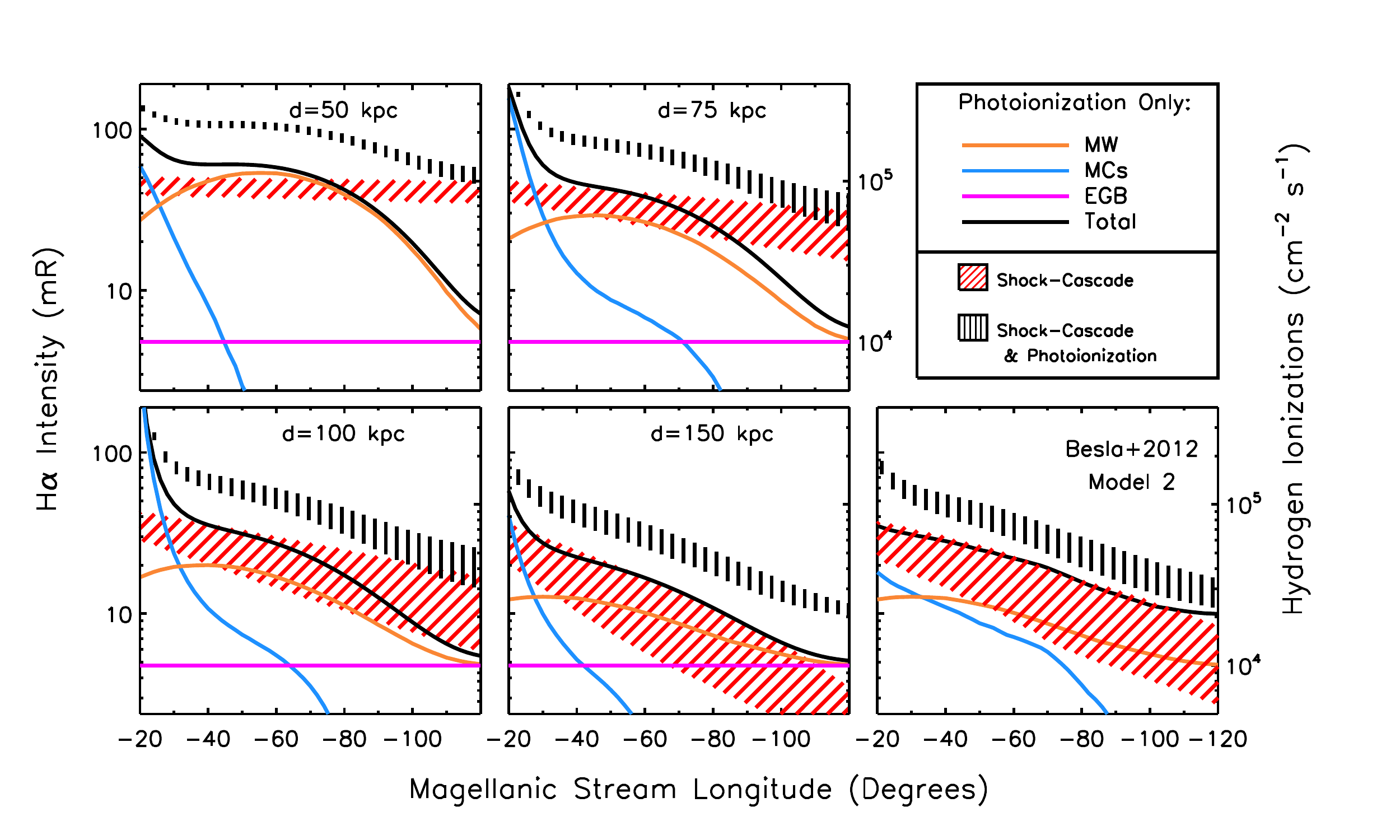}
\end{center}
\figcaption{A breakdown of the anticipated \ha\ intensity that results from each of the major sources of photoionization and collisional ionization for the five distance tracks shown in Figures~\ref{figure:photoionization_model} and~\ref{figure:position_ha}. The corresponding number of hydrogen ionization from these sources is indicated in the right vertical axis (Figure~\ref{eq:FHalpha}). The photoionization contribution from the MCs is shown by the blue lines \citep{2013ApJ...771..132B}, the Milky Way by the orange lines \citep{2005ApJ...630..332F}, and the EGB by the pink lines \citep{2001ApJ...561..559W}. The combined photoionization contribution from these three sources are the solid black lines. The estimated $I_{\rm H\alpha}$ that results from shock-cascade self interactions \citep{2015ApJ...813...94T} and halo-gas interactions (\citetalias{2013ApJ...778...58B}), respectively, for $n_{\rm halo}=2\times10^{-4}~\cm^{-3}$ are shown as the red envelopes. The projected $I_{\rm H\alpha}$ that results from both the photoionization and collisional ionization combined are indicated by the black envelopes. 
\label{figure:ionizing_contribution}}
\end{figure*}

In Figure~\ref{figure:position_ha}, we show how the strength of the $I_{\rm H\alpha}$ varies along the length of the Magellanic Stream. The horizontal axis indicates the angular distance from the LMC in Magellanic Stream coordinates (see \citealt{2008ApJ...679..432N}), where $l_{\rm MS}=0\arcdeg$ crosses through the center of the LMC, $l_{\rm MS}\approx-17\arcdeg$ passes through the center of the SMC, and the grey shaded region spanning $-33\arcdeg\lesssim l_{\rm MS}\lesssim-81\arcdeg$  corresponds to cone with a half-opening angle of $\theta_{1/2}\approx 25\arcdeg$ which flares out from the Galactic center and is centered on the SGP at $(l_{\rm MS},~b_{\rm MS})=(-57\fdg3,~+7\fdg5)$. The right-hand y-axis includes an estimate for the rate of hydrogen ionizations per surface area of the Stream (see Equation~\ref{eq:FHalpha}). To test how well photoionization from the MW, MCs, and EGB alone can reproduce the \ha\ emission in the Magellanic Stream, we illustrate the total $I_{\rm H\alpha}$ that is predicted from these ionization sources for our five position tracks shown in Figure~\ref{figure:photoionization_model}. We show the ionizing contribution for the MW, MCs, and EGB separately in Figure~\ref{figure:ionizing_contribution}. The ionizing contribution  from the MCs drops off steeply between $-40\arcdeg\le l_{\rm MS}\le-20\arcdeg$, the MW's contribution dominates above $l_{\rm MS}\ge-40\arcdeg$, while ionization from the EGB only produces a meager $I_{\rm H\alpha}\approx5~{\rm mR}$ increase along the Stream. For all of the assumed distances, the average \ha\ emission is roughly $50~{\rm mR}$ higher than anticipated from photoionization from these sources alone. \\

\noindent\textit{This leads us to a clear conclusion: photoionization from the MW, MCs, and the EGB alone is insufficient for producing the observed ionization in the Magellanic Stream.} 

\begin{figure*}
\begin{center}
\includegraphics[scale=0.5,angle=90]{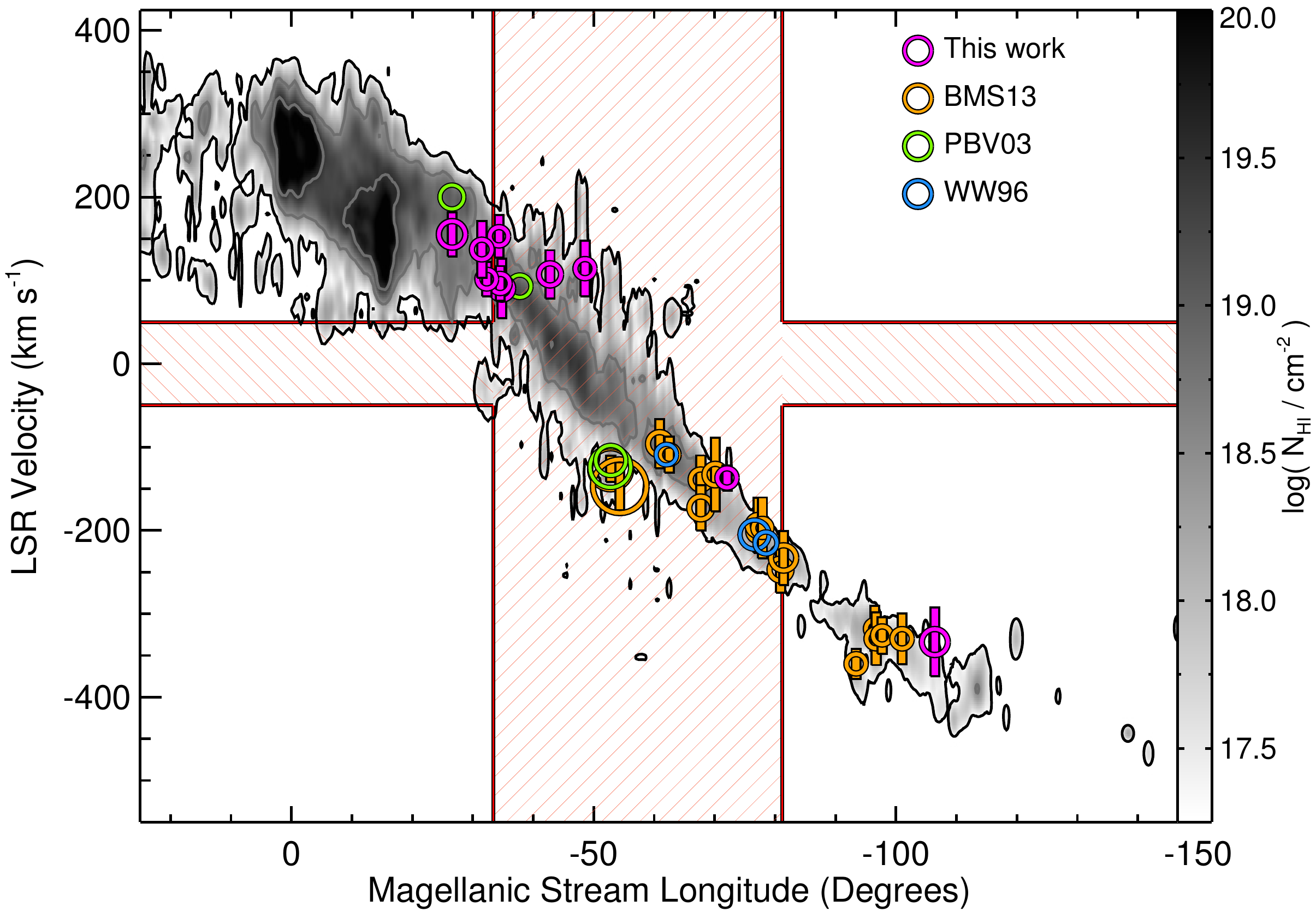}
\end{center}
\figcaption{Position-velocity diagram of the \ha\ detections along the Magellanic Stream. Symbol colors represent different studies, as in Figure~\ref{figure:position_ha}. The size of the circles scale with the strength of the \ha\ emission. The extent of the vertical colored bars signifying the width of the line (see Table~\ref{table:intensities}), assuming a Gaussian distribution. The background map displays the \hi\ column density of the Stream as a function of LSR velocity from the \citet{2008ApJ...679..432N} Gaussian decompositions of the LAB survey. The vertical red shaded region spans the locations of the Stream positioned below the Galactic pole at $b\leq-65\arcdeg$ for $b_{\rm MS}=0\arcdeg$. The horizontal purple shaded region marks the velocities where confusion between the Magellanic Stream and MW is greatest.
\label{figure:position_vel}}
\end{figure*}

\subsection{Collisional Ionization}\label{section:Collisional_Ionization}

Collisional ionization from interactions with the Galactic halo and the Stream itself might be the cause of some of the elevated \ha\ emission. Figure~\ref{figure:position_vel} illustrates how the \ha\ emission varies along the Magellanic Stream with longitude and LSR~velocity.\footnote{Two \ha\ detections from WW96 at $l_{\rm MS}<-100\arcdeg$ have been excluded from this study because their velocities differ from the Stream's by more than $200~\kms$ for their Magellanic Stream Longitude and are therefore likely unassociated with the Stream.} The size of the multicolored circles in this figure scales with the $I_{\rm H\alpha}$ strength and the vertical bars signify the kinematic extent of the emission. The sight lines that are positioned spatially or kinematically off the two main \hi\ filaments of the Stream are roughly $25~{\rm mR}$ brighter on average, excluding the brightest detections at $l_{\rm MS}\approx-55\arcdeg$ that will be discussed in Section~\ref{section:Flare} below. These sight lines probe locations on the Stream that are less shielded from their environment and are more exposed to the surrounding coronal gas and radiation field. In addition, limb brightening may result in a slightly higher $I_{\rm H\alpha}$ for sight lines that are projected spatially on the edge of the Stream. 

There is now general concordance that the orbit of the Magellanic System is highly elliptical, which would place all but the portions of the Stream that are nearest to the MCs much further away (e.g., \citealt{2007ApJ...668..949B,2008MNRAS.383.1686J,2014MNRAS.444.1759G}). Through ram-pressure modeling of the \hi\ density profile of LMC's disk, \citet{2015ApJ...815...77S} found that ${\rm n}_{\rm halo} \approx 10^{-4}~\cm^{-3}$ at $d\approx50~\kpc$ would reproduce the observed compression in the disk's leading edge. This density presumably drops with distance from the Milky Way, which will decrease the collisional interaction rate between the gas in the Stream with the surrounding coronal gas. Following the procedure employed in \citetalias{2013ApJ...778...58B}, we predict that halo-gas interactions will only marginally elevate the ionization in the Stream such that the $I_{\rm H\alpha}$ increases by $\le3~{\rm mR}$ (see their Equations~13 and A7-A12) when we spatially smooth their model to match the $1\arcdeg$ resolution of WHAM for a Stream positioned $d=75~\kpc$ below the SGP and assume ${\rm n}_{\rm halo} = 2\times10^{-4}~\cm^{-3}$ at $d\approx50~\kpc$. 

\citet{2007ApJ...670L.109B} predict that in addition to direct ionization through halo-gas interactions, self interactions amongst the gas in the Stream could also result in ionization. These authors show that a slow ``shock cascade'' arises if sufficient gas is stripped from the Stream as it flows through the halo such that the trailing clouds collide with the ablated material. \citet{2015ApJ...813...94T} presented an updated shock-cascade model for Stream gas at distance $d=75~\kpc$ from the Galactic center. We have slightly modified the \citet{2015ApJ...813...94T} model to account for the distance gradient along the length of the Stream. By interpolating this model for our five distance tracks, we find that if the trailing gas in the Magellanic Stream lies at $d=75~\kpc$ below the SGP, then a shock cascade could elevate the \ha\ emission by upwards of $50~{\rm mR}$ at the start of the Stream at $l_{\rm MS}=-20\arcdeg$ and by $20~{\rm mR}$ at its tail at $l_{\rm MS}=-120\arcdeg$ as illustrated in Figure~\ref{figure:ionizing_contribution} for a spatial resolution of $1\arcdeg$ to match the WHAM beam size. The red envelope in Figure~\ref{figure:ionizing_contribution} encloses our shock-cascade solutions for thermal ratios of the virial ``temperature'' of the dark matter halo to the halo-gas temperature ($\tau_{\rm thermal}=T_{\rm DM}/T_{\rm halo}$) of $0.5$, $1.0$, and~$1.5$ at $n_{\rm halo}=2\times10^{-4}~\cm^{-3}$. The black envelope in this figure illustrates the total ionization that is predicted for photoionization from the MW, MCs, EGB and collisional ionization from the shock-cascade and halo-gas interactions for the $50$, $75$, $100$, and $150~\kpc$ distance tracks.  

The additional ionization from shock cascade self ionization---along with the photoionization from the MW, MCs, and EGB---comes close to matching the general underlying \ha\ emission trend for the sight lines that are on the Magellanic Stream \hi\ filaments for the $d=75~\kpc$ position track as shown in Figure~\ref{figure:position_ha} as a red envelope. Additionally, \citet{2014ApJ...792...43F} found that many of the cloudlets that fragmented off of the main body of the Magellanic Stream had a head-tail structure that pointed in random directions. The orientation of the head-tail clouds could been randomized from Kelvin-Helmholtz instabilities that are generated as the Stream rubs against the surrounding coronal gas, which is consistent with a shock cascade scenario. However, the portion of the Stream that spans over $-90\arcdeg\le l_{\rm MS}\le-75\arcdeg$ deviates by $30~{\rm mR}$ on average for the combined photoionization and collisional ionization predictions. As the explored photoionization and collisional ionization models have large uncertainties, this is a minor deviation. 

\subsection{Galactic Center Ionization}\label{section:Flare}

Directly below the SGP, the observed $I_{\rm H\alpha}$ is \textit{much} higher than the rest of the Stream (see Figure~\ref{figure:position_ha}) with some sight lines more than $\sim200$ to $500~{\rm mR}$ brighter. Their position below the SGP suggests that energetic processes associated with the Galactic center could be ionizing the gas. As seen in the \citetalias{2014ApJ...787..147F} UV absorption-line study, the Stream has unusual ionization characteristics (e.g., Si\textsc{~iii}/Si\textsc{~ii}, C\textsc{~iv}/C\textsc{~ii}) along its length that require photons with up to $50~{\rm eV}$ energies. If this ionization is coming from the Galactic center, than the cone axis would need to be tilted by $15\arcdeg$ with respect to the SGP in the opposite direction from the LMC to reproduce these ratios (J.~Bland-Hawthorn et al. 2017, in preparation). This tilt has not been detected in the Fermi Bubbles, which appear to lie along the Galactic polar axis. A putative jet that originates from the Galactic Center does lie along this tilted projection as seen in both radio and X-ray (\citealt{1998ApJ...496L..97B, 2012ApJ...753...61S, 2012ApJ...758L..11Y}). A tilted cone could account for the hard ionization seen in UV absorption along the length of the Magellanic Stream (\citetalias{2014ApJ...787..147F}; J.~Bland-Hawthorn et al. 2017, in preparation). 

\citetalias{2013ApJ...778...58B} explored how well Seyfert activity could produce this elevated ionization. Large-scale bipolar bubbles from the Galactic Center have been observed in hard X-rays \citep{2003ApJ...582..246B} and gamma rays \citep{2010ApJ...724.1044S}. Most models and observations now agree that the AGN event that drove the bubbles took place in the last $2-6$~million years (e.g., models: \citealt{2012ApJ...756..181G, 2016ApJ...829....9M}; observations: \citealt{2015ApJ...799L...7F, 2017ApJ...834..191B}). The luminosity of MW's central black hole is a fraction of the Eddington limit with bursts in luminosity arising from stochastic accretion events \citep{2006ApJS..166....1H}. \citetalias{2013ApJ...778...58B} show that a Seyfert flare with an AGN spectrum that is 10\% of the Eddington luminosity ($f_{\rm Edd} = 0.1$) for a $4\times10^{6}~{\rm M}_\odot$ black hole can easily produce sufficient UV radiation to ionize the Magellanic Stream if it crosses $50 \le d \le 100~\kpc$ below the SGP. As Sgr A* is quiescent today, the \ha\ emission would have significantly faded since the flare. This is because the H\textsc{~ii} recombination with electrons is faster than the gas cooling time for Magellanic Stream densities (${\rm n}_{\rm e}\approx 0.1$ to~$1~\cm^{-3}$; \citetalias{2014ApJ...787..147F}) and metallicity ($Z \approx 0.1~Z_{\odot}$; \citealt{2000AJ....120.1830G, 2010ApJ...718.1046F, 2013ApJ...772..110F}). \citetalias{2013ApJ...778...58B} found that the event must have happened within the last few million years to be consistent with jet-driven models of the $10~\kpc$ bipolar bubbles. This timescale includes the crossing time (the time for the flare radiation to reach the Stream) plus the recombination timescale.

However, an important cautionary note is that these bright \ha\ detections all lie within $\sim2\arcdeg$ of each other as illustrated in Figure~\ref{figure:greg_1_2}. The two bright observations from \citetalias{2003ApJ...597..948P} have angular resolutions between $5-10\arcmin$ and the bright observations from \citetalias{2002ASPC..254..256W} and \citetalias{2003ASSL..281..163W} have an angular resolution of $25\arcmin$. The \citetalias{2003ApJ...597..948P} observations and most of the \citetalias{2002ASPC..254..256W}, and \citetalias{2003ASSL..281..163W} observations lie within the much larger $1\arcdeg$ WHAM beam of sight line~(1). The two bright \citetalias{2013ApJ...778...58B} observations at sight lines~(1) and~(2) confirm that this region is much brighter in \ha\ than sight lines $l_{\rm MS}\gtrsim\lvert5\rvert\arcdeg$ away. These sight lines lie on the kinematic edge of the Stream (see Figure~\ref{figure:position_vel}), which may slightly elevate their emission due to limb brightening and their increased exposure to their environment, but not enough to produce \ha\ emission that is this bright. Although this \textit{small} $\sim2\arcdeg$ region of the Stream is sampled well, it may not be representative of the $I_{\rm H\alpha}$ below the SGP. Mapped \ha\ observations of this region are vital to ascertain the distribution of this elevated ionization to constrain its source. Mapped multiline observations could further aid in identifying the source by determining the hardness of the ionization. 

\section{Summary}\label{section:summary}
 
We observed the emission from the warm ionized gas in the Magellanic Stream with WHAM toward $39$ sight lines in \ha, 19 sight lines in \sii\ and \nii, and a handful of sight lines in other lines (see Tables~\ref{table:obs} and~\ref{table:multiline_intensities}). We detected \ha\ emission in $26$ of these sight lines, \sii\ in four sight lines, and \nii\ in three sight lines. We further observed two sight lines with extremely different metallicities (FAIRALL~9 at position (a) with $Z/Z_{\odot}=0.5$; RBS~144 at position (e) with $Z/Z_{\odot}=0.1$) in seven additional lines, detecting \oi\ in both, \hb\ in one of these directions; \oii$~\lambda7320$, \oiii$~\lambda5007$, \nii$~\lambda5755$, and \hei$~\lambda5876$ was not detected. These kinematically resolved observations span over $100\arcdeg$ along the Stream and substantially increase the number of detections of warm ionized gas along the Magellanic Stream. We also compared our observations of the warm ionized gas phase with \ha\ observations from other studies and with \hi\ 21-$\cm$ emission from the LAB and GASS \hi\ surveys. We finish with the main conclusions of our study: 

\begin{enumerate}
 
\item{\bf Ionized Gas Morphology.} The ionized gas spatially tracks the \hi\ emission well with $80\%$ (or $22/27$) of the 21-cm emitting sight lines detected in \ha\ emission. The strength of the neutral and ionized gas emission are uncorrelated (see Figure~\ref{figure:column_intensity}). The \ha\ emission often extends many degrees off the two main \hi\ filaments. Five of our $39$ sight lines are only detected in \ha\ and with velocities that are consistent with the Magellanic Stream at their position. 
\item{\bf Ionized Gas Kinematics.} The \ha\ line centers for sight lines positioned on the main \hi\ Magellanic Stream filaments agree within $10~\kms$ of the \hi\ emission, but the sight lines located on the edge or a few degrees away from the \hi\ line have profiles that are misaligned from the \hi\ gas by as much as $\sim30~\kms$ (see Figures~\ref{figure:compare_vel} and \ref{figure:position_vel}). The offset velocity centers, combined with \hi\ morphologies, suggests that the low \hi\ column density gas could be more exposed to the surrounding sources of ionization and rapidly evaporating into the Galactic halo.
\item{\bf Ionization Fraction.} The Magellanic Stream contains a substantial amount of ionized gas. Using the $I_{\rm [O\textsc{~i}]}/{I_{\rm H\alpha}}$ of sight lines, we find that regions with $\log N_{\rm H\textsc{~i}}/\cm^{-2}\approx19.5-20.0$ have $\chi_{\rm H\textsc{~ii}}\approx0.16-0.67$. These ionization fractions are in close agreement with those found by the \citetalias{2014ApJ...787..147F} absorption-line study for the same directions and provide the first direct comparison between values inferred from emission and absorption studies. Due to this similarity, we conclude that the ionization conditions of the gas change very little from the small pencil beam scales to the $1\arcdeg$ scales for these compared directions. 
\item{\bf Ionization Source.} We explored the ionizing contribution of photoionization and collisional ionization along five different distance tracks along the Magellanic Stream that ranged from $50$ to $150~\kpc$ below the SGP. For all the assumed distances, most of the \ha\ detections are much higher than expected if the primary ionization source is photoionization from MW \citep{2005ApJ...630..332F}, MCs \citep{2013ApJ...771..132B}, and the EGB \citep{2001ApJ...561..559W} as shown in Figure~\ref{figure:position_ha}. We find that although halo-gas interactions interactions likely affect the morphology of the Stream, they only produce a negligible amount of ionization that would result in a $I_{\rm H\alpha}<3~{\rm mR}$ increase. However, the Stream may become self ionized through a shock-cascade process that results from ram-pressure stripped gas colliding with trailing gas down stream. For a Stream positioned at $75~\kpc$ above the SGP, we find that this process could elevate the \ha\ emission by upwards of $50~{\rm mR}$ near the LMC and by $20~{\rm mR}$ at $l_{\rm MS}=-120\arcdeg$ (see Figure~\ref{figure:ionizing_contribution}), which could produce enough ionization to match the \textit{underlying} \ha\ emission along the entire Stream. The elevated \ha\ emission directly below the SGP suggests that this region is susceptible to other energetic processes associated with the Galactic center, such as a short lived Seyfert flares (e.g., \citealt{2013ApJ...778...58B}). However, only $\sim2\arcdeg$ region below the SGP has been sampled well and mapped observations are needed to ascertain the distribution of this elevated ionization to constrain its source. 
\end{enumerate}

\acknowledgments

We thank Ben Weiner for sharing \ha\ intensities from \citetalias{2002ASPC..254..256W} and \citetalias{2003ASSL..281..163W}. We thank the anonymous referee for their comments and suggestions, which strengthened the paper.  The IDL MPFIT routines are available at http://purl.com/net/mpfit. The National Science Foundation has supported WHAM  through grants AST~0204973, AST~0607512, AST~1108911, and AST~1203059. Barger received additional supported through NSF Astronomy and Astrophysical Postdoctoral Fellowship award AST~1203059. Wakker was supported by \textit{HST} grant GO-12604.01.

\facility{WHAM}
\software{MPFIT \citep{2009ASPC..411..251M}}.
\bibliographystyle{aasjournal} 
\bibliography{$HOME/WHAM/References} 

\section{Appendix}\label{section:appendix}

\begin{figure*}
\begin{center}
\includegraphics[trim=0 0 0 0,clip,scale=0.55,angle=90]{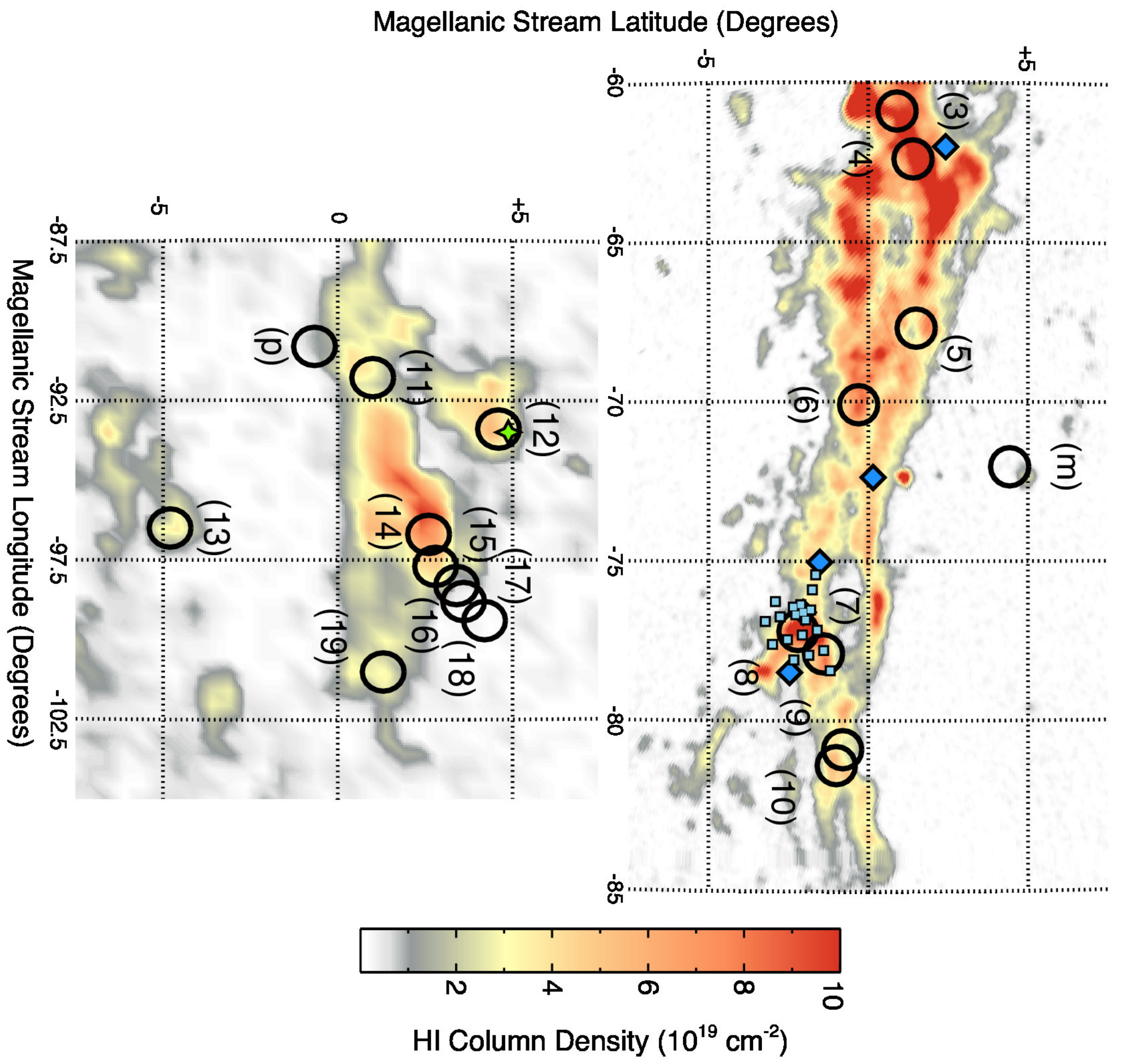}
\end{center}
\figcaption{The H\textsc{~i} column density map shows the neutral gas distribution from $-60\arcdeg\le l_{\rm MS}\le-85\arcdeg$ \text{(Top)} and from $-87.5\fdg5\le l_{\rm MS}\le-105\arcdeg$ \text{(Bottom)} over $-7\fdg5\le b_{\rm MS}\le+7\fdg5$. The map and spectra labels coincide with the ones used in Figure~\ref{figure:hi_map} and Table~\ref{table:obs}. The black circles in the H\textsc{~i} map represent the positions and coverage area of pointed WHAM observations. The top map displays the \hi\ column density over the range $-375\le\vlsr\le-175~\kms$ with \hi\ from GASS H\textsc{~i} Survey. The locations of WW96 observations are shown as dark blue diamonds and the positions of the \citetalias{2002ASPC..254..256W} and \citetalias{2003ASSL..281..163W} observations are mark with light blue squares. The bottom map over the range $-250\le\vlsr\le-75~\kms$ with \hi\ from the LAB H\textsc{~i} Survey and the green star marks the location of an \citetalias{2003ApJ...597..948P} observation.
\label{figure:high_lon_map}}
\end{figure*}

The following figures include two zoomed in \hi\ maps of the $-60\arcdeg\le l_{\rm MS}\le-85\arcdeg$ and $-87.5\fdg5\le l_{\rm MS}\le-105\arcdeg$ regions with the location and the size of the WHAM beam indicated (Figure~\ref{figure:high_lon_map}) and of the LAB \hi\ and WHAM \ha\ spectra for each of the $39$ sight lines in our study. The locations within the maps and the spectra are labeled according to IDs in Tables~\ref{table:obs} and~\ref{table:intensities}. Each of the \hi\ spectra shown was produced by averaging all of the spectra enclosed within the same $1\arcdeg$ angular coverage as the WHAM observations. In these figures, the vertical dashed and dotted lines mark the centroid positions of \hi\ and \ha\ Magellanic Stream components, respectively (see Table~\ref{table:intensities}). Spectra with a vertical dashed line and no corresponding \hi\ emission mark the average velocity for that region of the Stream (see Figure~\ref{figure:position_vel}). To reduce small variations in intensity, we increased the bin size of the WHAM spectra by a factor of two. For sight line HE0056-3622 at position (5), we display the GASS \hi\ survey spectrum (increased bin size by almost three), which enabled us to detect \hi\ from the Stream emission at this location due to higher sensitivity of this survey compared to the LAB \hi\ Survey.

Our observations include three sight line pairs that substantially overlap with each other at only $0\fdg1$ apart (small compared to the $1\arcdeg$ beam size): (4a \& 4b), (5a \& 5b), and (7a \& 7b). The \hi\ distribution of sight lines pairs is pretty constant, as is the \ha\ distribution for (7a \& 7b) for the $\sim100~\kms$ that they overlap. Along the (4a \& 4b) sight lines, the ``a'' spectrum is enhanced at $\vlsr\approx-180$ and $-75~\kms$. Although the \ha\ spectrum at position (4a) seems to trace emission from the Magellanic Stream over the $-150\le\vlsr\le-50~\kms$ range, the large difference between the \hi\ and \ha\ spectra at $-250\le\vlsr\le-150~\kms$ preclude us from confidently identifying this as emission. For this reason, we only report an upper limit of ${\rm I}_{\rm H\alpha}<40~{\rm mR}$ for the (4a) sight line.

Sight line (5b) has a similar \ha\ enhancement at $\vlsr=-175~\kms$ compared to its (5a) pair. These sight lines lie $\sim5\arcdeg$ away and $-45~\kms$ from the (4a \& 4b) pair. Unlike the (4a \& 4b) pair, the \ha\ between $-200\le\vlsr\le-100~\kms$ overlaps substantially with the \hi\ emission at $-175\le\vlsr\le-125~\kms$. Therefore we treat the emission along both of the 5a and 5b sight lines as real with aligned \hi\ emission at $\vlsr\approx-156~\kms$ and misaligned \ha\ emission at $\vlsr\approx-139~\kms$ and $\vlsr\approx-173~\kms$, respectively. This velocity offset in these nearby sight lines suggest that there is small scale variation in the Stream. However, \hi\ and \ha\ match up very well for sight lines (7a \& 7b), positioned another $10\arcdeg$ down stream. To confidently probe the small scale variation of this tidal remnant, higher spatial resolution observations are needed.

\clearpage

\begin{figure*}
\begin{center}
\includegraphics[trim=0 38 65 30,clip,scale=0.45,angle=0]{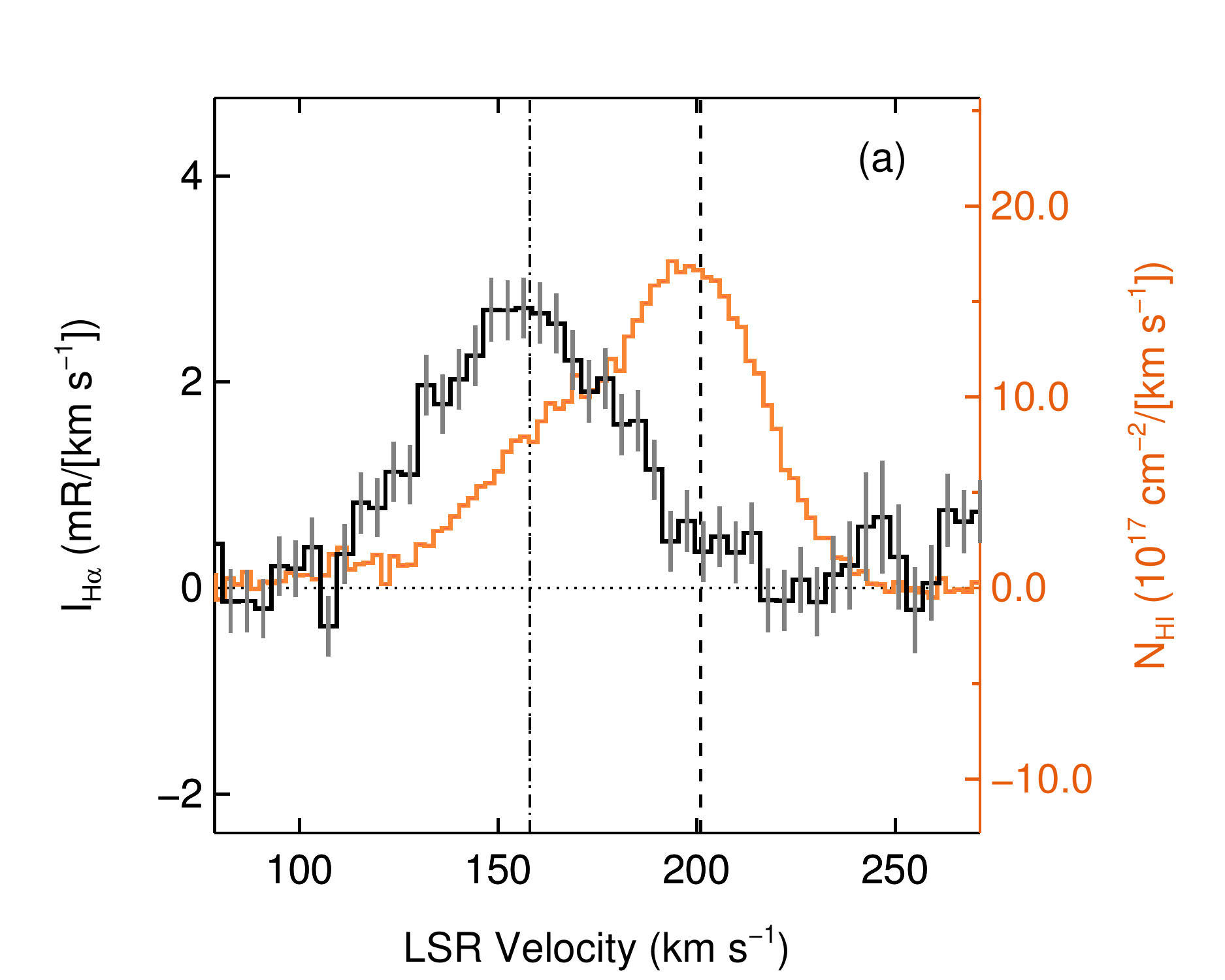}\includegraphics[trim=45 38 0 30,clip,scale=0.45,angle=0]{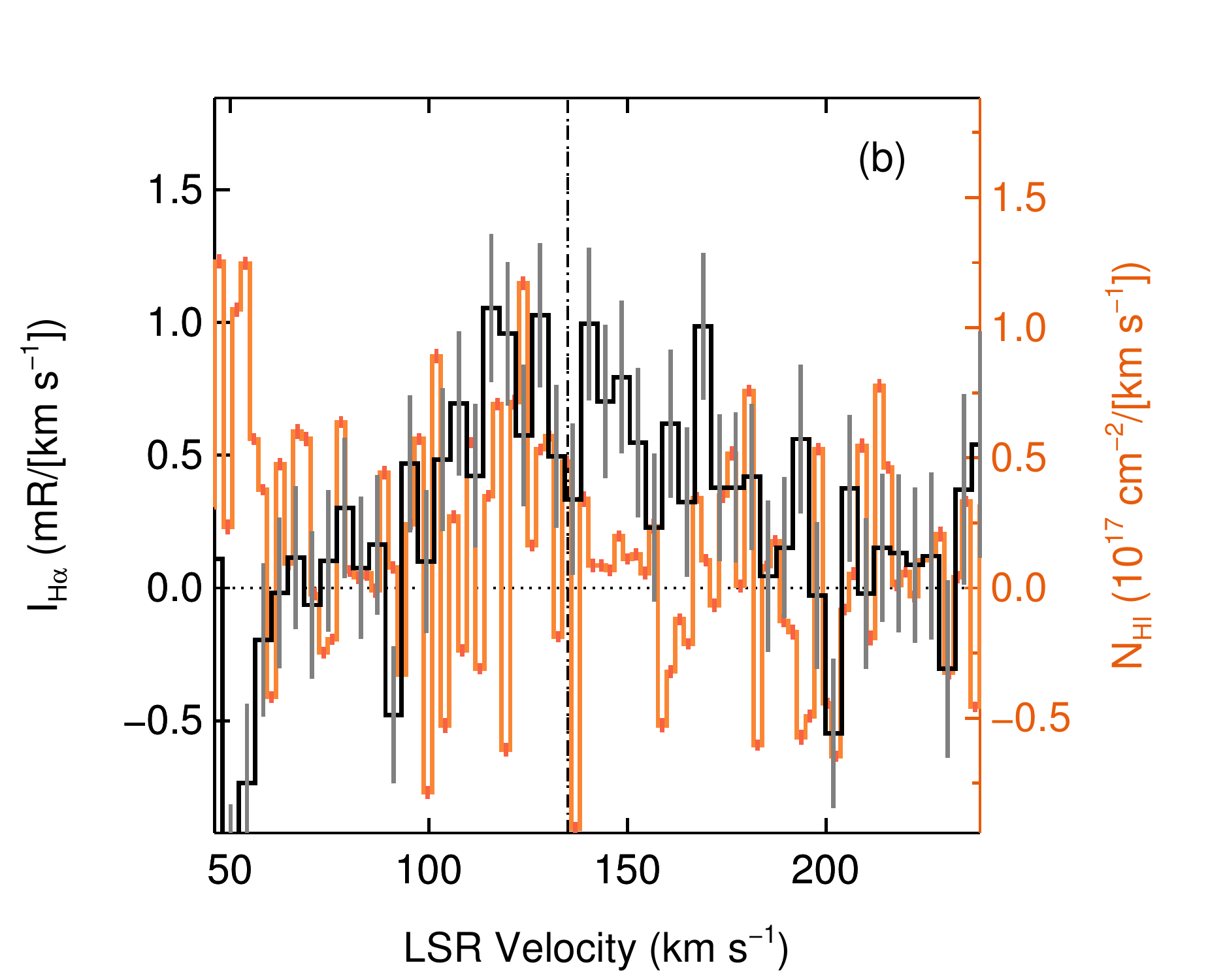} \\
\includegraphics[trim=0 38 65 30,clip,scale=0.45,angle=0]{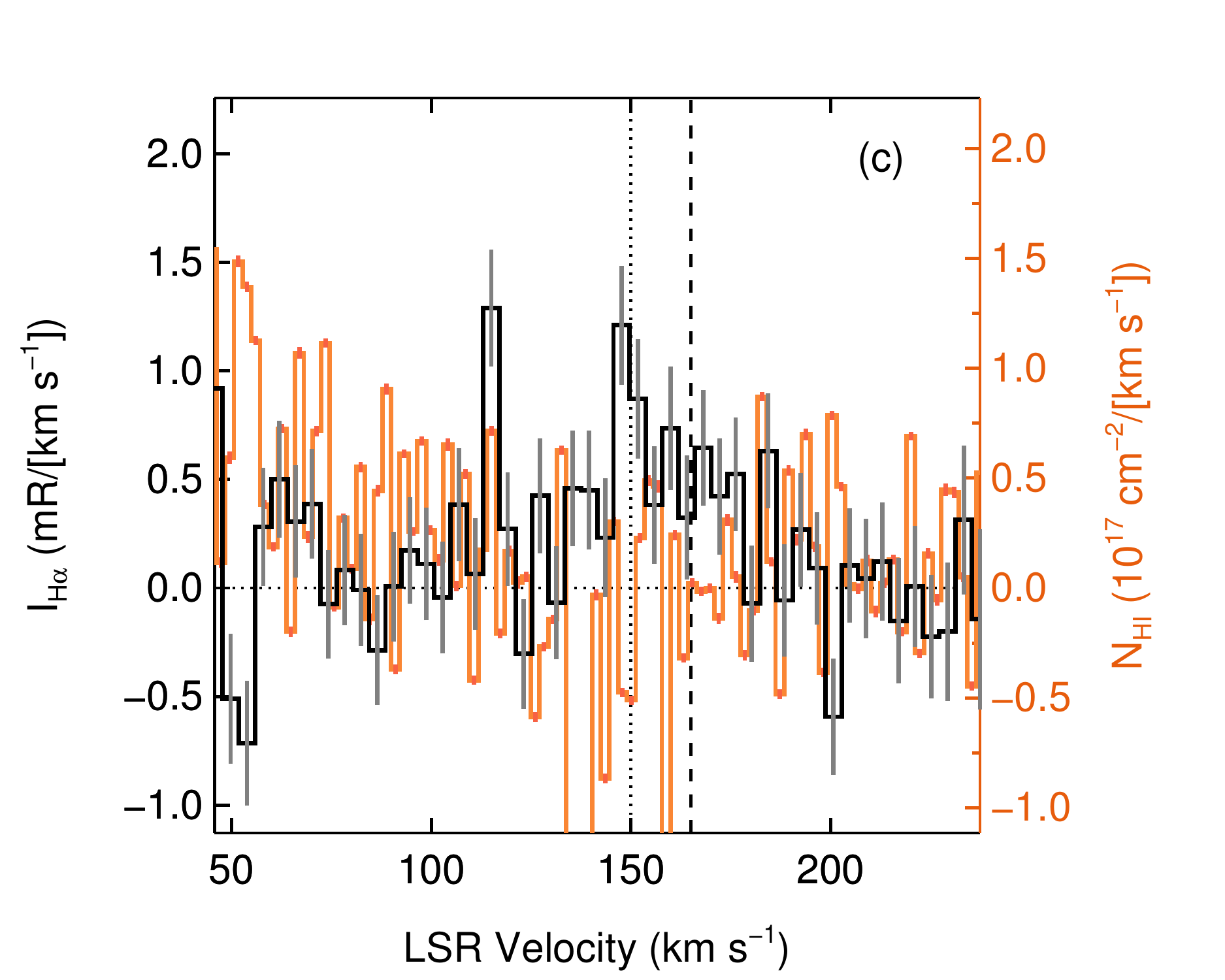}\includegraphics[trim=45 38 0 30,clip,scale=0.45,angle=0]{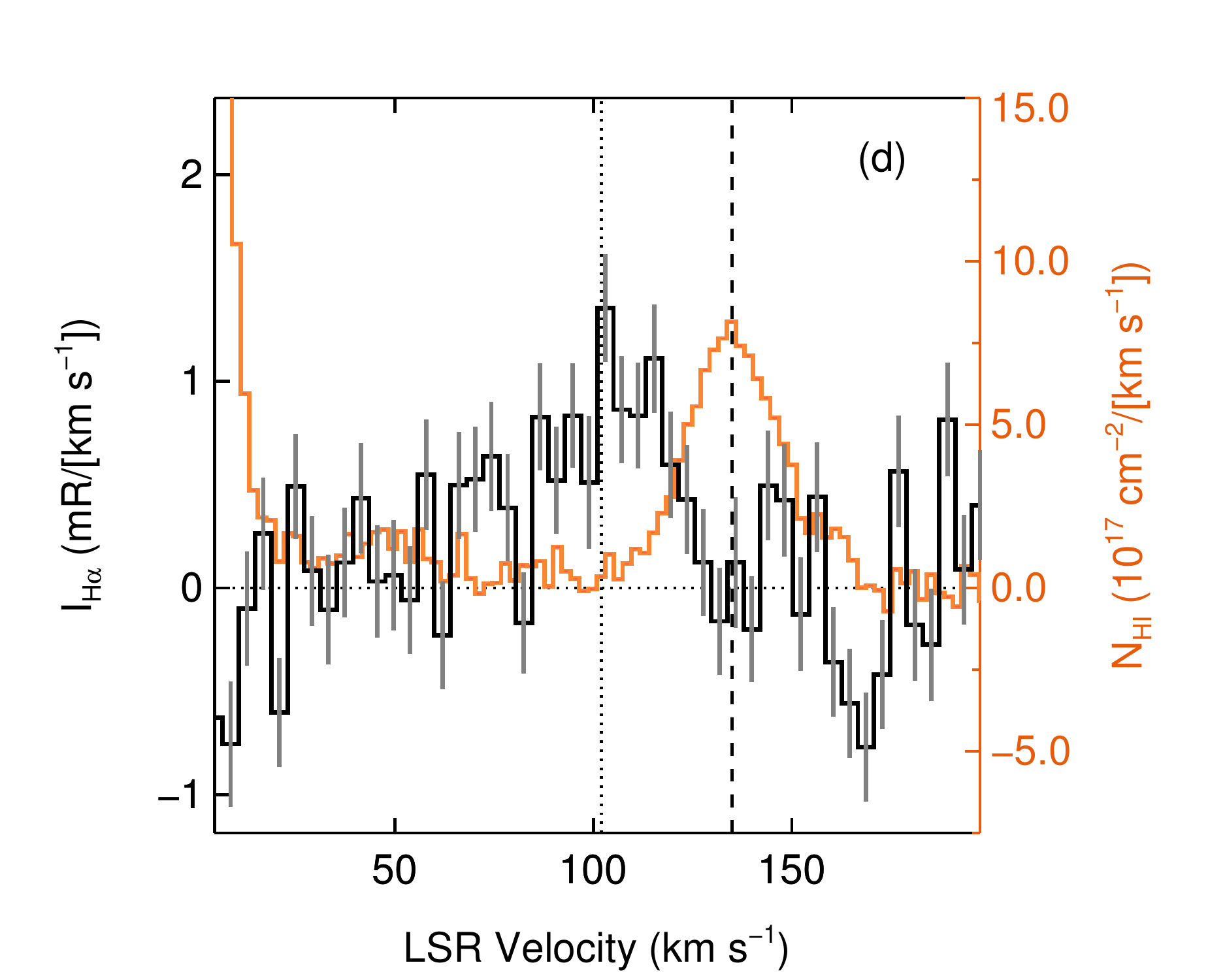} \\
\includegraphics[trim=0 38 65 30,clip,scale=0.45,angle=0]{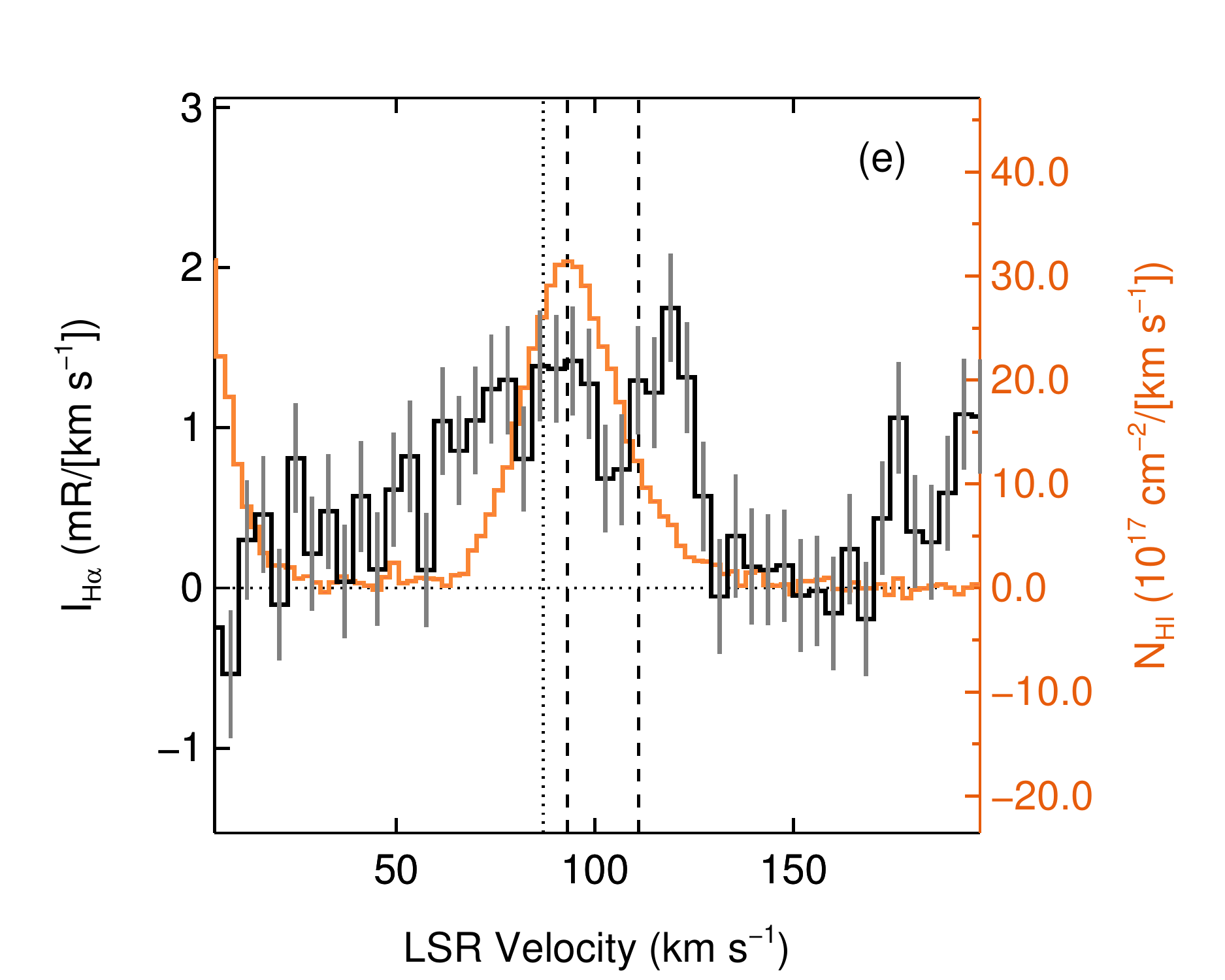}\includegraphics[trim=45 38 0 30,clip,scale=0.45,angle=0]{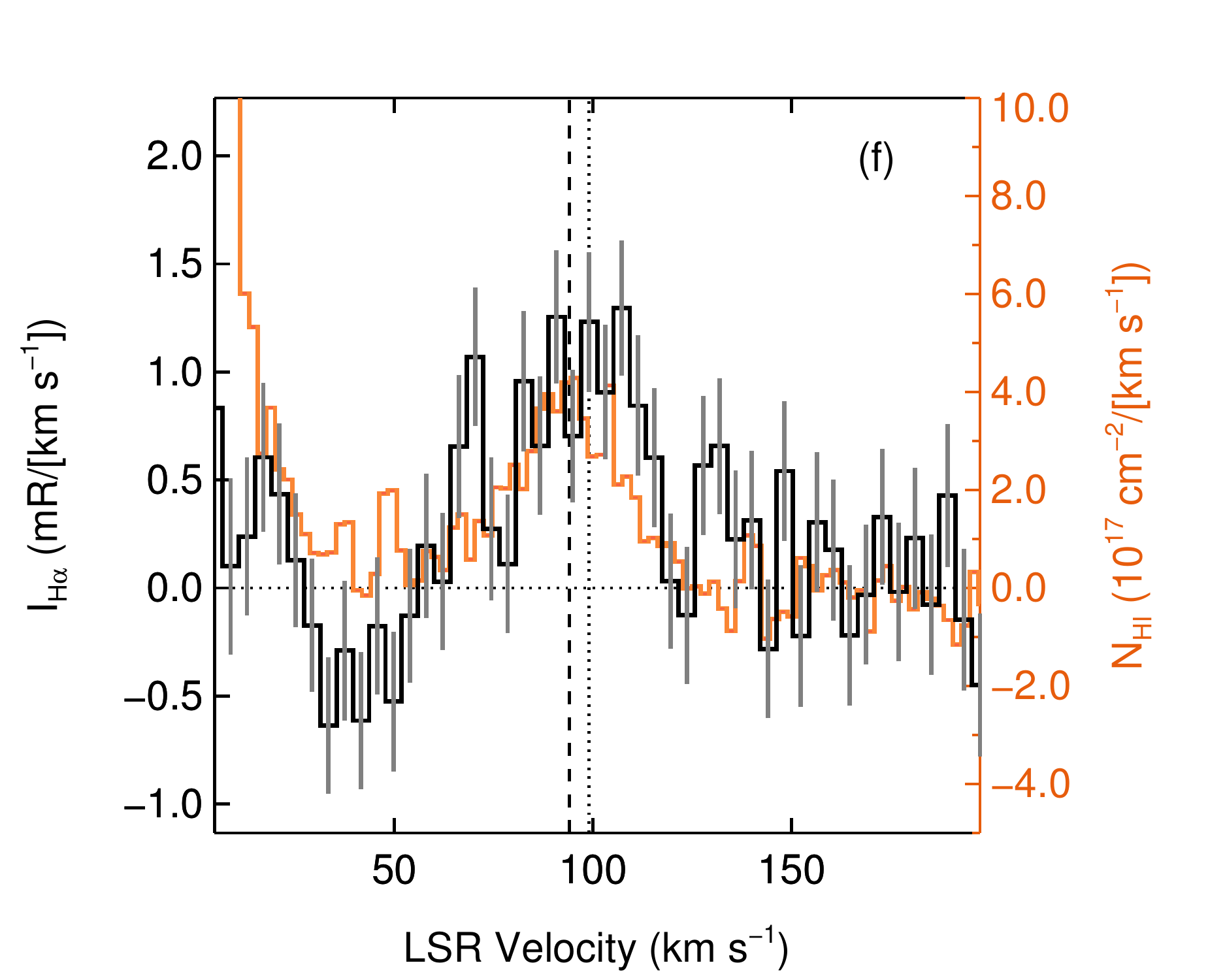} \\
\includegraphics[trim=0 0   65 30,clip,scale=0.45,angle=0]{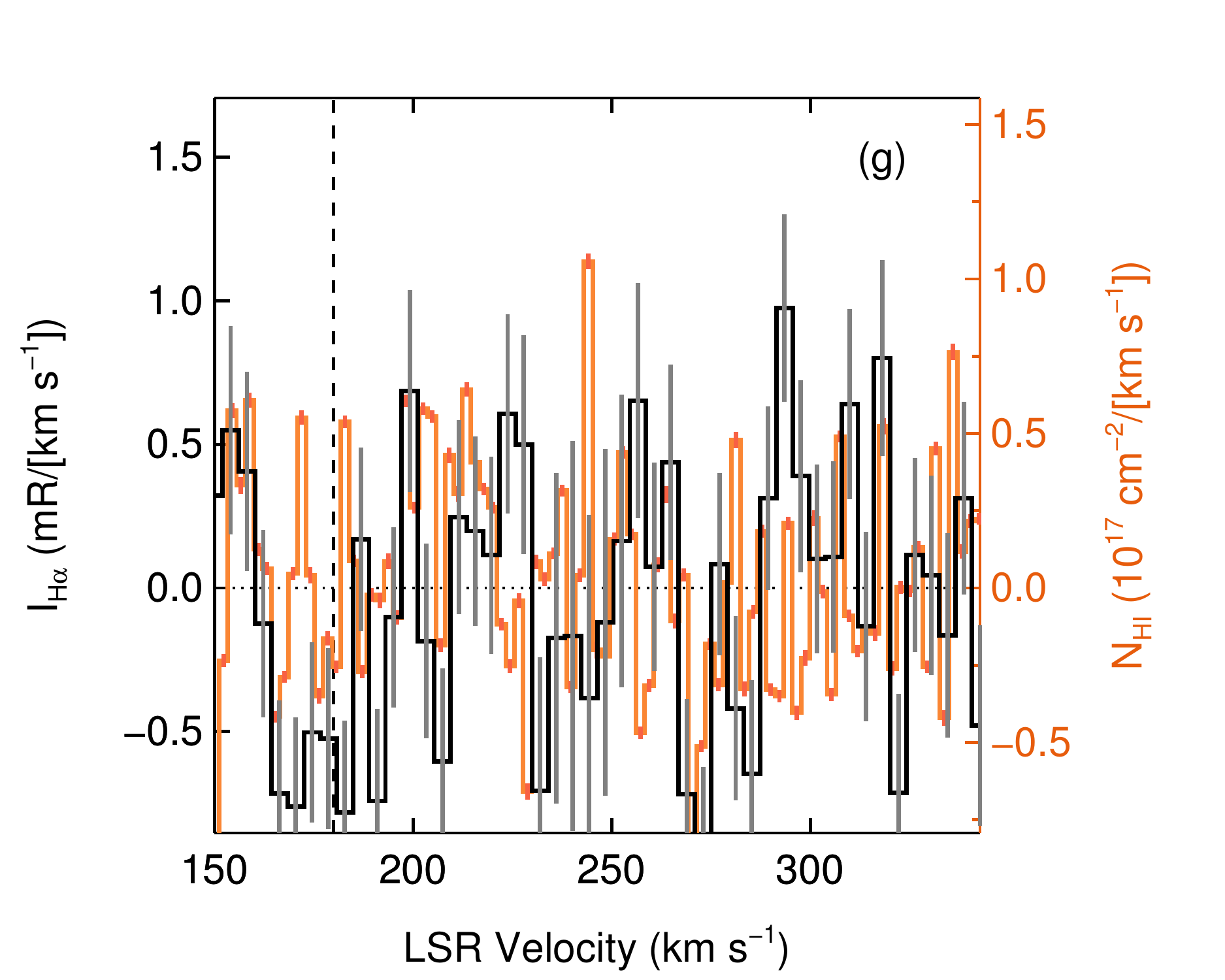}\includegraphics[trim=45   0 0 30,clip,scale=0.45,angle=0]{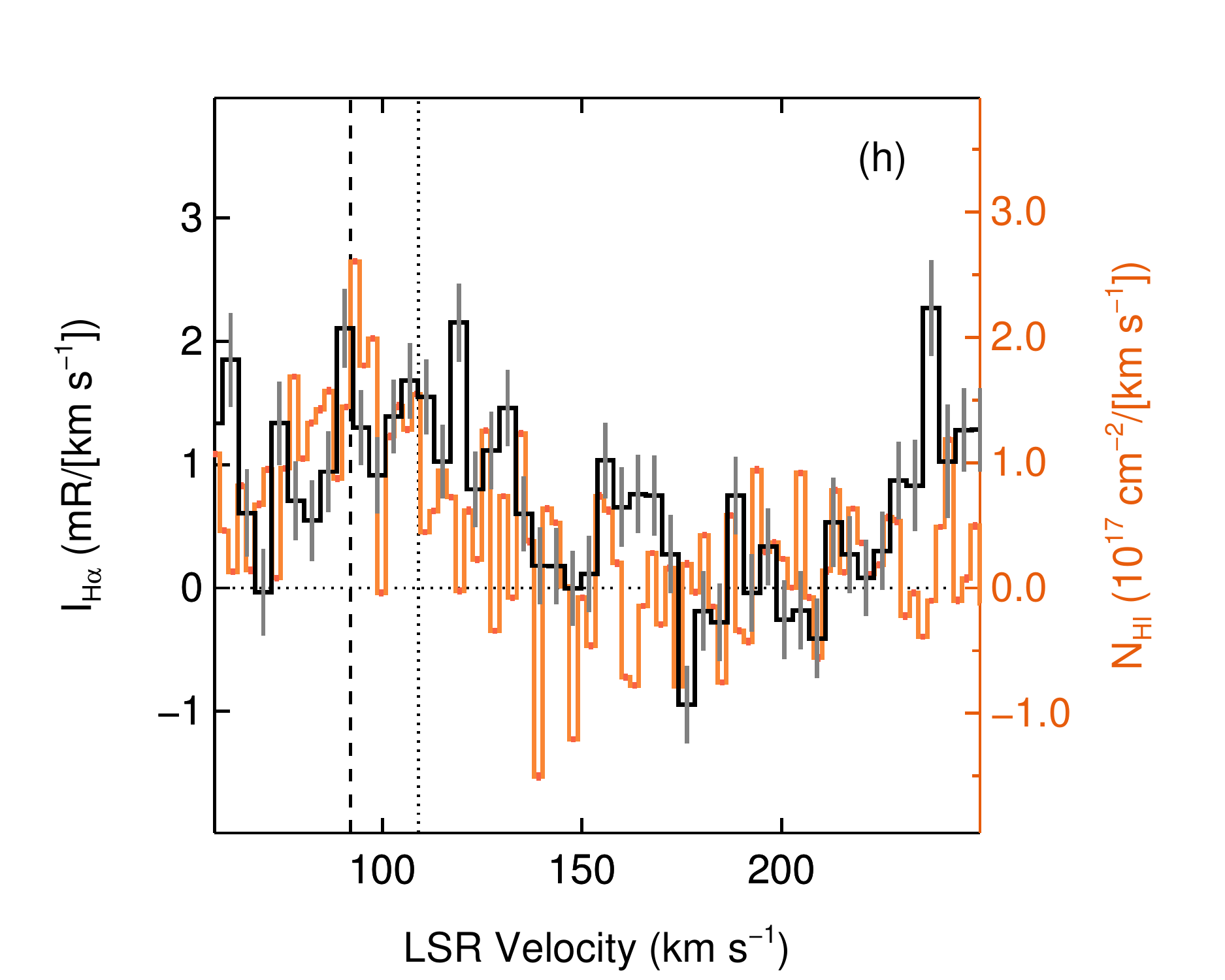} 
\end{center}
\figcaption{The LAB \hi\ and WHAM \ha\ spectra for all $39$ sight lines, where the letters and numbers coincide with the sight line identifiers listed in Table~\ref{table:obs}. The vertical dash and dotted lines mark the center of the \hi\ and \ha\ emission, respectively. To reduce small variations in \hi\ column density and the \ha\ intensity, we increased the bin size of the spectra by a factor of two. }
\end{figure*}

\begin{figure*}
\begin{center}
\includegraphics[trim=0 38 65 30,clip,scale=0.45,angle=0]{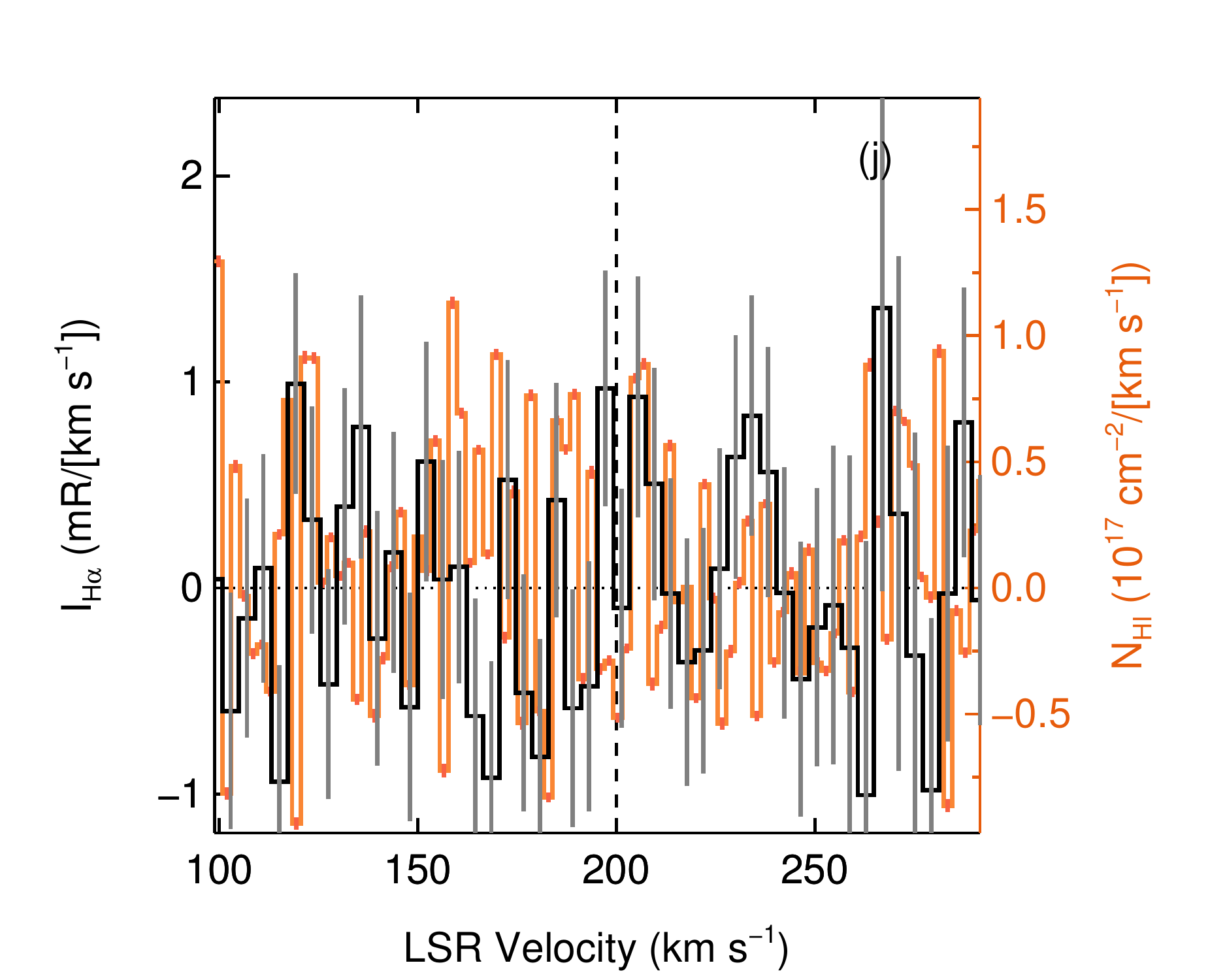}\includegraphics[trim=45 38 0 30,clip,scale=0.45,angle=0]{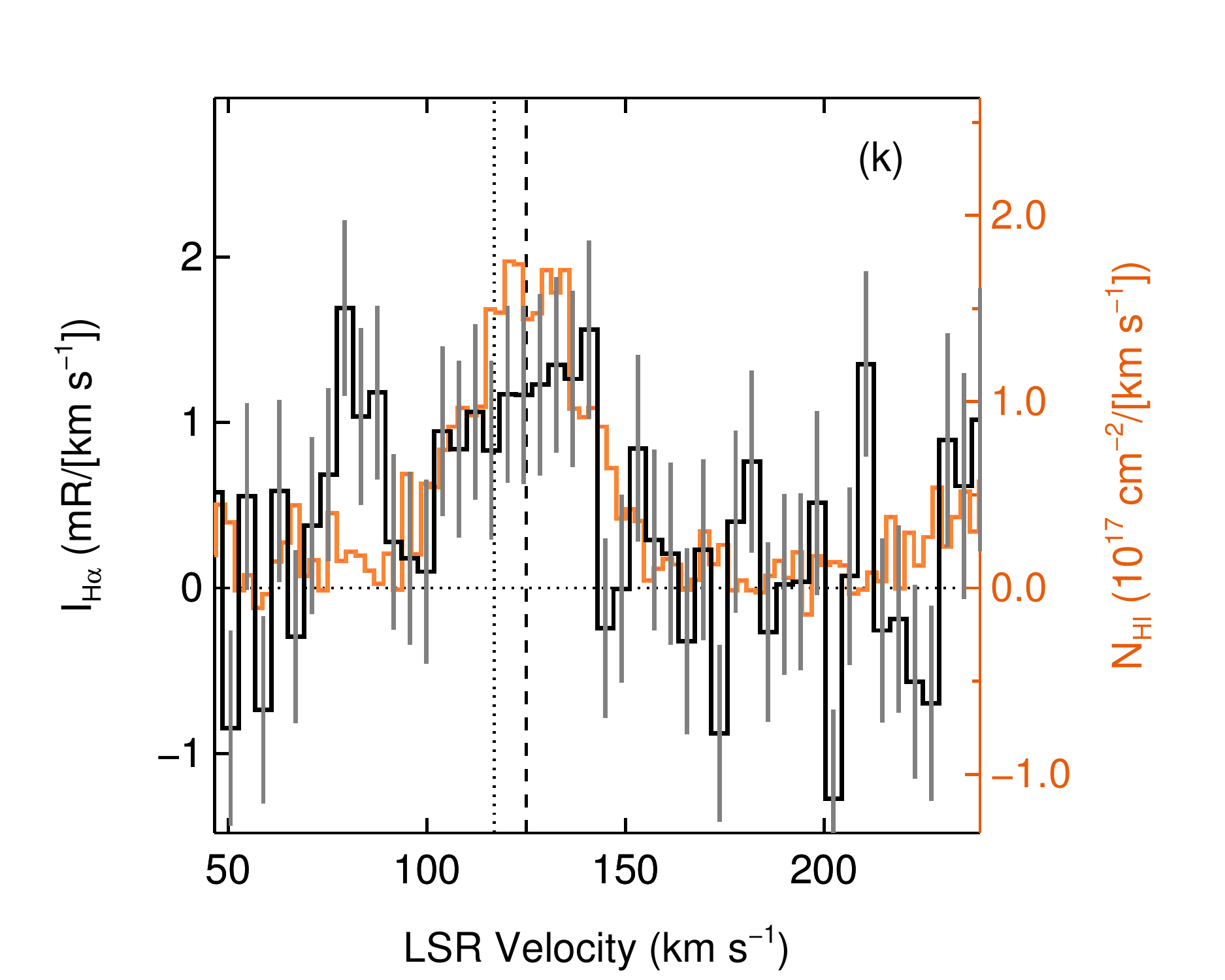} \\
\includegraphics[trim=0 38 65 30,clip,scale=0.45,angle=0]{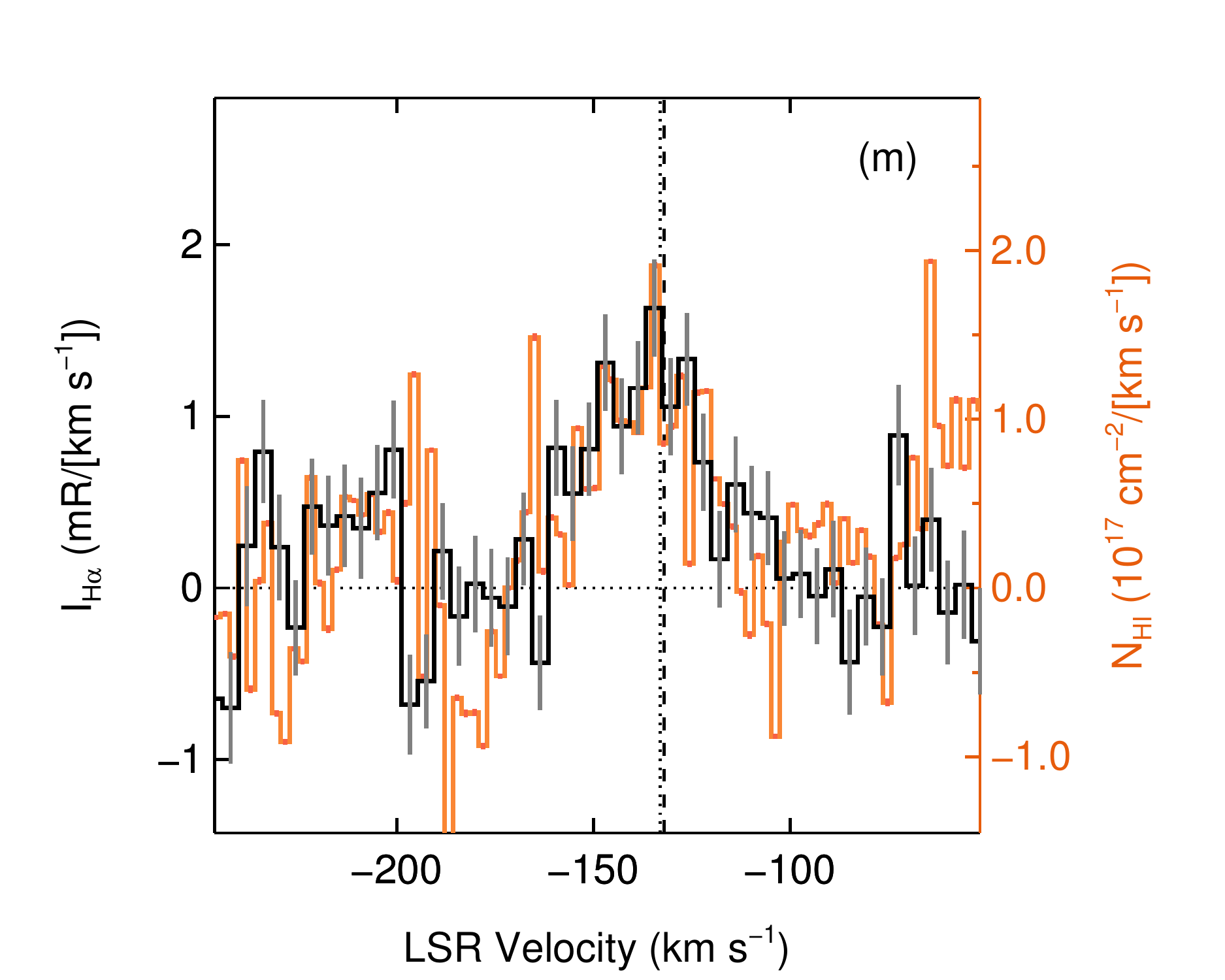}\includegraphics[trim=45 38 0 30,clip,scale=0.45,angle=0]{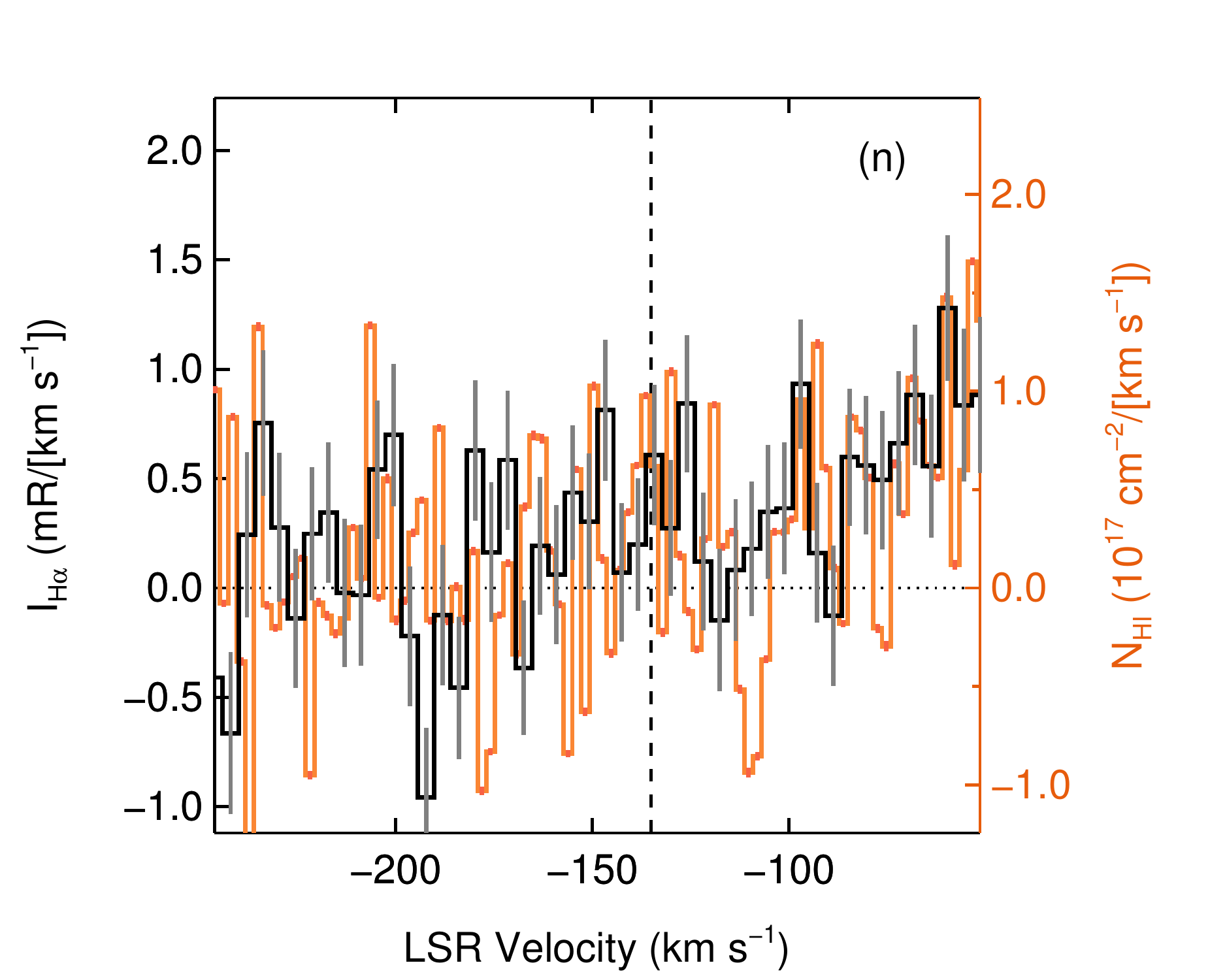} \\
\includegraphics[trim=0 38 65 30,clip,scale=0.45,angle=0]{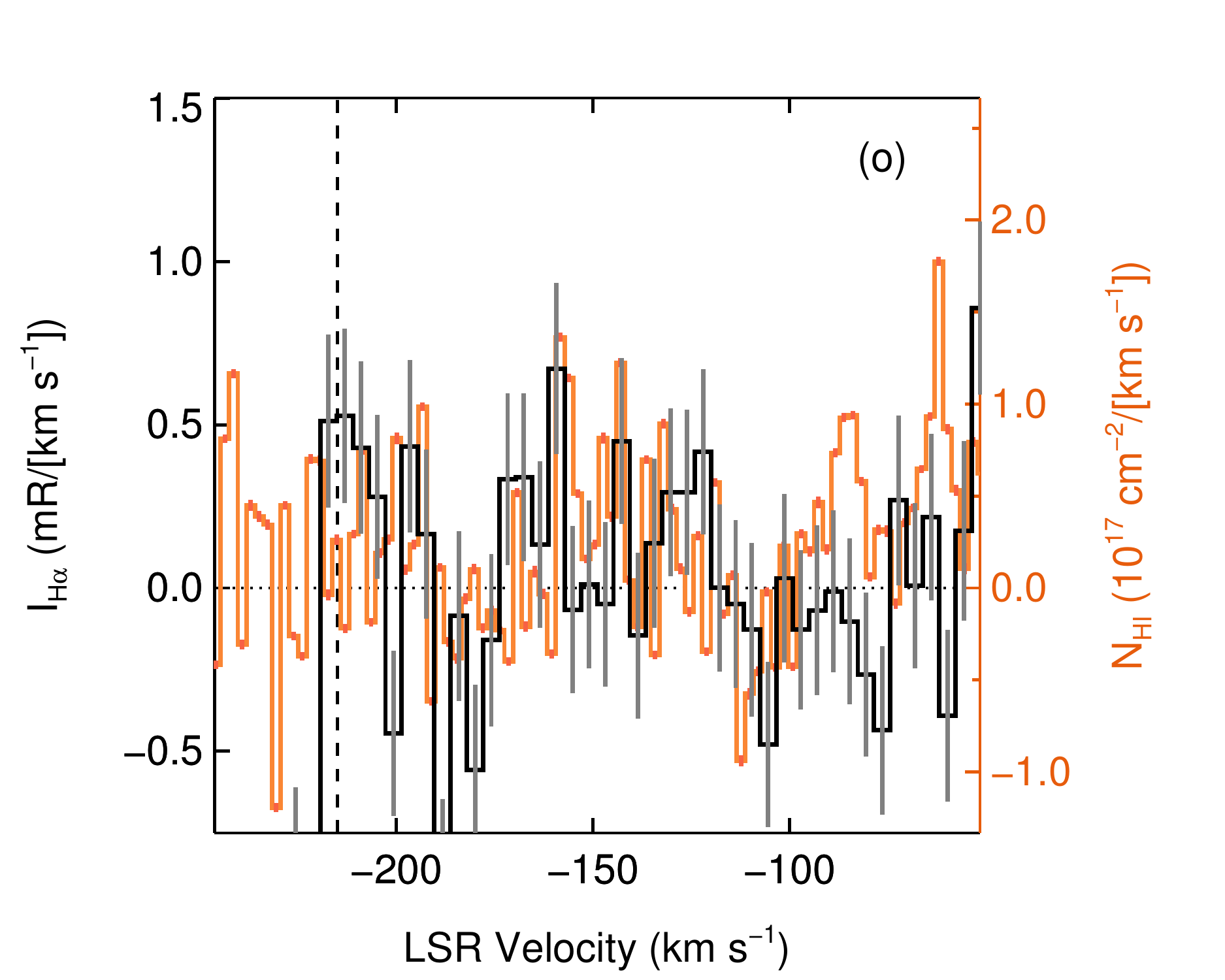}\includegraphics[trim=45 38 0 30,clip,scale=0.45,angle=0]{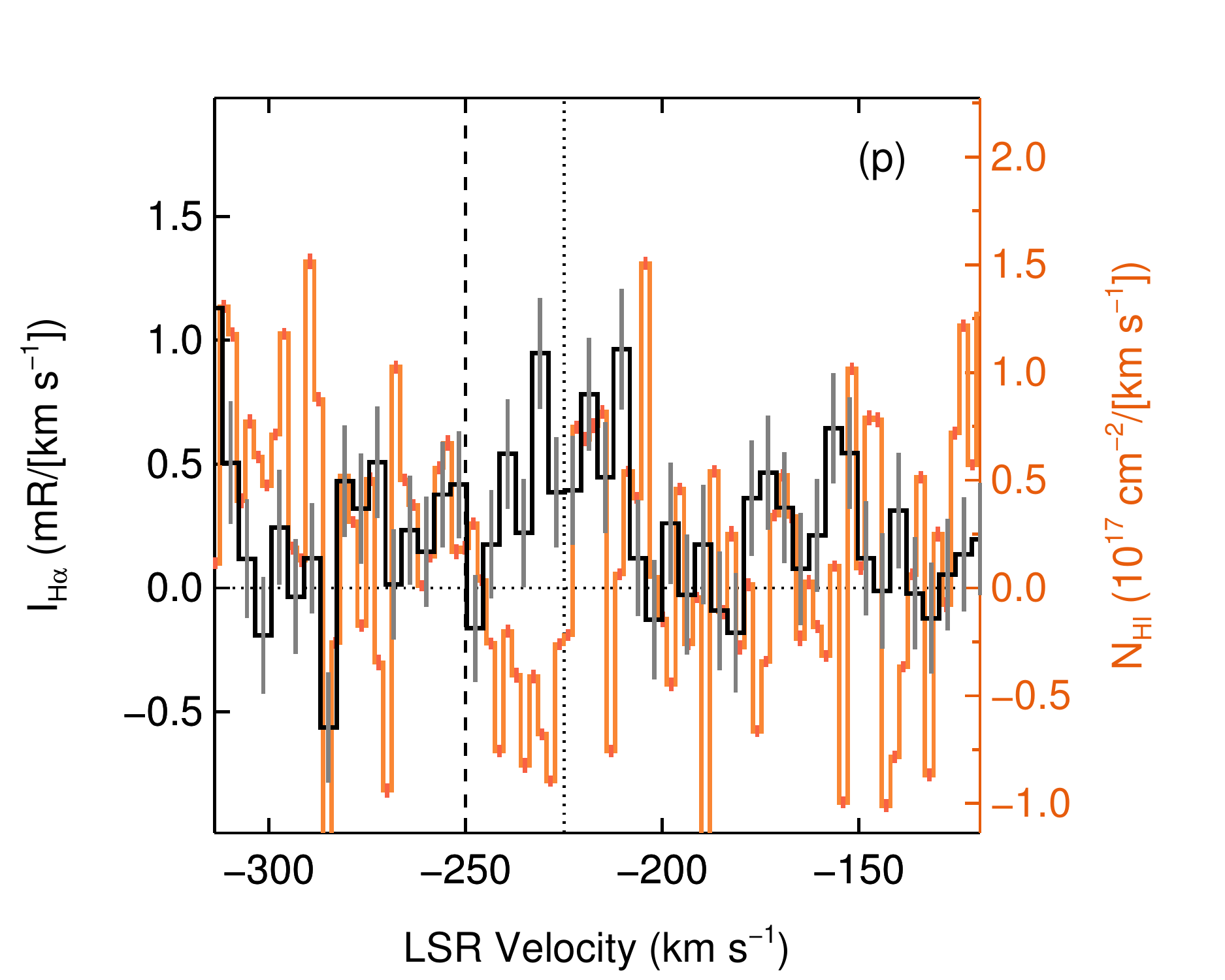} \\
\includegraphics[trim=0   0 65 30,clip,scale=0.45,angle=0]{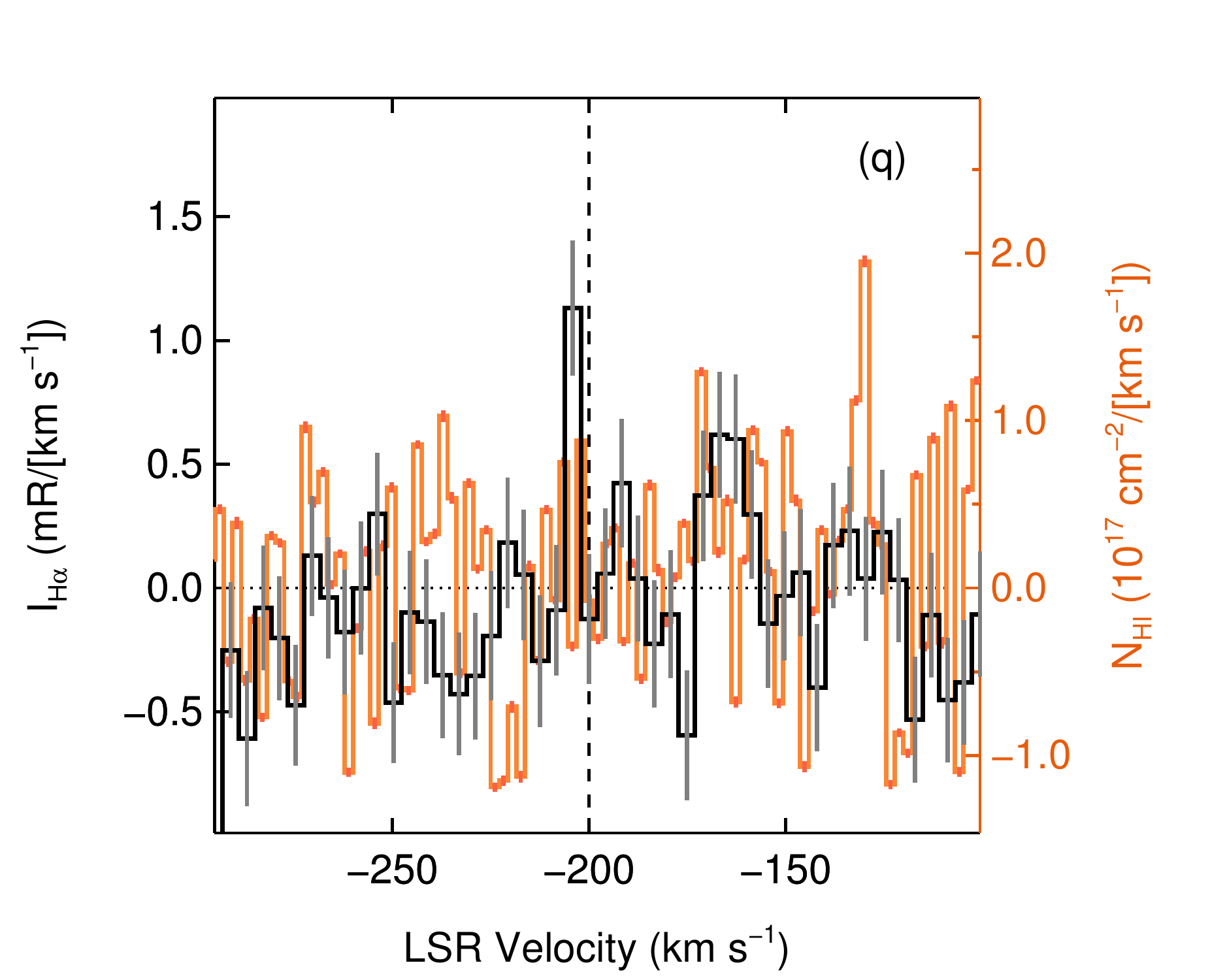}\includegraphics[trim=45   0 0 30,clip,scale=0.45,angle=0]{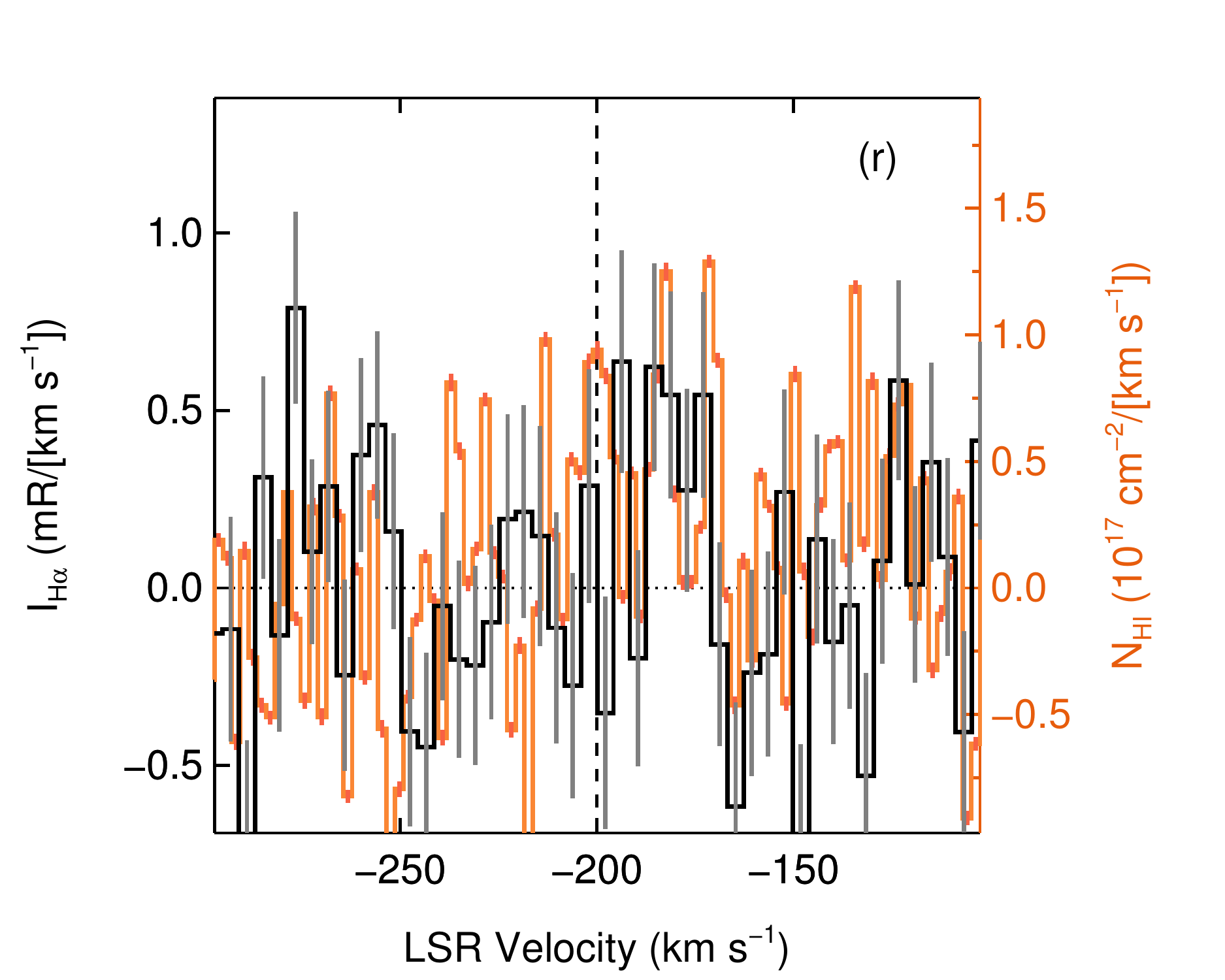} 
\end{center}
\end{figure*}

\begin{figure*}
\begin{center}
\includegraphics[trim=0 38 65 30,clip,scale=0.45,angle=0]{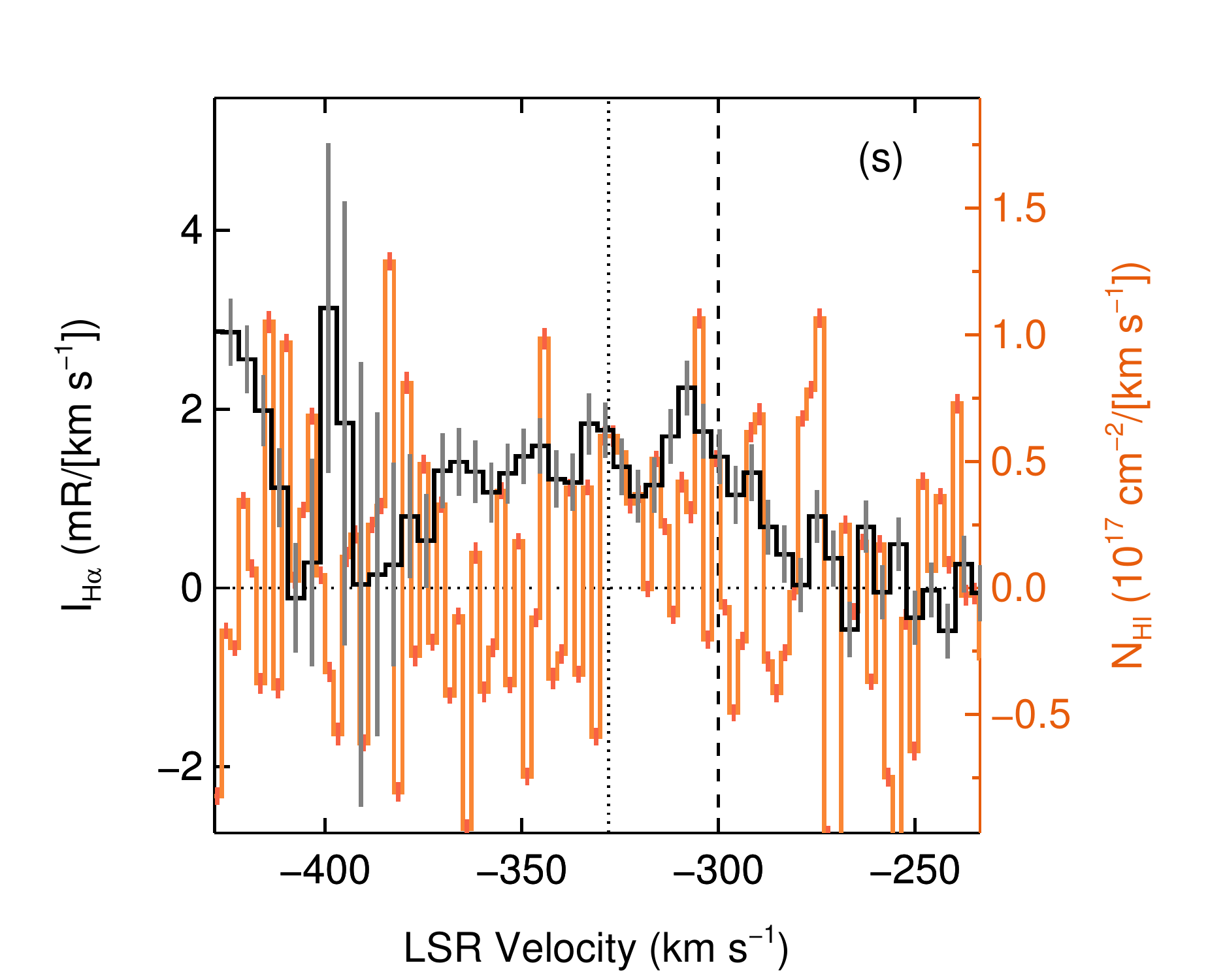}\includegraphics[trim=45 38 0 30,clip,scale=0.45,angle=0]{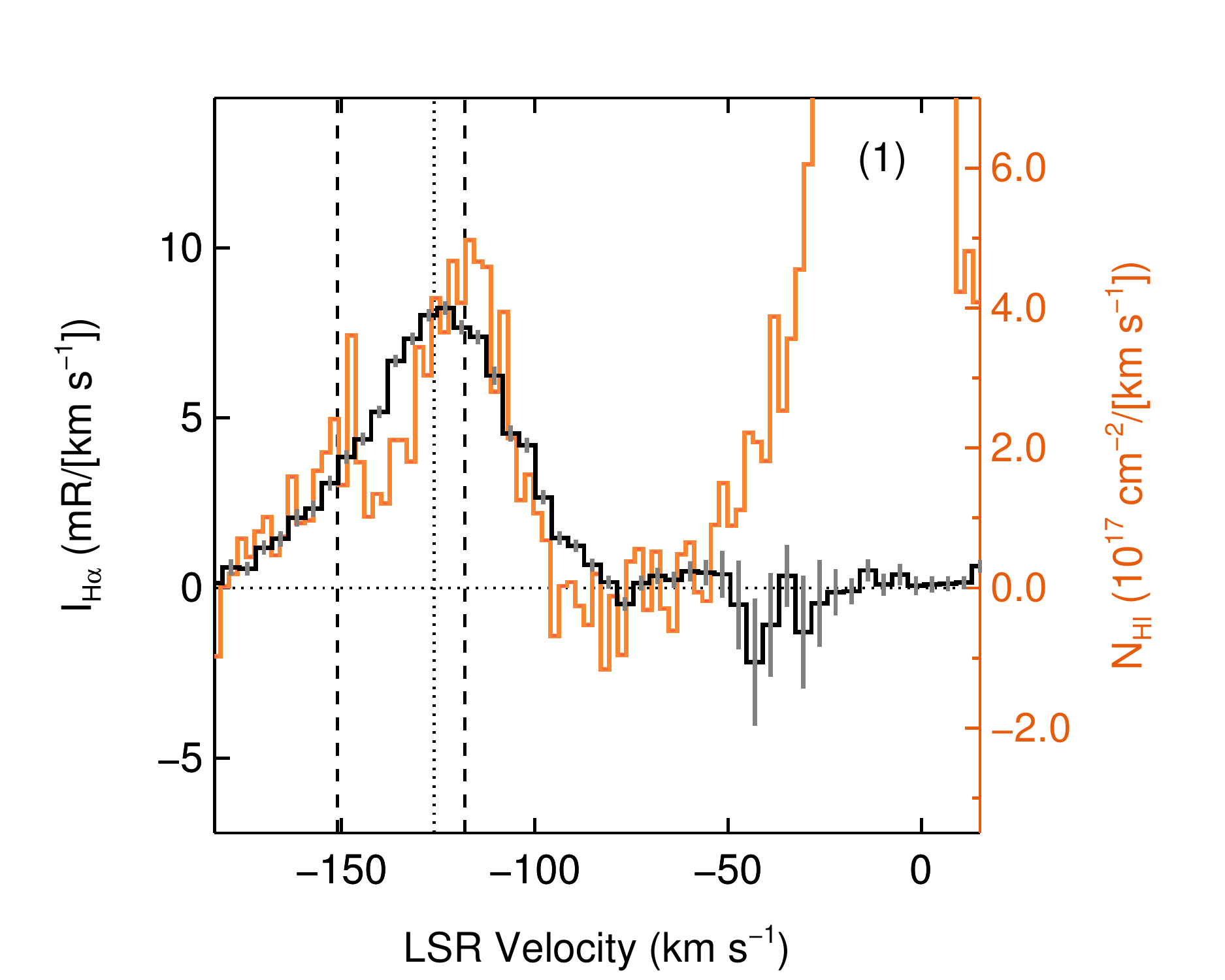} \\
\includegraphics[trim=0 38 65 30,clip,scale=0.45,angle=0]{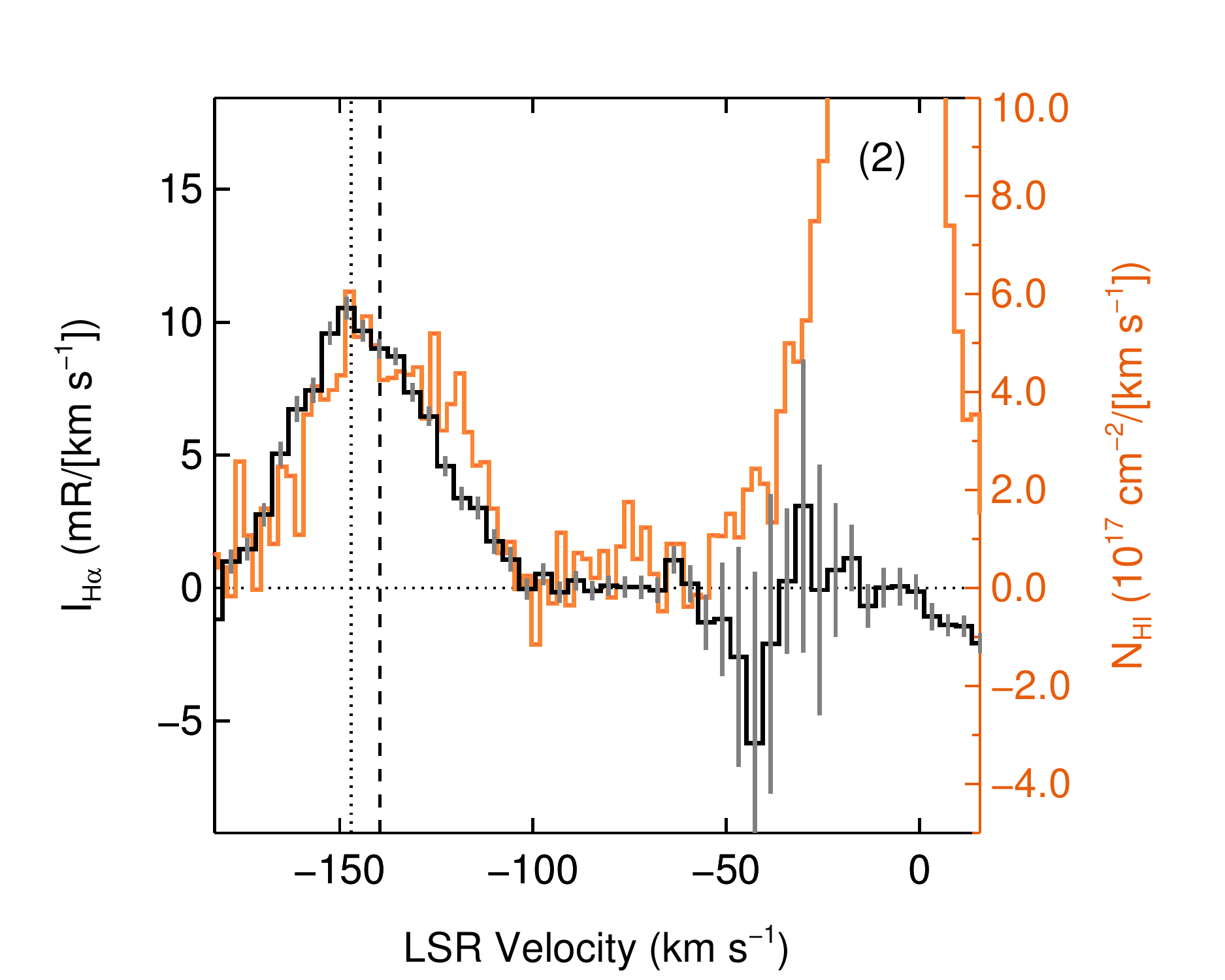}\includegraphics[trim=45 38 0 30,clip,scale=0.45,angle=0]{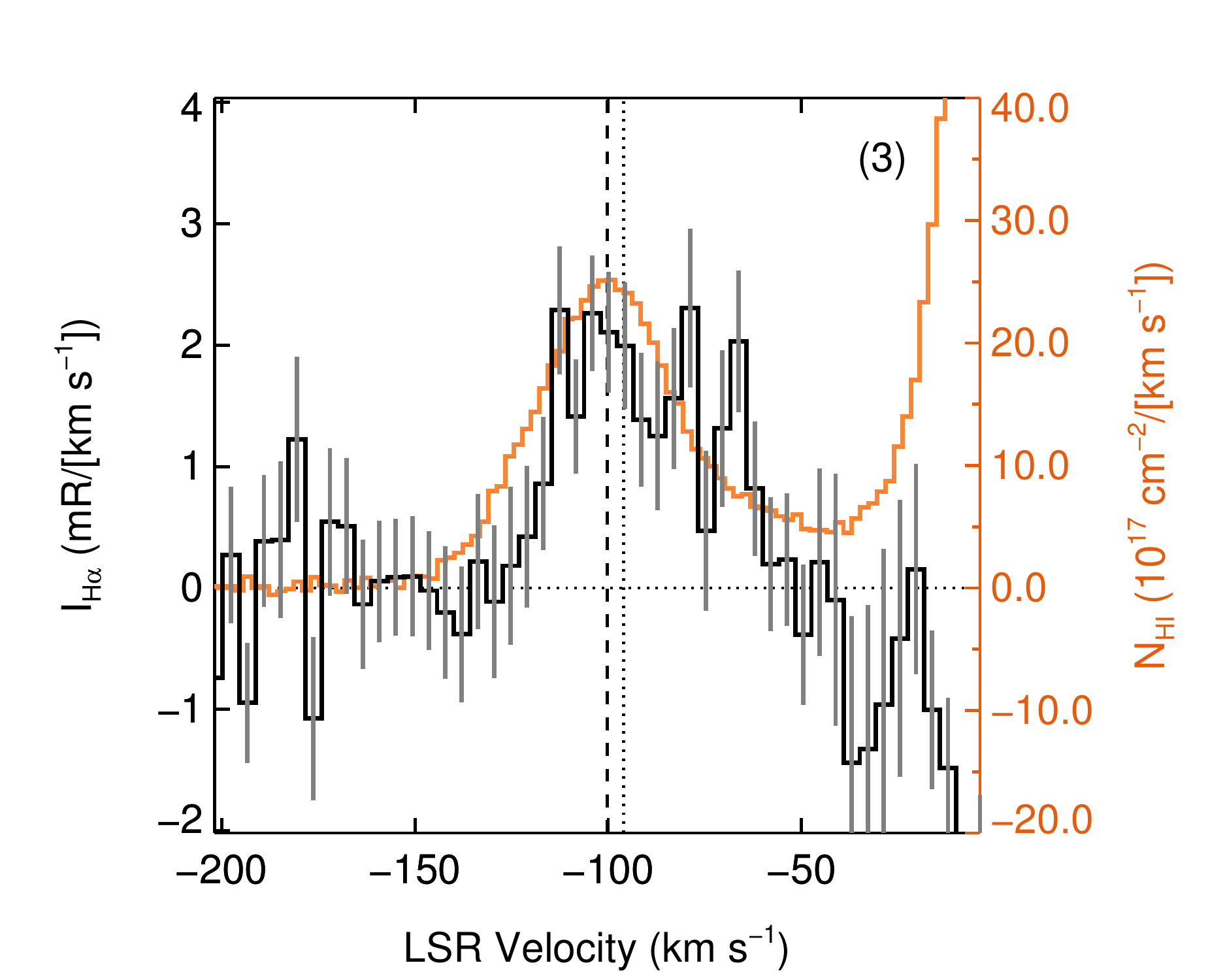} \\
\includegraphics[trim=0 38 65 30,clip,scale=0.45,angle=0]{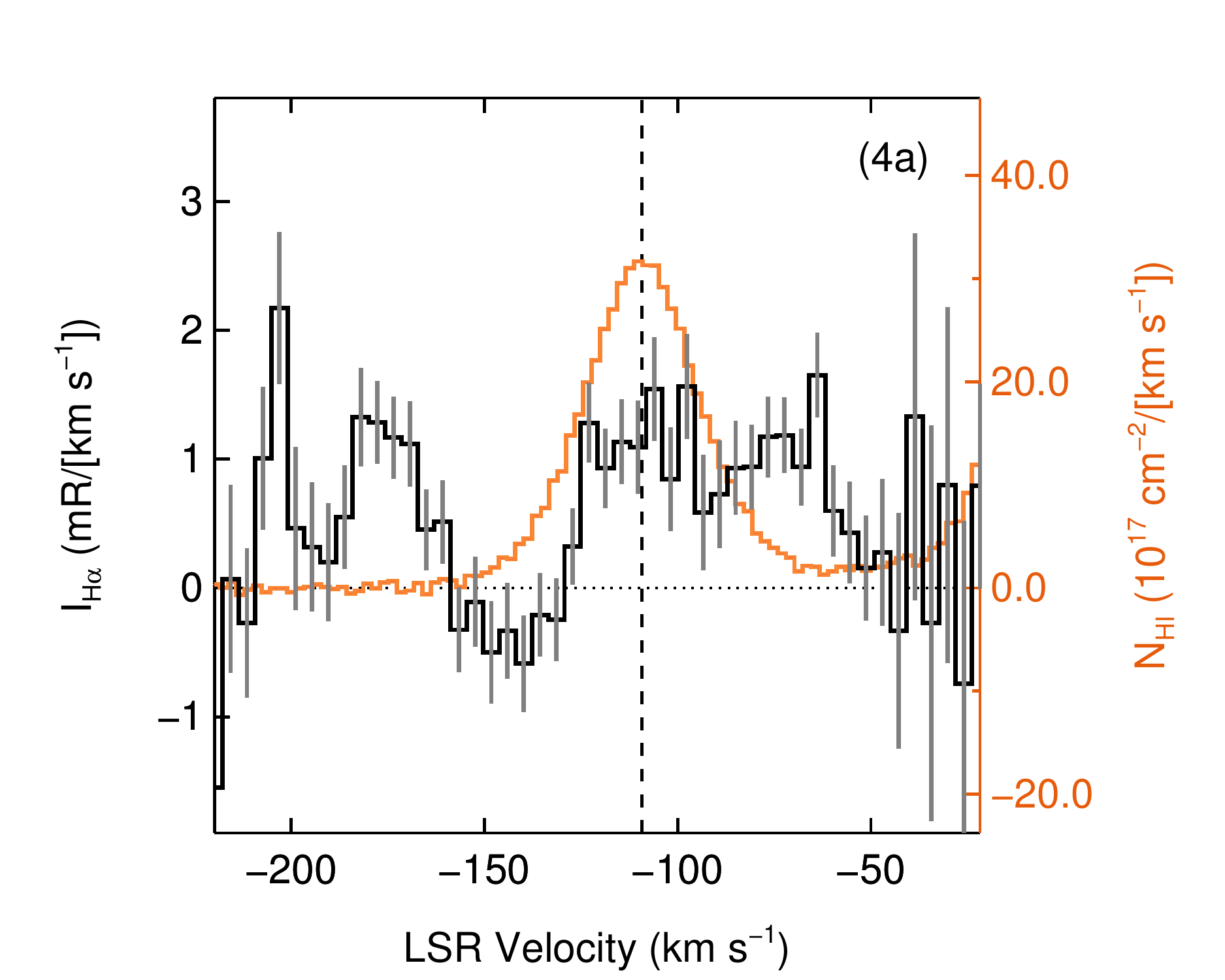}\includegraphics[trim=45 38 0 30,clip,scale=0.45,angle=0]{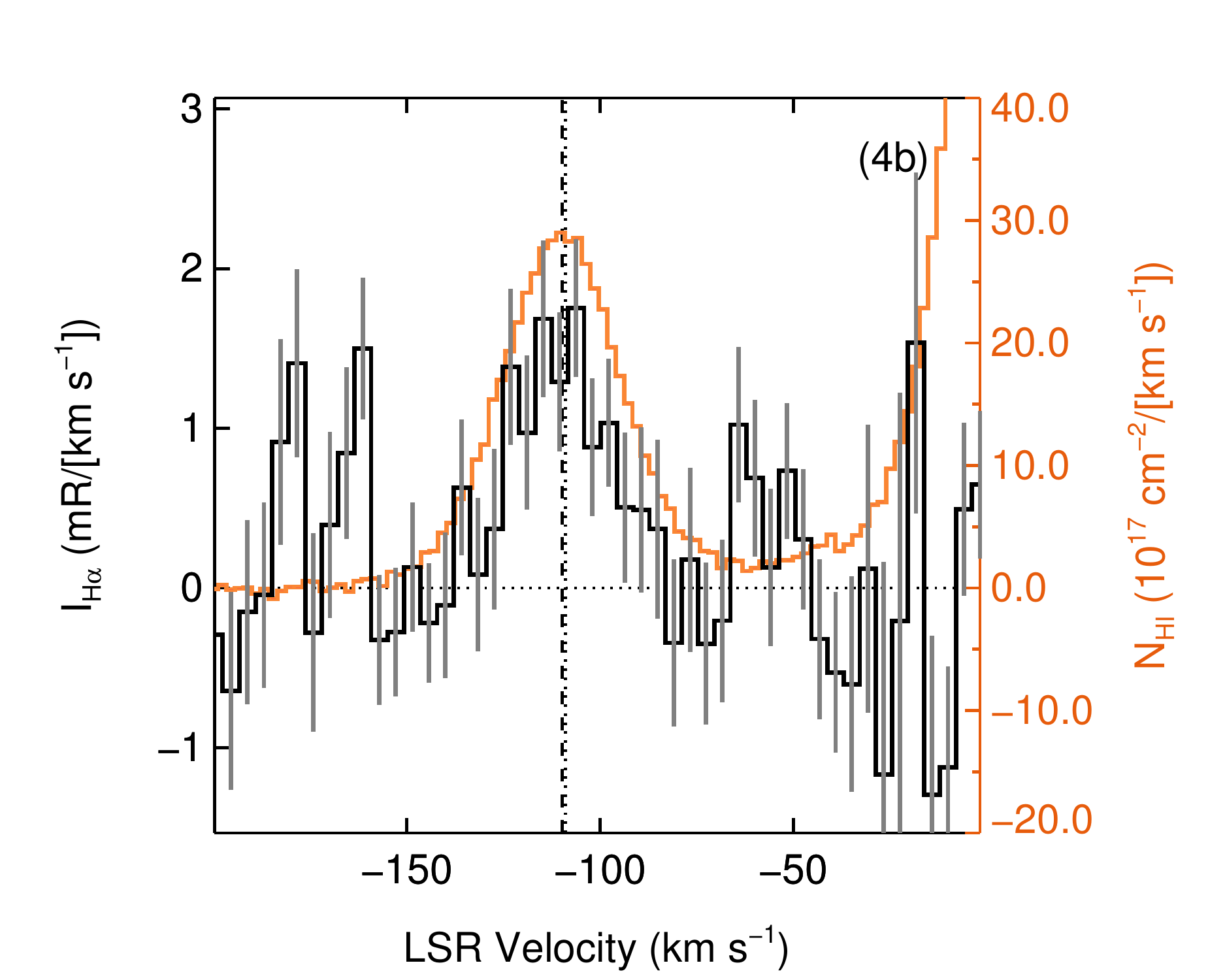} \\
\includegraphics[trim=0   0 65 30,clip,scale=0.45,angle=0]{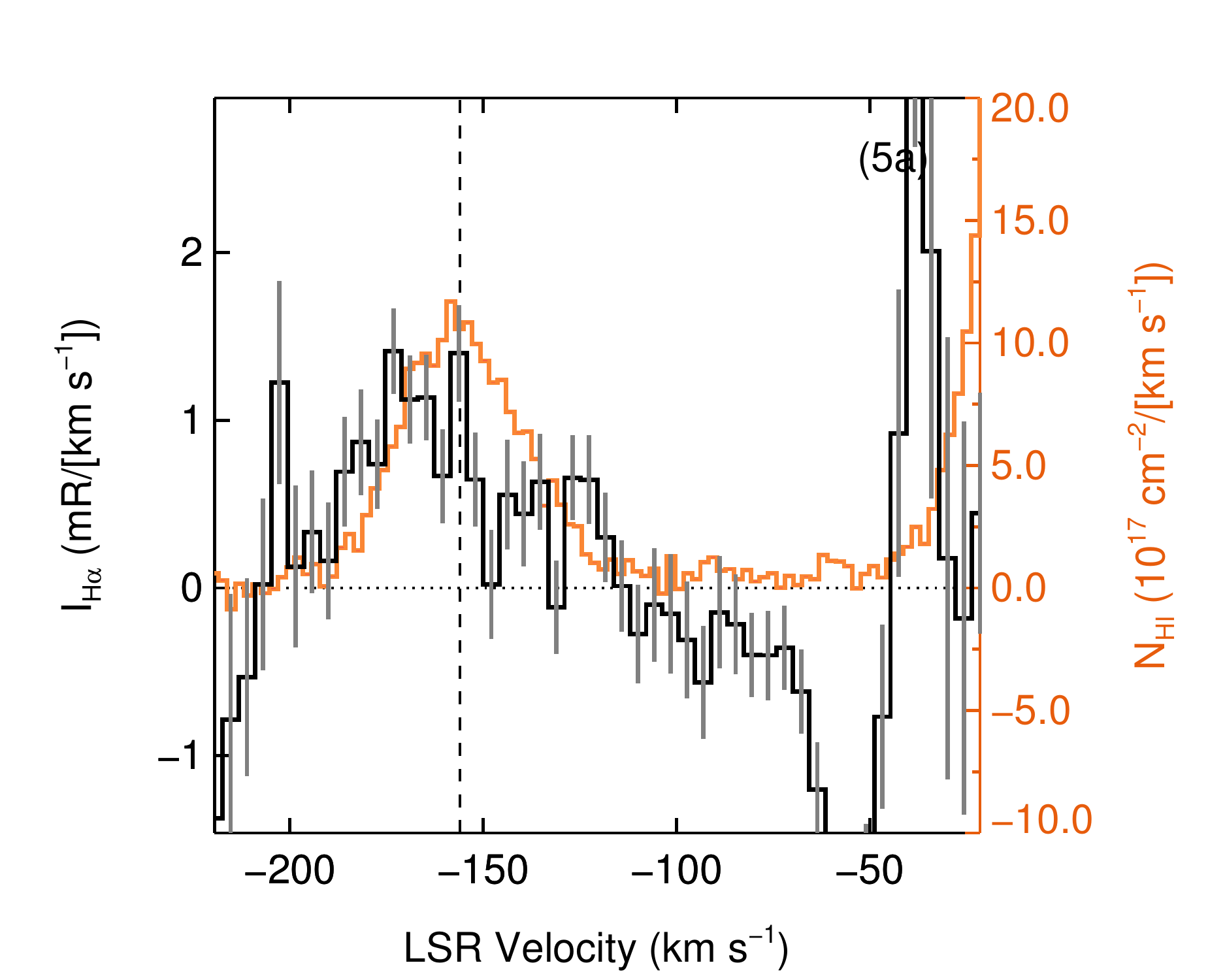}\includegraphics[trim=45   0 0 30,clip,scale=0.45,angle=0]{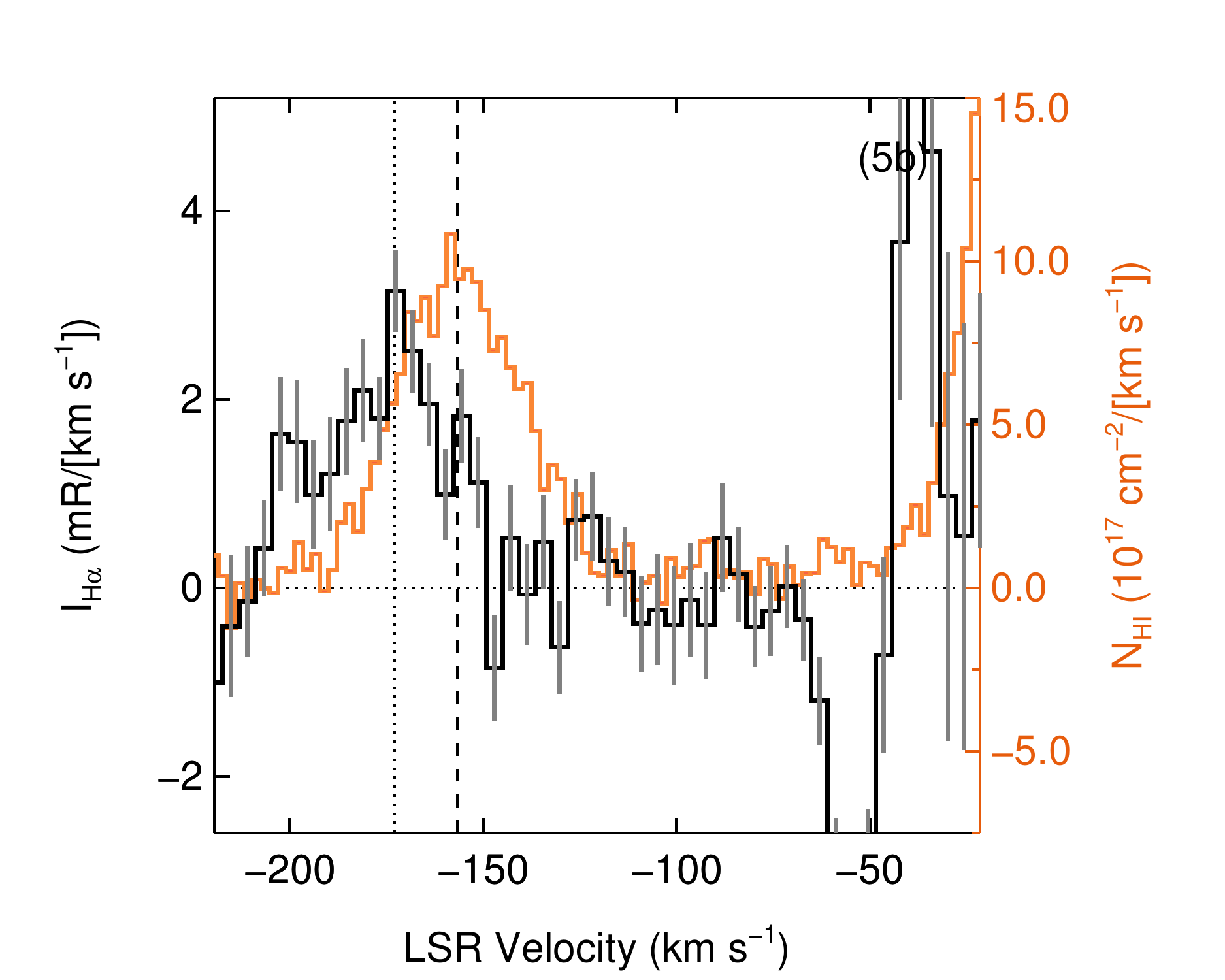} 
\end{center}
\end{figure*}

\begin{figure*}
\begin{center}
\includegraphics[trim=0 38 65 30,clip,scale=0.45,angle=0]{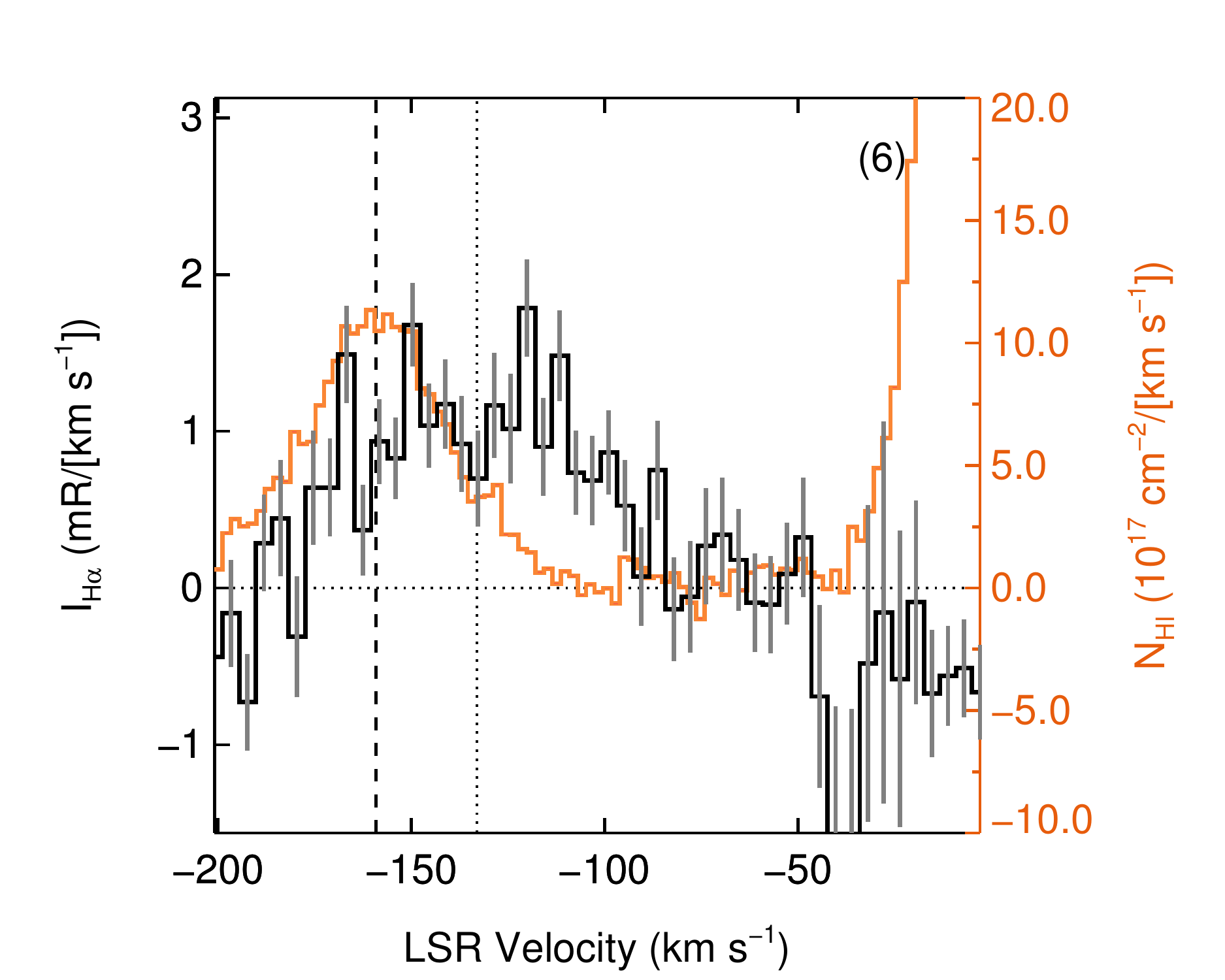}\includegraphics[trim=45 38 0 30,clip,scale=0.45,angle=0]{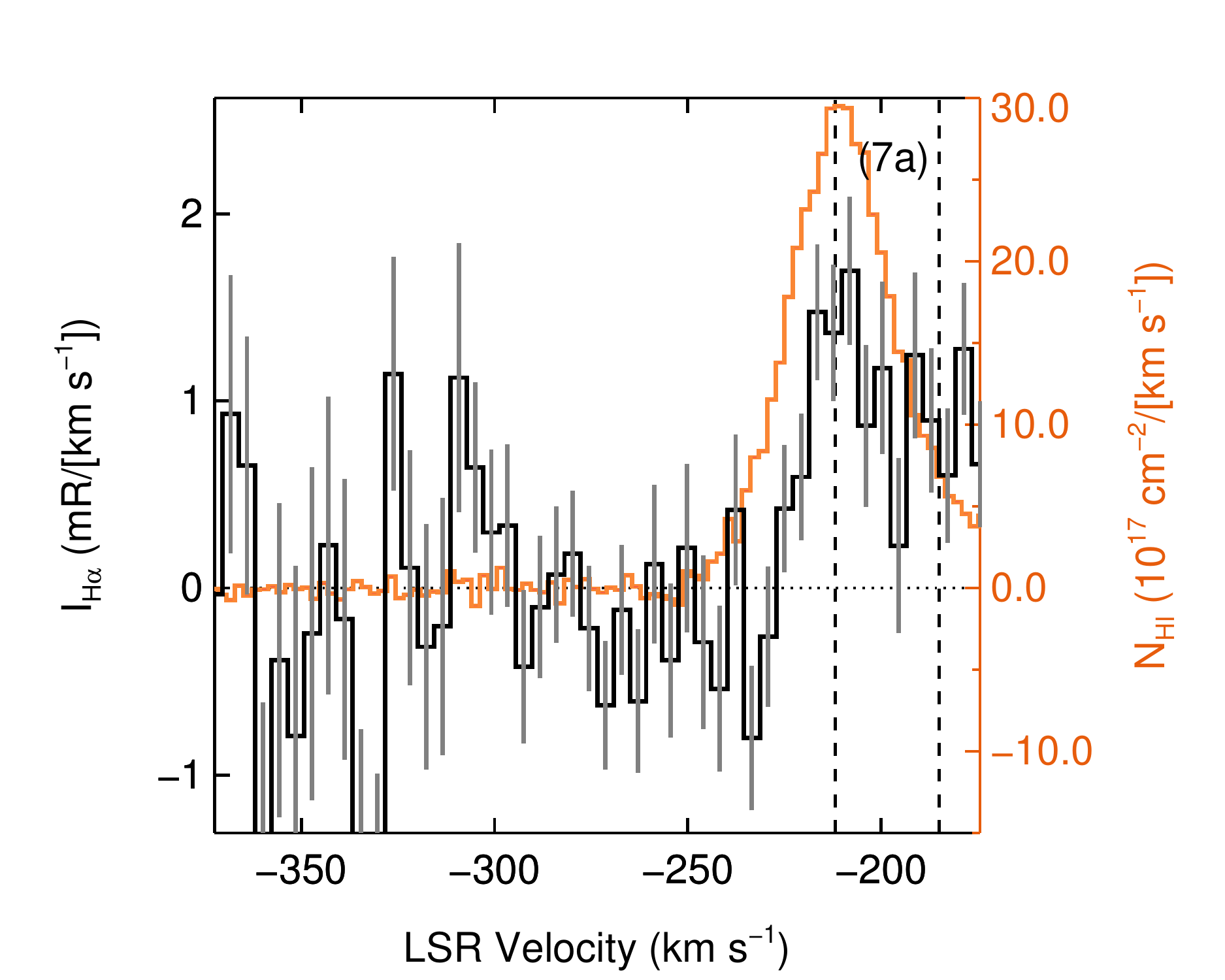} \\
\includegraphics[trim=0 38 65 30,clip,scale=0.45,angle=0]{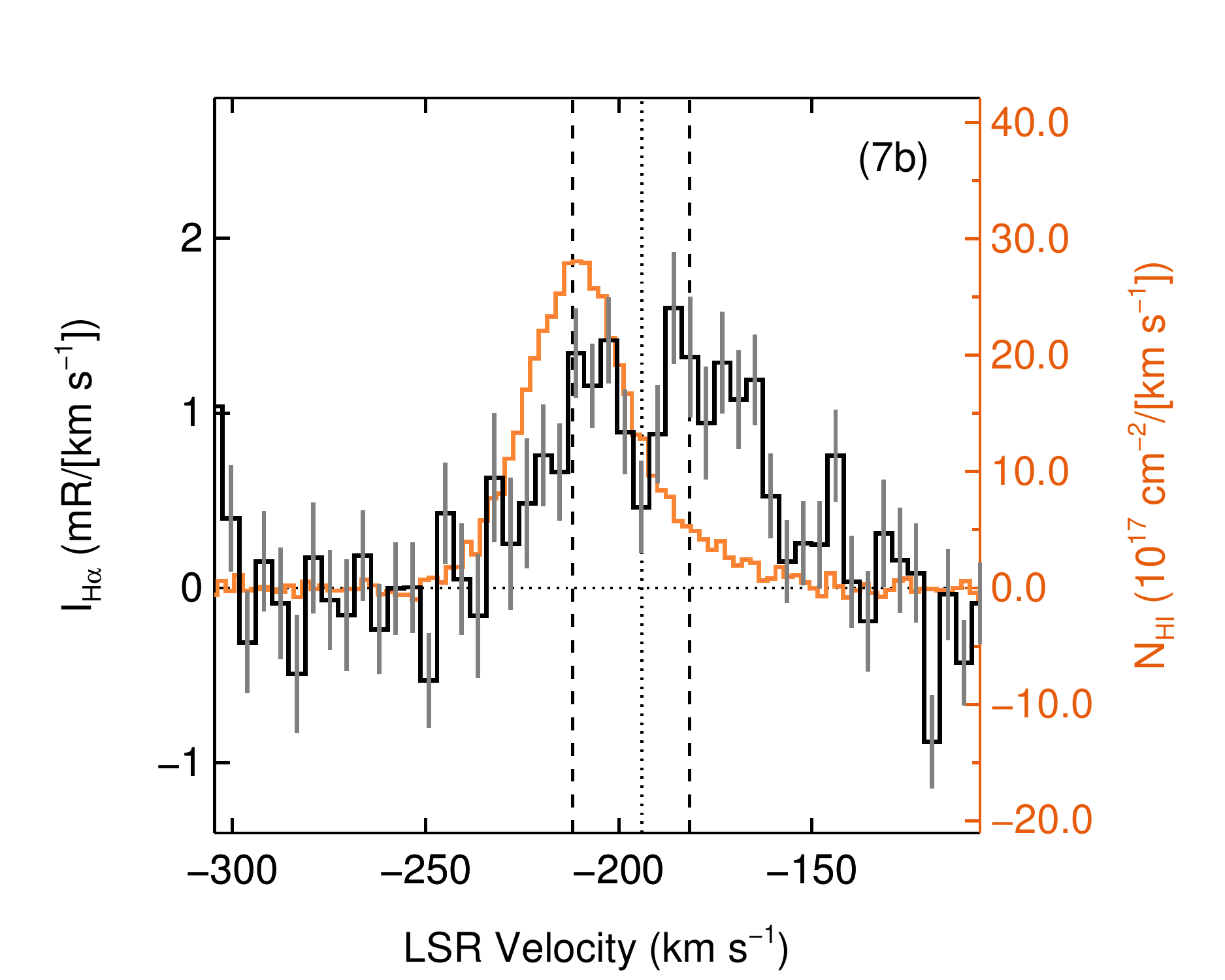}\includegraphics[trim=45 38 0 30,clip,scale=0.45,angle=0]{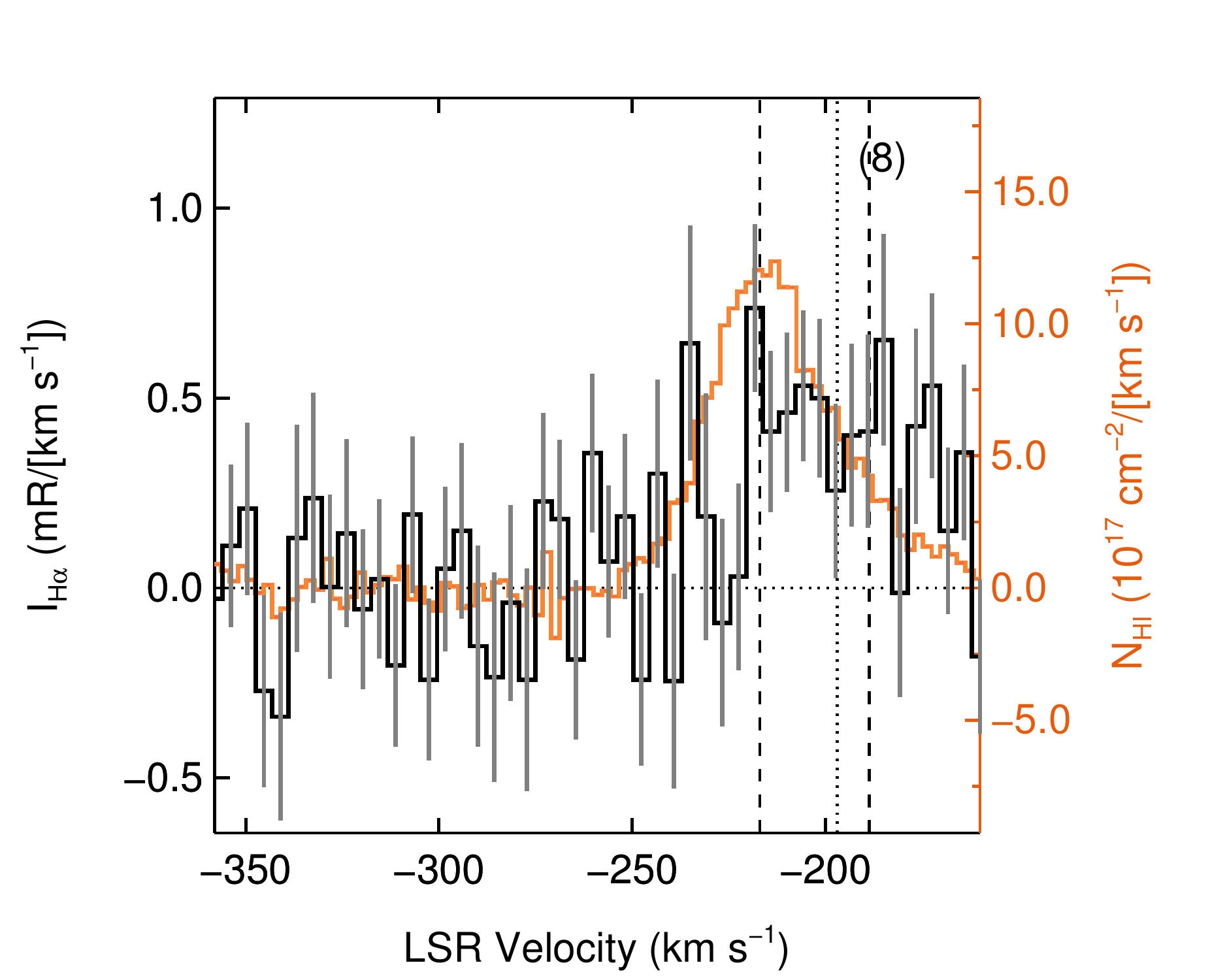} \\
\includegraphics[trim=0 38 65 30,clip,scale=0.45,angle=0]{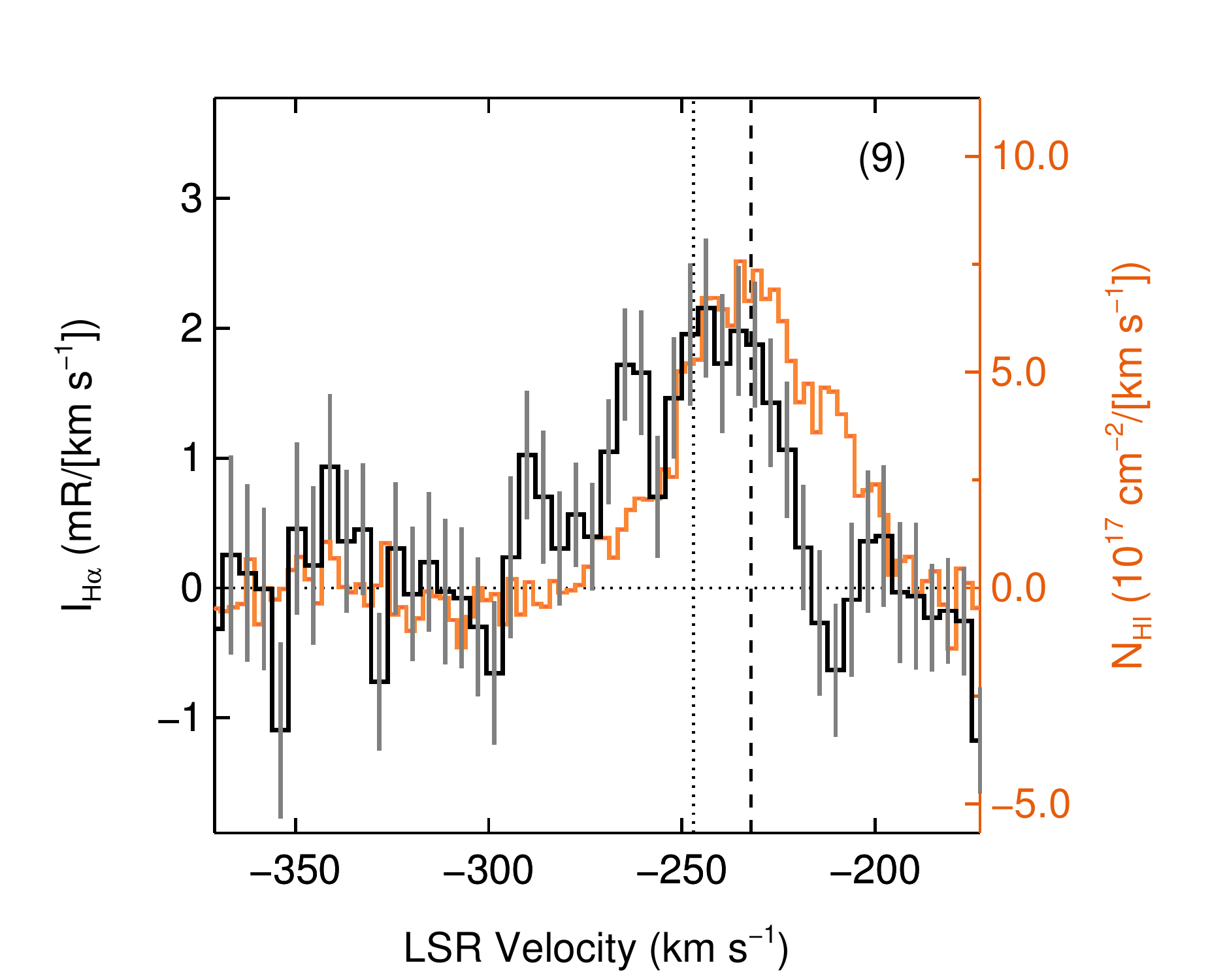}\includegraphics[trim=45 38 0 30,clip,scale=0.45,angle=0]{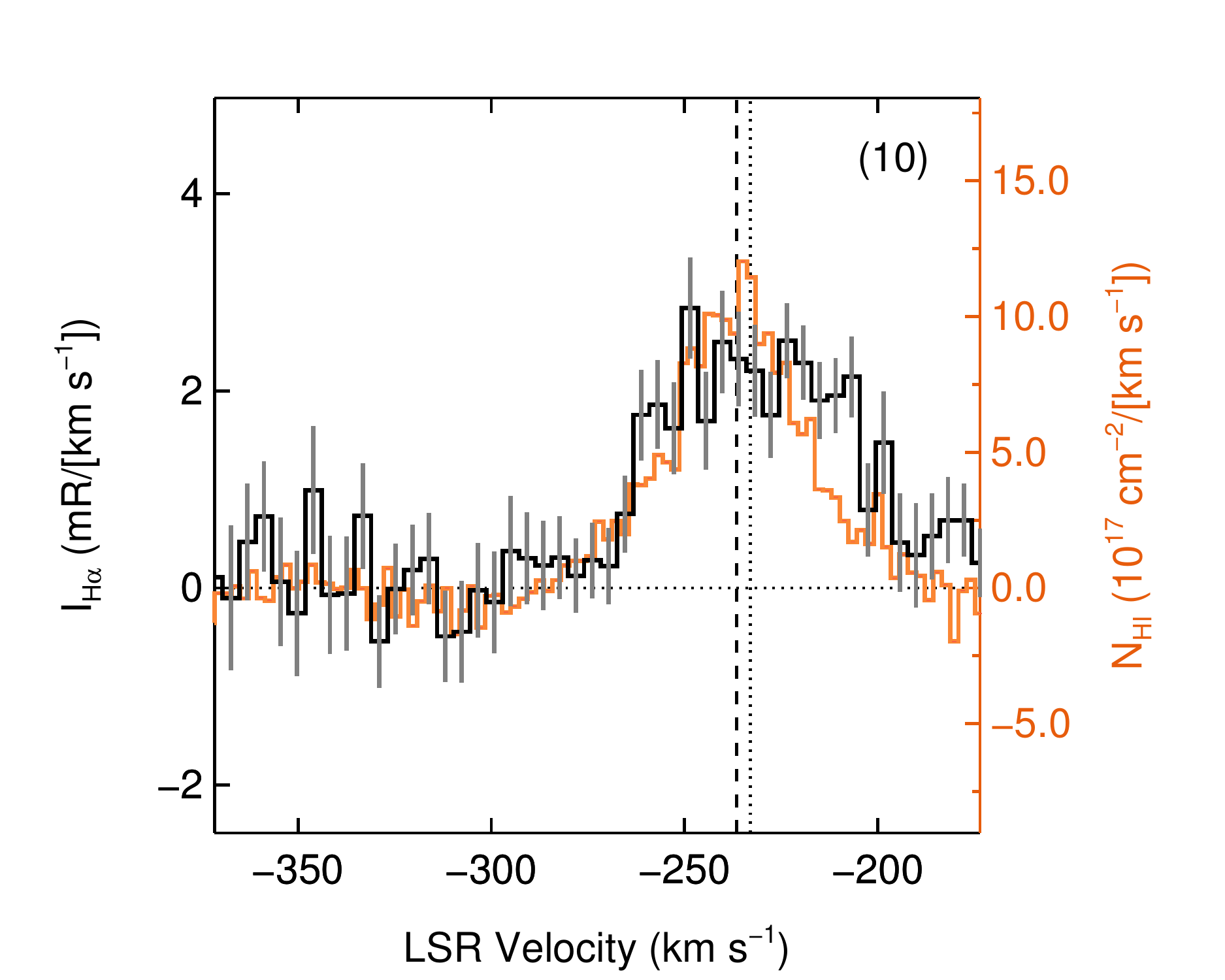} \\
\includegraphics[trim=0   0 65 30,clip,scale=0.45,angle=0]{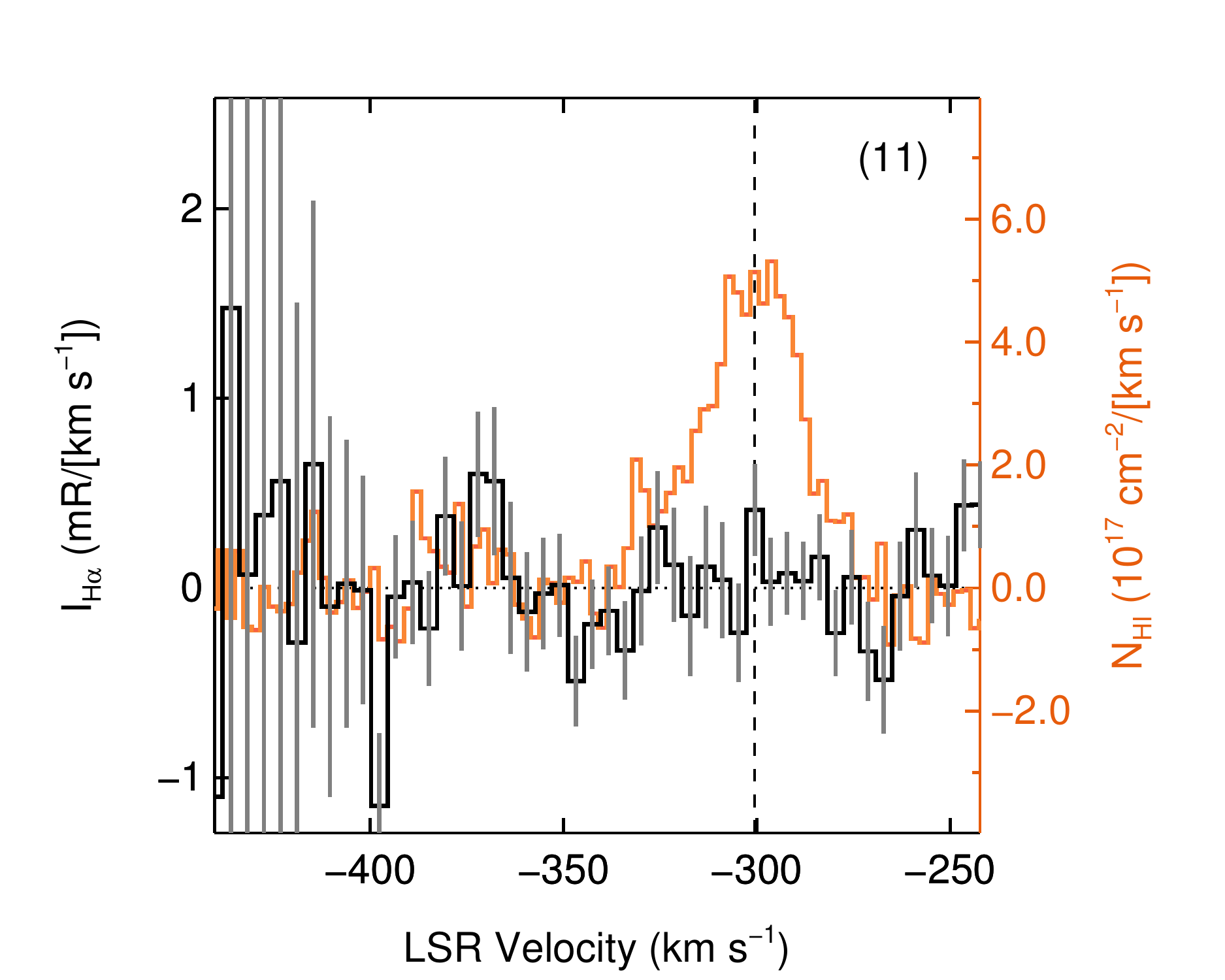}\includegraphics[trim=45   0 0 30,clip,scale=0.45,angle=0]{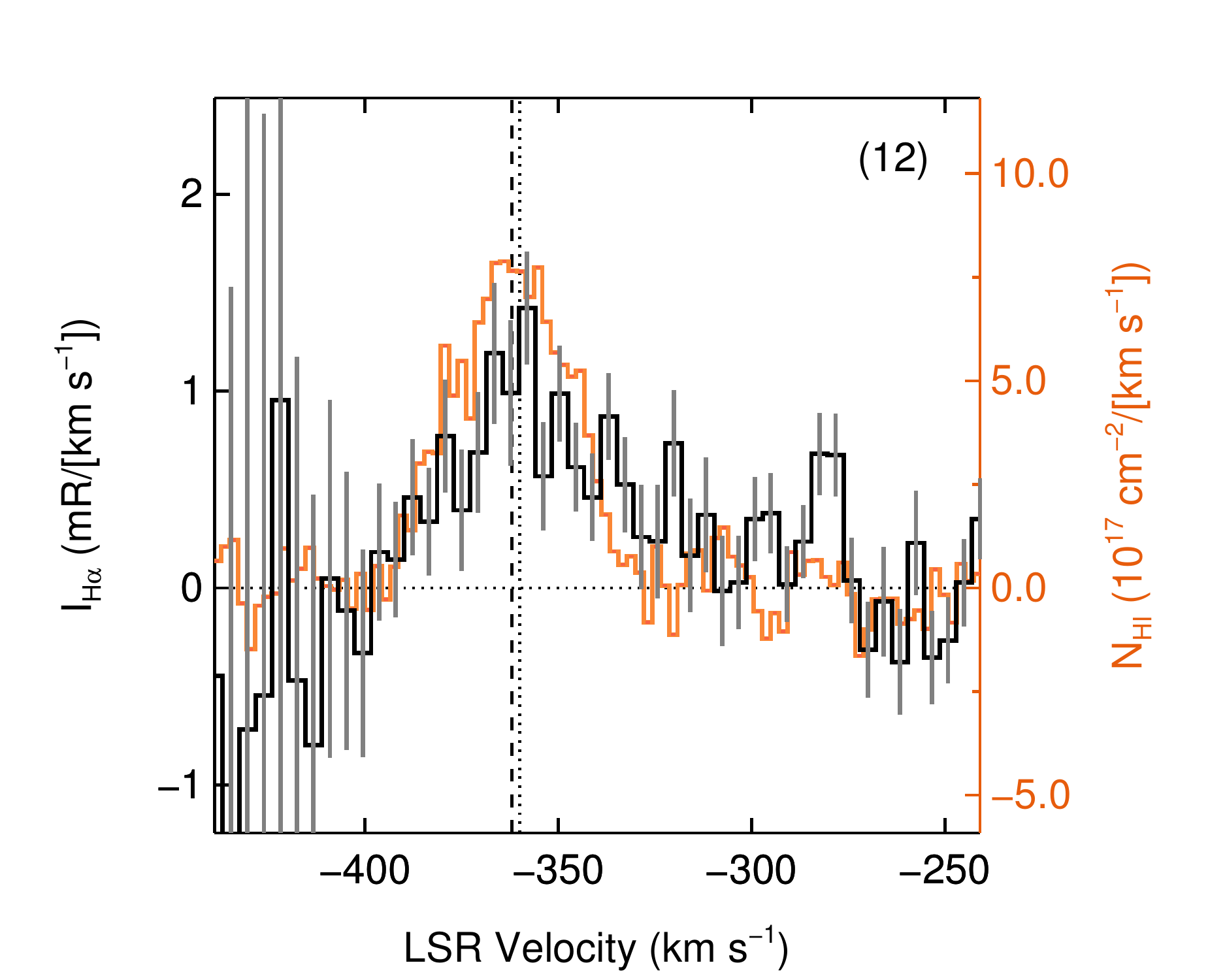} 
\end{center}
\end{figure*}

\begin{figure*}
\begin{center}
\includegraphics[trim=0 38 65 30,clip,scale=0.45,angle=0]{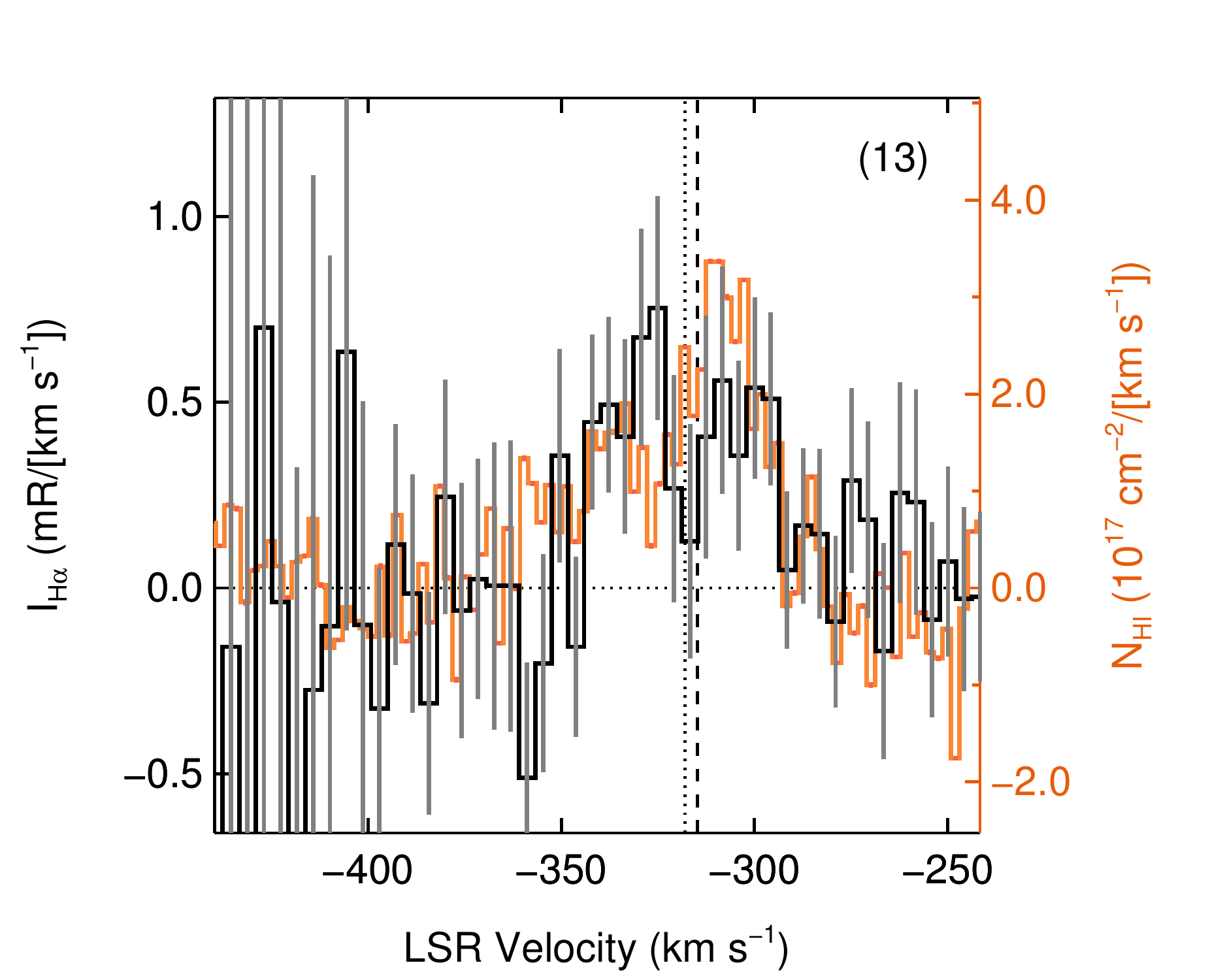}\includegraphics[trim=45 38 0 30,clip,scale=0.45,angle=0]{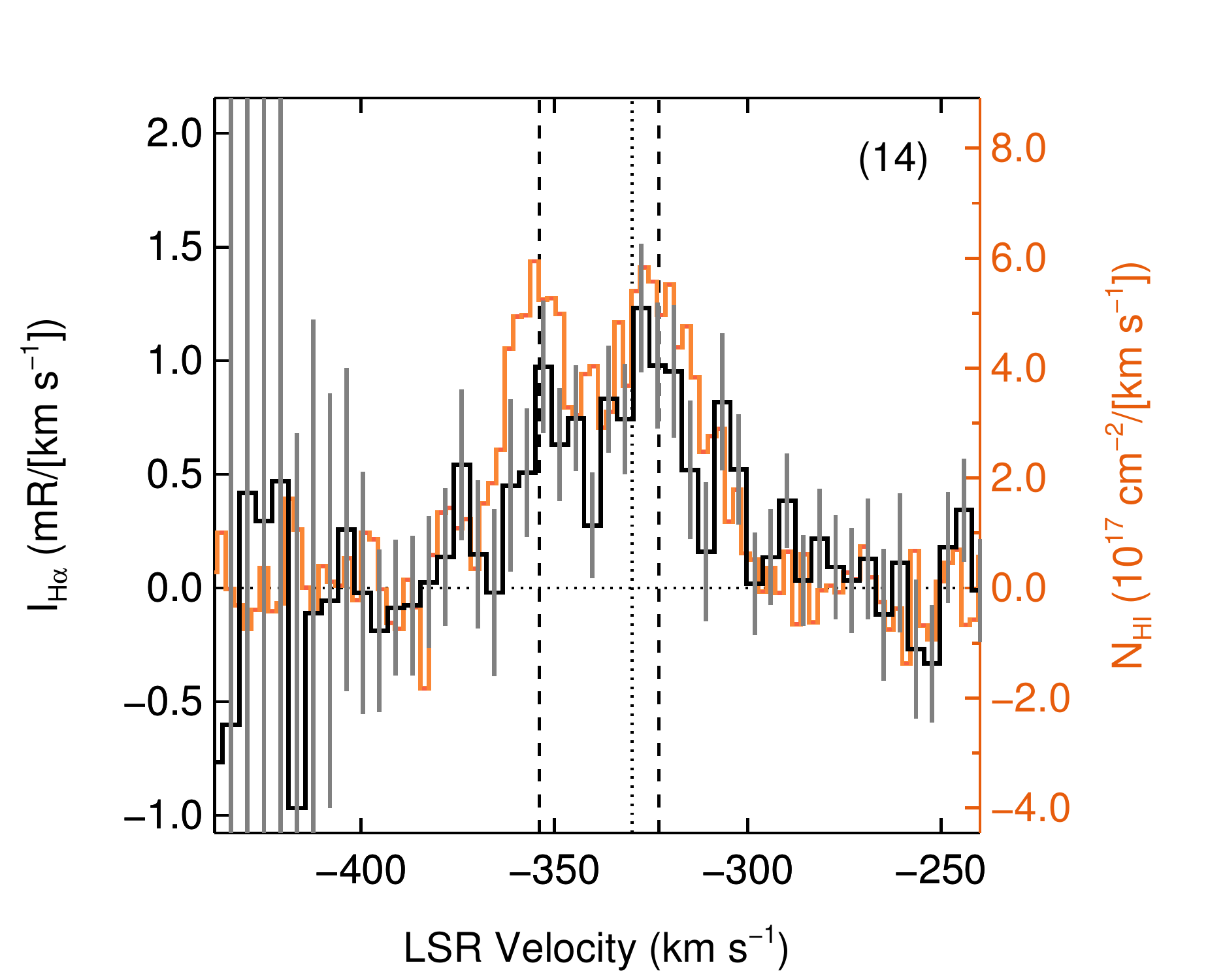} \\
\includegraphics[trim=0 38 65 30,clip,scale=0.45,angle=0]{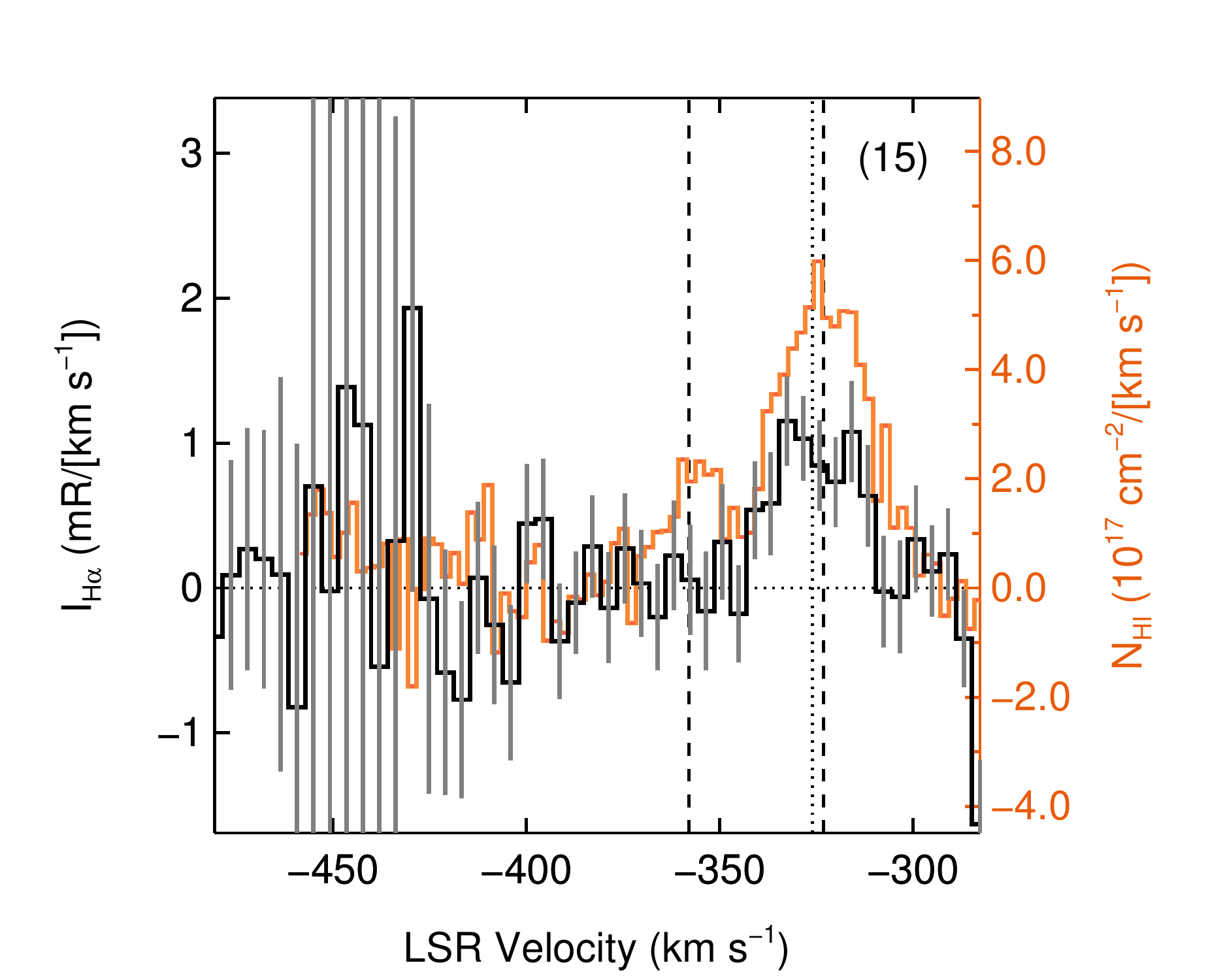}\includegraphics[trim=45 38 0 30,clip,scale=0.45,angle=0]{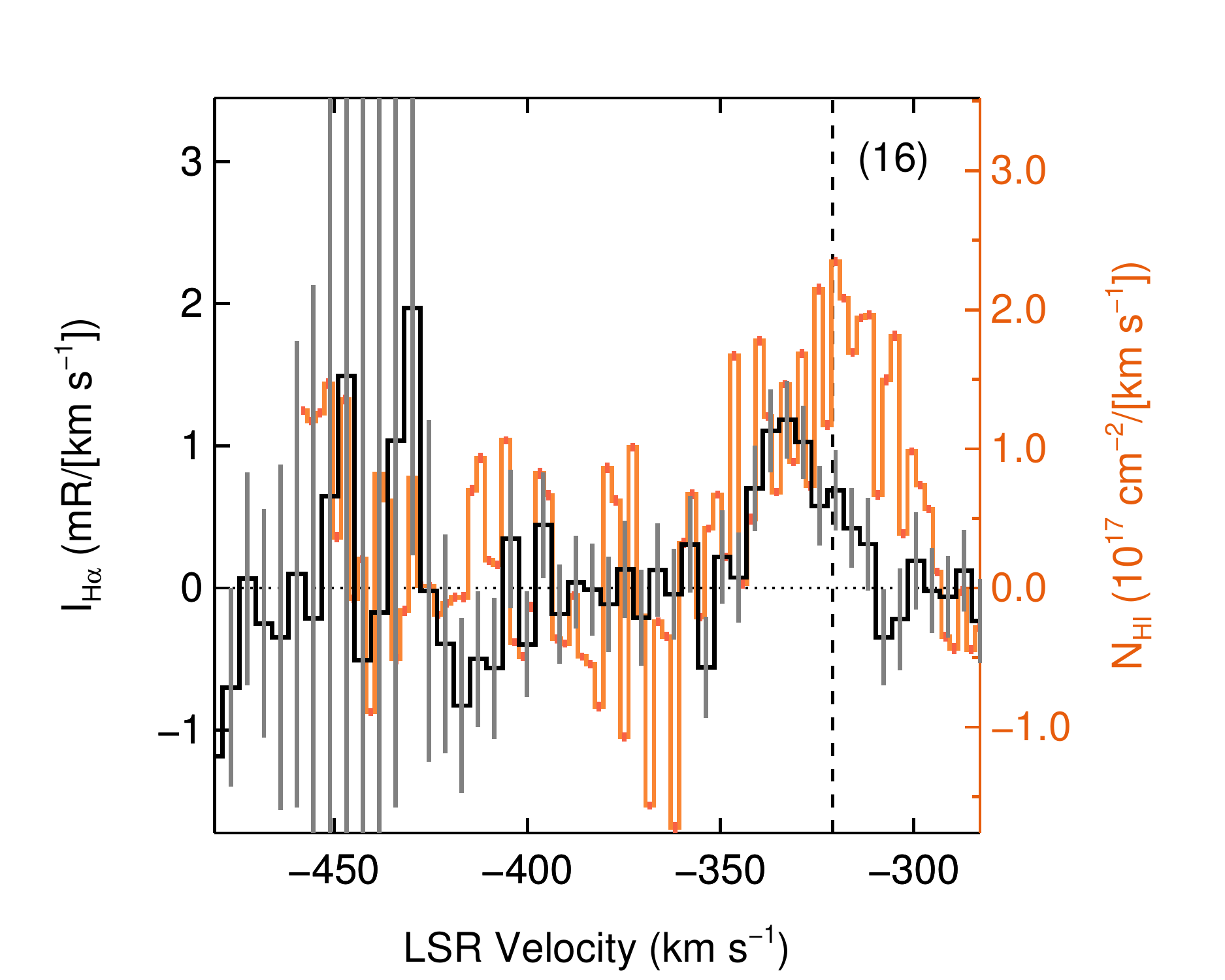} \\
\includegraphics[trim=0 38 65 30,clip,scale=0.45,angle=0]{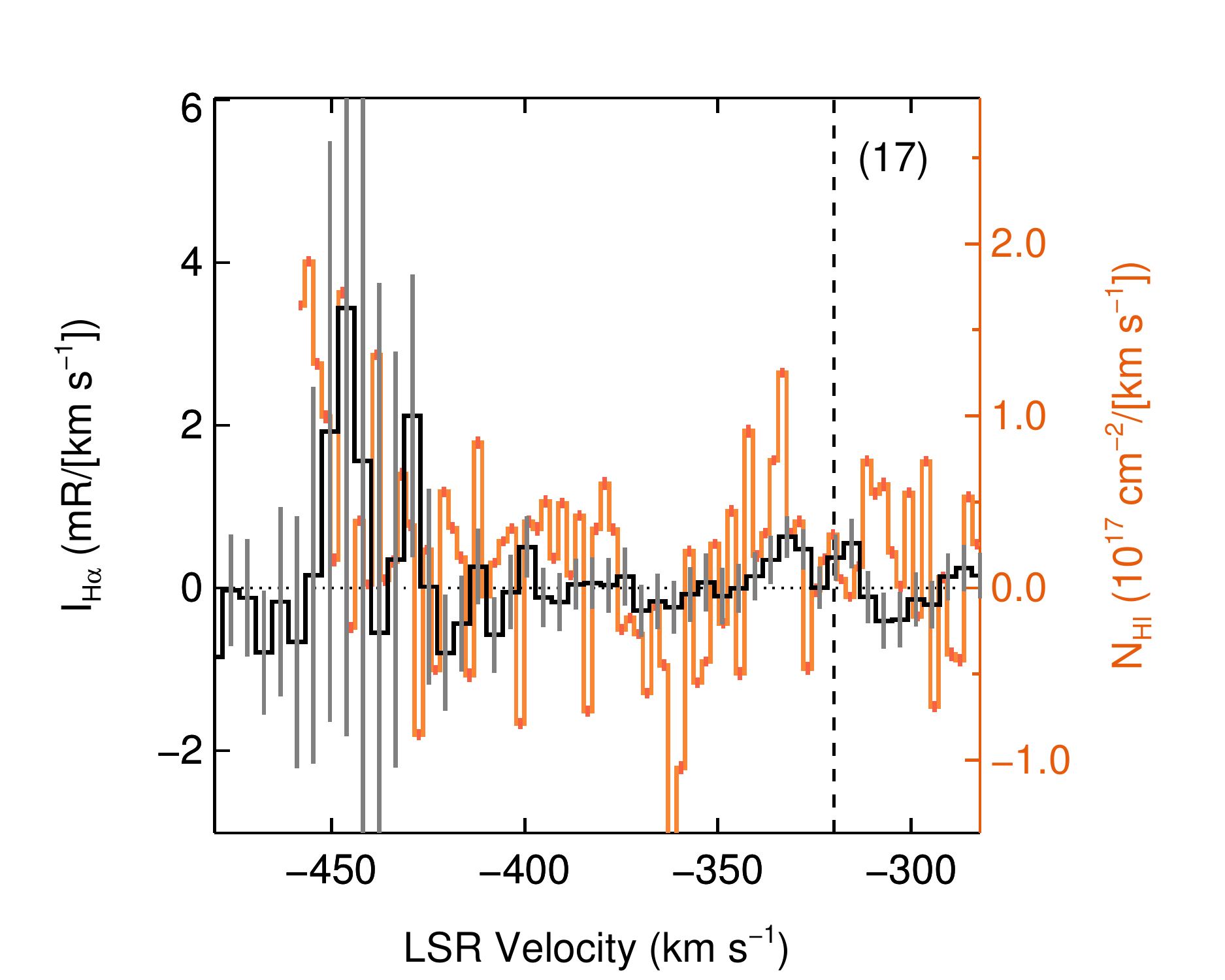}\includegraphics[trim=45 38 0 30,clip,scale=0.45,angle=0]{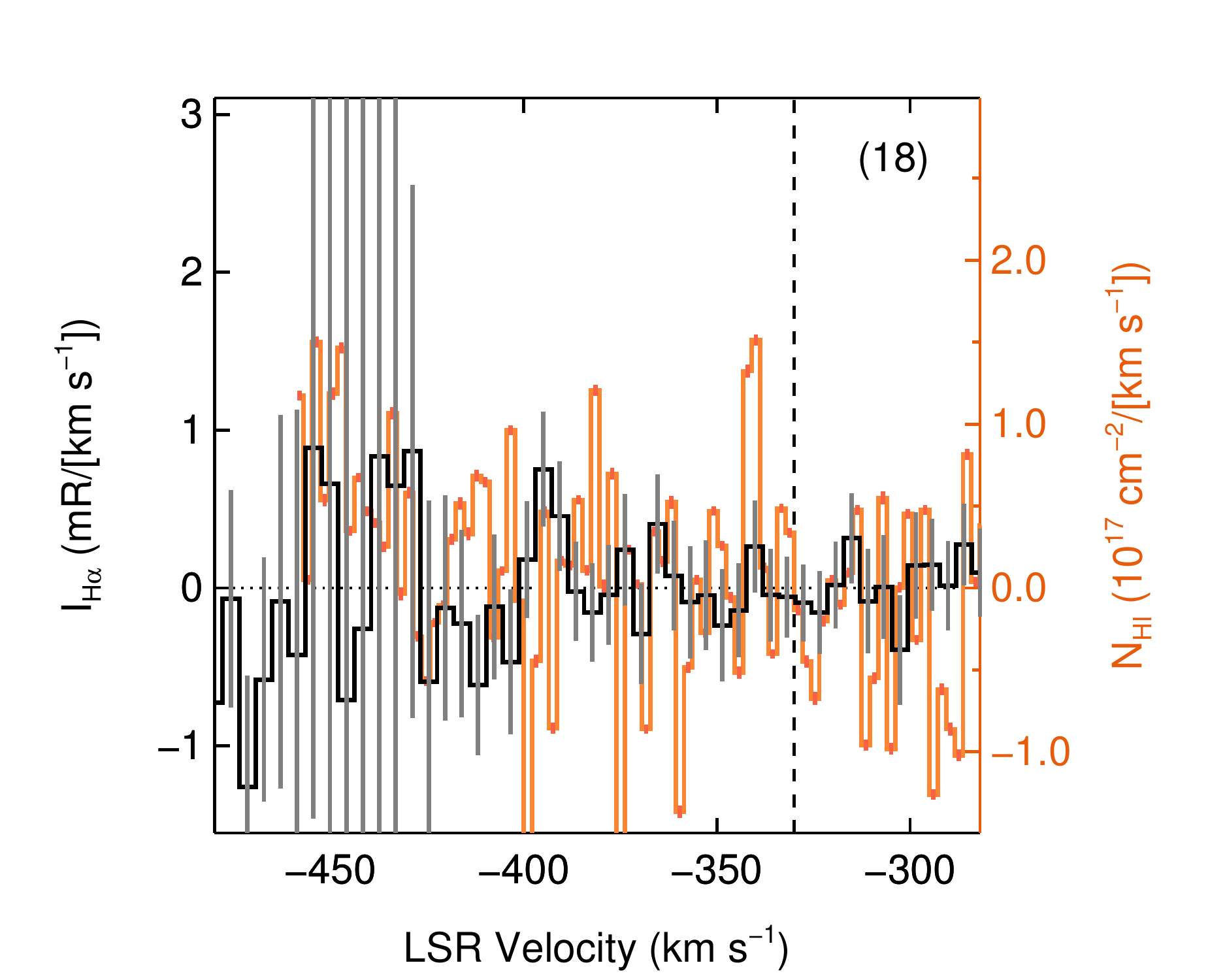} \\
\includegraphics[trim=0   0 0 30,clip,scale=0.45,angle=0]{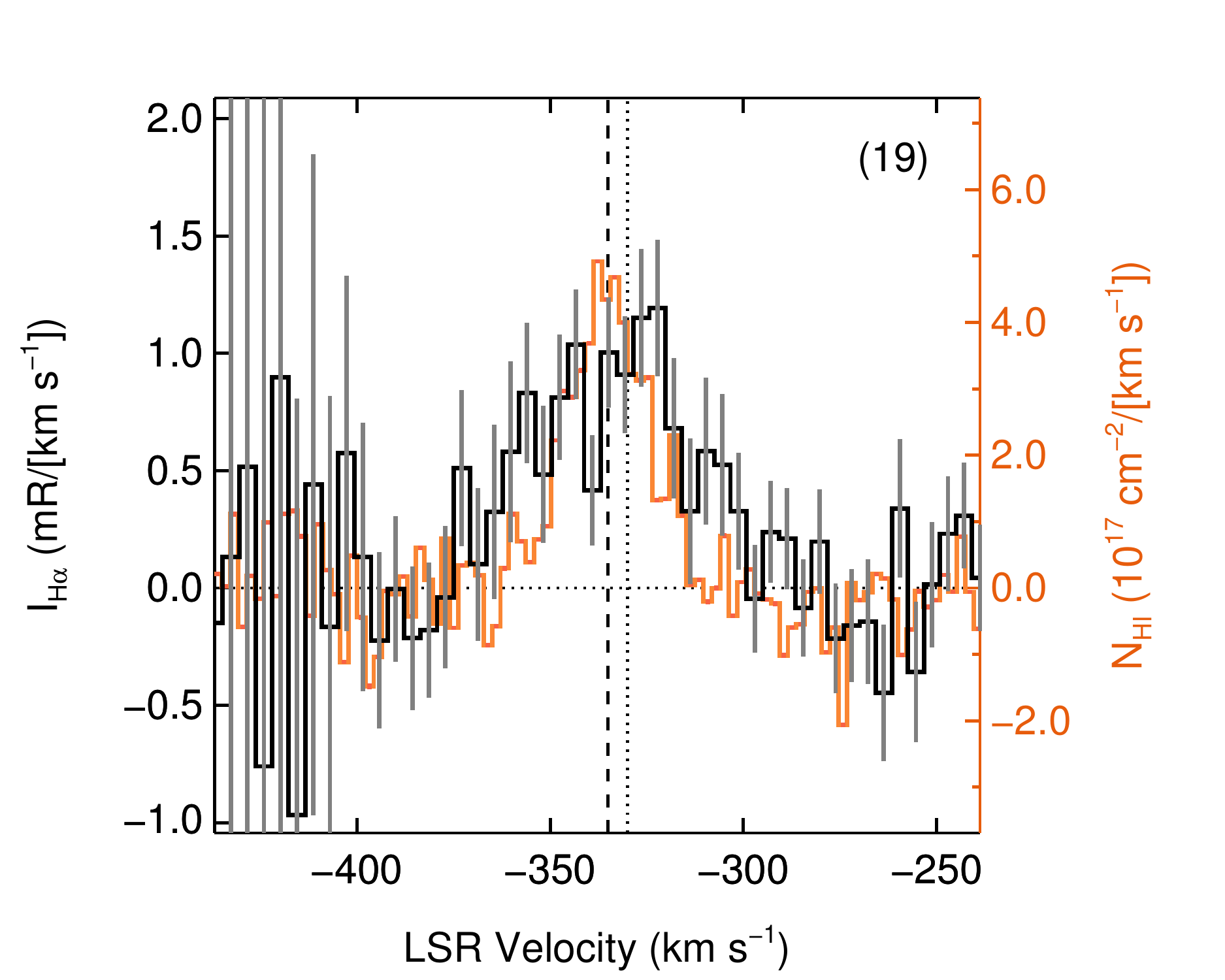}
\end{center}
\end{figure*}

\end{document}